\documentclass[a4paper,oneside,12pt]{article}
\usepackage{setspace}
\usepackage{amsthm}
\usepackage{amsfonts}
\usepackage{amssymb}
\usepackage{amsmath}

\usepackage[dvipsnames,table]{xcolor}
\usepackage[colorlinks]{hyperref}
\definecolor{NavyBlue}{HTML}{000080}

\usepackage{longtable}

\usepackage{pdflscape}
\usepackage{booktabs, siunitx}
\usepackage[tableposition=above]{caption}
\usepackage{tabularx}
\usepackage{enumitem}

\usepackage{lmodern}

\usepackage{sectsty}
\sectionfont{\centering}
\usepackage{dblfloatfix}
\usepackage{array}

\usepackage[demo]{graphicx}
\usepackage{booktabs, makecell, tabularx}
\usepackage{rotating}
\usepackage[export]{adjustbox}

\usepackage{booktabs}
\newcommand{\ra}[1]{\renewcommand{\arraystretch}{#1}}

\usepackage{bookmark}
\bookmarksetup{numbered}

\usepackage{adjustbox}

\DeclareSymbolFont{matha}{OML}{txmi}{m}{it}% txfonts

\DeclareMathSymbol{\varv}{\mathord}{matha}{118}

\usepackage{eurosym}
\usepackage[utf8]{inputenc}
\usepackage{bbm}
\usepackage{url}
\usepackage[stable]{footmisc}
\usepackage{t1enc}
\usepackage{supertabular}
\usepackage{lscape}
\usepackage[dvips]{epsfig}
\usepackage{epsf} 
\usepackage{fancyhdr}
\usepackage{caption}
\usepackage{setspace}
\usepackage{dsfont}
\usepackage[T1]{fontenc}
\usepackage{verbatim}
\usepackage{bbding}
\usepackage{mathrsfs}
\usepackage{bm}
\usepackage{pst-node}
\usepackage{pst-plot}
\usepackage{bigdelim}
\usepackage{multirow}
\usepackage{caption}
\usepackage{pdfsync}
\usepackage{booktabs}
\usepackage[flushleft]{threeparttable}
\usepackage[toc,page]{appendix}

\usepackage[
citestyle=authoryear-comp,
bibstyle=authoryear,
maxbibnames=5,		% maximum number of names printed in bibliography before truncation with ``et al.'' is used
minbibnames=1, 		% number of authors displayed if truncation happens
maxcitenames=2,		% maximum number of names printed in citation before et al. is used
mincitenames=1,		% number of authors displayed if truncation happens
datezeros=false,	% no leading 0 if dates are printed
date=long,
natbib=true,			% enable natbib-compatibility
backend=bibtex		
]{biblatex}
\renewbibmacro*{cite}{%
  \printtext[bibhyperref]{%
    \iffieldundef{shorthand}
      {\ifthenelse{\ifnameundef{labelname}\OR\iffieldundef{labelyear}}
         {\usebibmacro{cite:label}%
          \setunit{\printdelim{nonameyeardelim}}}
         {\printnames{labelname}%
          \setunit{\printdelim{nameyeardelim}}}%
       \usebibmacro{cite:labeldate+extradate}}
      {\usebibmacro{cite:shorthand}}}}

\renewbibmacro*{citeyear}{%
  \printtext[bibhyperref]{%
    \iffieldundef{shorthand}
      {\iffieldundef{labelyear}
         {\usebibmacro{cite:label}}
         {\usebibmacro{cite:labeldate+extradate}}}
      {\usebibmacro{cite:shorthand}}}}

\renewbibmacro*{textcite}{%
  \ifnameundef{labelname}
    {\iffieldundef{shorthand}
       {\printtext[bibhyperref]{%
          \usebibmacro{cite:label}}%
        \setunit{%
          \global\booltrue{cbx:parens}%
          \printdelim{nonameyeardelim}\bibopenparen}%
        \ifnumequal{\value{citecount}}{1}
          {\usebibmacro{prenote}}
          {}%
        \printtext[bibhyperref]{\usebibmacro{cite:labeldate+extradate}}}
       {\printtext[bibhyperref]{\usebibmacro{cite:shorthand}}}}
    {\printtext[bibhyperref]{\printnames{labelname}}%
     \setunit{%
       \global\booltrue{cbx:parens}%
       \printdelim{nameyeardelim}\bibopenparen}%
     \ifnumequal{\value{citecount}}{1}
       {\usebibmacro{prenote}}
       {}%
     \usebibmacro{citeyear}}}

\renewbibmacro*{cite:shorthand}{%
  \printfield{shorthand}}

\renewbibmacro*{cite:label}{%
  \iffieldundef{label}
    {\printfield[citetitle]{labeltitle}}
    {\printfield{label}}}

\renewbibmacro*{cite:labeldate+extradate}{%
  \printlabeldateextra}

\usepackage{hyperref}
\hypersetup{
  hyperfootnotes=true,
  colorlinks   = true,
  allcolors    = NavyBlue,
  linkcolor    = NavyBlue, 
}

\usepackage{geometry}
\usepackage{subfigure}
\usepackage{pgfplots}
\usepackage{tikz}
\usetikzlibrary{arrows.meta}
\usetikzlibrary{bending}

\tikzset{%
  my node/.style={rectangle,sloped,fill=black,minimum size=0mm,inner sep=0pt},
  draw my path/.style={draw,bend angle=90,thick,#1},
}

\NewDocumentCommand{\statcirc}{ O{#2} m }{%
    \begin{tikzpicture}
    \fill[#2] (0,0) circle (1.0ex); % Fill circle with base colour (arg#2)
    \fill[#1] (0,0) -- (180:1ex) arc (180:0:1ex) -- cycle; % Fill a half circle filled with second colour (arg#1), if specified
    \end{tikzpicture}
}

\usepackage{comment}
\newcommand*{\matminus}{%
  \leavevmode
  \hphantom{0}%
  \llap{%
    \settowidth{\dimen0 }{$0$}%
    \resizebox{1.1\dimen0 }{\height}{$-$}%
  }%
}

\usepackage{multirow}

\newcommand{\qedd}{\nobreak \ifvmode \relax \else
      \ifdim\lastskip<1.5em \hskip-\lastskip
      \hskip1.5em plus0em minus0.5em \fi \nobreak
      \vrule height0.75em width0.5em depth0.25em\fi}

\usepackage{mathtools}
\usepackage{rotating}

\DeclareMathOperator*{\argmin}{arg\,min}

\DeclareMathOperator{\E}{\mathbb{E}}
\newcommand{\Var}{\mathbb{V}}

\DeclareMathOperator\MSE{MSE}
\DeclareMathOperator*{\minimize}{minimize}
\DeclareMathOperator*{\maximize}{maximize}
\DeclareMathOperator{\sign}{sign}

\newcommand{\norm}[1]{\left\lVert#1\right\rVert}

\newtheorem*{assumption*}{\assumptionnumber}
\providecommand{\assumptionnumber}{}


\usepackage{changepage}

\begin{document}
\title{
{\bf Optimized Inference in \\ Regression Kink Designs} \\
\vspace{0.5cm} }

\linespread{1.2} 

\author{\large {\bf Majed Dodin}
\thanks{First Version: Januar 5, 2020. This Version: \today. I want to thank my supervisor Christoph Rothe for his patient guidance throughout this project. The author acknowledges support by the state of
Baden-Württemberg through bwHPC and the German Research Foundation (DFG) through grant INST 35/1134-1 FUGG.
Contact information: Department of
Economics, University of Mannheim, 68131 Mannheim, Germany, email: majed.dodin@gess.uni-mannheim.de. }}
\date{}

\renewcommand\footnotemark{}

\maketitle

\linespread{1.5} 
\vspace*{-0.5cm}
\begin{abstract}
We propose a method to remedy finite sample coverage problems and improve upon the efficiency of commonly employed procedures for the construction of nonparametric confidence intervals in regression kink designs.
The proposed interval is centered at the half-length optimal, numerically obtained linear minimax estimator over distributions with Lipschitz constrained conditional mean function. Its construction ensures excellent finite sample coverage and length properties which are demonstrated in a simulation study and an empirical illustration. Given the Lipschitz constant that governs how much curvature one plausibly allows for, the procedure is fully data driven, computationally inexpensive, incorporates shape constraints and is valid irrespective of the distribution of the assignment variable. 
\end{abstract}
\linespread{1.5}

\vspace{-1ex} \hspace{2.8cm}
\normalsize
\newpage

\section{Introduction}

%% Attractiveness of discontinuity designs

Research designs that warrant a causal interpretation of model parameter estimates based on observational data are becoming increasingly popular in the empirical social sciences. The quality of such observational studies is often assessed in terms of how well the assumptions that allow for the estimation of counterfactual quantities can be justified. In this context, methods that exploit discontinuities arising naturally from institutional rules are particularly attractive, as they allow the researcher to be precise about the source of variation utilized in the estimation of the parameter and discuss the assumptions that allow for a causal interpretation.

%% Kink Designs // Relevance

The Regression Kink Design (RKD) can be employed to estimate causal parameters when the variable of interest is a kinked function of an assignment variable.
Analogous to the better-known Regression Discontinuity Design (RDD),  which estimates the effect of a variable that changes its level discontinuously at a threshold, kink designs utilize discontinuous changes in the slope of the policy variable, effectively exposing units on each side of the threshold to different incentives. Such kinks arise naturally whenever marginal rates change discontinuously or benefit formulas involve maxima or minima and can be exploited to address important questions that are often difficult to be addressed experimentally, e.g. questions regarding the optimal design of unemployment insurance policies (see \cite{card2015effect}; \cite{landais2015assessing}; \cite{kolsrud2018optimal}).

%% Kink Design // Example 

For example, a common feature of unemployment insurance systems is that unemployment benefits, typically a function of income in some base period, are capped at some maximum level. Consequently, the incentives for reemployment differ at each side of the threshold defined by the cap, and the (local) causal effect of the benefit level on unemployment duration can be estimated by comparing how the duration outcome changes with prior income at each side of the cutoff relative to the size of the kink. In practice, this amounts to estimating the jump in the first derivative of the conditional expectation function (CEF) of the duration outcome given the assignment variable at the kink point, and dividing the estimate by the kink size. If unobserved confounders vary smoothly at the discontinuity point, this corresponds up to scale to the (local) elasticity of unemployment duration with respect to the benefit level, a key parameter in dynamic labor supply models and corresponding welfare optimal benefit formulas.

%% Kink Designs // Sufficient statistics
Since the welfare effects of policy changes can often be expressed in terms of elasticities, kink designs can be used in a sufficient statistics approach to policy evaluation, avoiding the need for parametric assumptions and the estimation of structural primitives of the model, combining credible identification with the ability to make welfare predictions \citep{chetty2009sufficient}.

%% Smoothing bias
\noindent
A practical challenge for researchers utilizing discontinuity designs is to choose estimation and inference procedures that preserve the credibility of their findings stemming from the nonparametric identification of the causal parameter.
The standard approach to inference in regression kink designs relies on local polynomial regression, that is fitting a polynomial model of order $p \geq 1$ using only observations within a prespecified window of length $2h$ around the threshold $c$ by weighted least squares. The theoretical properties of local polynomial estimators have been studied extensively (e.g. \cite{ruppert1994multivariate}; \cite{fan1996local}; \cite{fan1997local}) and it is well understood that if the true CEF differs from a polynomial of order $p$ on $[c-h,c+h]$, the resulting estimator for the kink parameter is generally biased. The two standard approaches to inference in regression kink designs recognize this by combining an asymptotic normality result with an argument that adresses this smoothing bias.

%% Undersmoothing	

One approach is to choose the bandwidth $h$ sufficiently small such that for a given sample size $n$ the resulting bias is (hopefully) negligible relative to the estimator's standard deviation.
This strategy is referred to as undersmoothing in the nonparametric regression literature and uses plug-in estimates of bandwidth sequences that shrink faster than the asymptotic mean squared error (MSE) optimal sequence, eliminating the bias from the asymptotic approximation invoked for inference. In practice, the bandwidth is chosen by multiplying a regularized estimate of the pointwise, that is evaluated at the unknown true CEF, asymptotic MSE optimal bandwidth by $n^{- \delta}$ for some small $\delta > 0 $ (cf. \cite{imbens2008regression}).

%% RBC

A second approach relies on estimating the leading bias term using a higher order polynomial and constructing the interval around a bias corrected point estimate, taking into account the additional variance introduced by bias estimation in the asymptotic approximation. This approach is referred to as robust bias correction (\cite{calonico2014robust}) and is implemented using a pilot bandwidth tuned for bias estimation and a regularized estimate of the pointwise asymptotic MSE-optimal bandwidth for confidence interval (CI) construction.

%% Undercoverage of pointwise methods

In a recent paper, \citet{armstrong2020simple} demonstrate that both default approaches to nonparametric confidence interval construction can lead to severe undercoverage in finite samples when implemented using bandwidth selectors justified by pointwise asymptotics, as is common. They attribute this finding to the pointwise statistical guarantees underlying these procedures in general, and in particular to the fact that, irrespective of the global curvature of the CEF, the pointwise asymptotic MSE-optimal bandwidth selector can be arbitrarily large if the \linebreak $(p+1)$-th derivative of the CEF is close to zero at the threshold, resulting in large bandwidth choices for potentially highly nonlinear functions. While this problem was previously recognized, the regularization terms that are added to prevent the empirical bandwidth selectors from selecting too large bandwidths are sensitive to ad-hoc choices of tuning parameters that drive the finite sample coverage rates of the resulting intervals. As a solution, the authors propose to explicitly restrict the parameter space by placing a bound on the $(p+1)$-th derivative of the unknown CEF and select bandwidths according to a minimax criterion, avoiding the need for regularization and allowing for the construction of intervals that are honest, in the sense that they are valid uniformly over the considered parameter space. This approach is feasible, as the bound allows for the computation of the magnitude of the exact worst-case smoothing bias of the local polynomial estimator over the class of functions restricted by the bound, which is taken into account by adjusting critical values accordingly. It is shown that explicitly accounting for possible smoothing bias in this fashion can substantially sharpen inference relative to traditional methods.

%% Contribution

This paper proposes confidence intervals for regression kink designs that leverage this potential.  Our proposed method is optimal in the sense that it minimizes the interval length amongst all honest procedures that utilize linear estimators and thus improves upon the performance of intervals based on local polynomial regression in a minimax sense.  The efficiency gains in terms of the length of two-sided 95\% intervals relative to the uniform MSE-optimal local linear\footnote{We focus on local linear estimators as the relevant benchmark as they are a popular choice in applied work and have leading bias proportional to the second derivative of the unknown function.} intervals for continuous designs are approximately 6 percent, and increase when the assignment variable is discrete. It follows from the arguments in \citet{armstrong2020simple} that the efficiency gains relative to undersmoothing and robust bias-correction under valid bandwidth choices are even larger, which we demonstrate in a simulation study and an empirical illustration. Our results suggest that the efficiency gains relative to traditional approaches to nonparametric inference in regression kink designs are substantial. These improvements are particularly valuable for applied work that utilizes RKDs, as power concerns are common and existing procedures that take into account the smoothing bias often yield uninformative confidence intervals despite graphical evidence suggesting the existence of kink effects \citep{card2017regression}. While optimally tuned honest intervals based on local polynomial regression attain on par efficiency in continuous designs, the optimization approach to inference allows us to flexibly incorporate shape constraints that can further shrink the confidence set. We extend the optimization approach to fuzzy designs by employing the inversion strategy proposed in \citet{noack2019bias} and state conditions that ensure that the optimized linear confidence intervals are honest in the sense of \citet{li1989honest}. Finally, we illustrate the utility of the procedure in the context of inference on the elasticity of unemployment duration with respect to the benefit level and provide software that implements the procedure.

%% Tuning
 
  The key tuning parameter of our proposed method is the bound on the second derivative of the CEF, which governs how much curvature one plausibly allows for. In contrast to undersmoothing and robust bias correction, our method requires that this bound is specified explicitly. However, as pointed out in \citet{armstrong2020simple}, these methods can not factually avoid this choice, as they must also implicitly restrict a derivative of order two or higher in order to maintain coverage over a given nonparametric class of functions.  Once this bound is specified, the proposed method is fully data driven and avoids the choices associated with local polynomial regression regarding the polynomial order, the kernel or the bandwidth. These advantages come at the cost of a closed-form expression for the estimator that we center our interval around, which is defined as the solution to a linear minimax problem over the space of distributions defined by the bound and obtained by numerical convex optimization.

% Linear Minimax RBCs / Related literature II
% \citet{donoho1991geometrizing} and
The study of linear minimax estimators for nonparametric regression problems (among other problems) goes back to \citet{donoho1994statistical}, who showed that, under broad conditions, the ratio of the linear minimax risk to the general minimax risk is bounded by 1.25, and that the minimax linear estimator for linear functionals over convex parameter spaces can be obtained via convex optimization. In the context of inference in discontinuity designs, \citet{armstrong2018optimal} first applied this result to RDDs over a class of functions proposed by \citet{sacks1978linear} that restricts the approximation error of a Taylor expansion about the discontinuity point\footnote{In this class, the minimax linear estimator can be obtained in closed form. However, since discontinuity designs are predicated on the assumption of continuity of the CEF away from the threshold it is conceptually less appealing than the Hölder class considered in \citet{imbens2019optimized}, \citet{armstrong2020simple} and this paper. Moreover, permitting functions that are discontinous away from the threshold also leads to inference that is too conservative at smooth functions, as the worst-case bias is larger in this class.}. In the same context, \citet{imbens2019optimized} impose a bound on the second derivative and propose a numerical optimization strategy to construct honest RDD intervals based on the linear minimax estimator under homoskedasticity, emphasizing the advantages of direct optimization in settings with discrete assignment variables and multivariate assignment rules. 

The method proposed in this paper contributes to the methodological literature on inference in regression kink designs (\cite{calonico2014robust}; \cite{card2015inference}; \cite{ganong2018permutation}; \cite{noack2019bias}).
In implementing our method, we rely on the discretization strategy of \citet{imbens2019optimized}, which is employed to the dual of the linear minimax problem for kink estimation. Conceptually, this paper builds on the "bias-aware" approach to nonparametric inference in particular (\cite{armstrong2018optimal}; \cite{kolesar2018inference}; \cite{imbens2019optimized}; \cite{ignatiadis2019bias}; \cite{noack2019bias}; \cite{rambachan2019honest}; \cite{schennach2020bias}; \cite{armstrong2020simple}) and a large methodological literature on inference in discontinuity designs in general (cf. \cite{lee2010regression}; \cite{cattaneo2019practical}).

%% Outline
The remainder of this paper is structured as follows: Section 2 introduces our notation, sketches the identification result underlying regression kink designs and defines the parameter of interest. Section 3 formally states the objective of this paper, explains how the bound permits a bias characterization that facilitates the implementation of the procedure, states assumptions under which optimized linear confidence intervals attain honesty and discusses practical questions related to their implementation. In Section 4, we conduct a simulation study to investigate the performance of optimized linear intervals relative to a variety of methods based on local linear regression. In Section 5, we demonstrate the utility of our method in a sensitivity analysis to \citet{landais2015assessing}. Section 6 concludes. The Appendix contains proofs and additional results.

\section{Framework and Notation}

%% Notation

\paragraph{2.1 Setup.} We observe a random sample of $n$ independent pairs $(X_i,Y_i)$, where $Y_i \in \mathbb{R}$ is the outcome of interest and $X_{i} \in \mathbb{R}$ is the assignment variable for unit $i$. We write the nonparametric regression model as
\begin{equation*}
Y_i = \mu(X_i) + u_i \qquad \E[u_i|X_i] = 0 \qquad \Var[u_i|X_i] = \sigma_i^2 \qquad i=1,\dots,n, 
\end{equation*}
\noindent
so that $\mu(X_i) = \E[Y_i | X_i]$ denotes the conditional expectation function of the outcome given the assignment variable. The subscript $N$ indicates a column vector of length $n$ where the $i$-th element corresponds to unit $i$, such that e.g. $X_{N} = (X_{1},X_{2},\cdots, X_{n})^{T}$. We normalize the threshold $c$ to zero and define an indicator function $D(x) = \mathbbm{1}_{[x \geq 0]}$. For a general function $f(x)$ we write $f_{+}(x) = f(x) D(x)$ and $f_{-}(x) = f(x) (1-D(x))$ and denote the $j$-th derivative of $f(x)$ with respect to $x$ by $f^{j}(x)$, where $f^{0}(x)$ is understood to mean the function itself. Moreover, let $f^{j}_{\pm}(0) =  \lim_{ \pm x \downarrow 0}  f_{\pm}^{j}(x)$ so that in particular $\mu^{1}_{+}(0) = \lim_{x \downarrow 0} \mu^{1}_{+}(x)$ and $\mu^{1}_{-}(0) = \lim_{x \uparrow 0}  \mu^{1}_{-}(x)$. Let $R_{X}$ denote the support of $X_{i}$, i.e. the smallest closed set in $\mathbb{R}$ such that $\Pr \lbrace X_{i} \in R_{X} \rbrace =1$. We assume that $R_{X}$ covers zero, is bounded and denote by $\underline{x}$ and $\bar{x}$ its minimal and maximal element. Let $\mathcal{X}_{-} = [\underline{x},0)$ and $\mathcal{X}_{+} = (0,\bar{x}]$. We assume that the CEF $\mu$ is a member of the class
\begin{align*}
\mathcal{F}(L) = \lbrace f : | f_{\pm}^{1}(x) - f_{\pm}^{1}(x') | \leq L |x-x'|, (x,x') \in \mathcal{X}_{\pm} \rbrace, 
\end{align*}
\noindent
which formalizes the notion that $\mu$ is two times differentiable on either side of the threshold, potentially discontinuous at the threshold, and has second derivative bounded by $L$ uniformly over $\mathcal{X}= \mathcal{X}_{+} \cup \mathcal{X}_{-}$. The bound $L$ effectively governs how much curvature one allows for, as values of $L$ close to zero imply that the members of $\mathcal{F}(L)$ are close to linear, while larger values of $L$ allow for increasing amounts of curvature. Given $L$ and $\sigma_{N}$, the method that is proposed in this paper is fully data-driven and we will assume throughout the derivation that both are known. The choice of $L$ and estimation of $\sigma_{N}$ are discussed separately in Section 3.7.

Since conditional expectation functions are unique only over the support of the conditioning variable, we follow \citet{kolesar2018inference} in that, assuming $\mu \in \mathcal{F}(L)$ is understood to mean that there exist $\mu$ in $\mathcal{F}(L)$ such that $\Pr\lbrace\mu(X_i) = \E[Y_i|X_i] \rbrace=1$. While the canonical regression kink design is predicated on continuous assignment variables, our setup therefore does not assume the assignment variable to be of any specific type.  This is advantageous in settings in which we only have coarse measurements at our disposal, in particular if the support of the assignment variable does not contain an open neighborhood around the threshold. In such situations the bound $L$ allows for extrapolation that ensures meaningful partial identification of  $\mu^{1}_{+}(0)$ and $\mu^{1}_{-}(0)$, as discussed in Section 3.6.

%% Structural Model
% \footnote{The deterministic relationship between the typically endogenous assignment variable and the policy variable pose a problem for standard instrumental variable approaches to the estimation of $\tau_{RKD}$, as the required independence assumption is likely to be violated.}
\paragraph{2.2 Parameter of Interest.} Regression Kink Designs consider structural models of the form
\begin{align*}
Y = Y(T,X,E) \qquad T=T(X),
\end{align*}
\noindent
where the function $Y: \mathbb{R}^{3} \mapsto \mathbb{R}$ describes how the outcome is produced, the random variable $E \in \mathbb{R}$ aggregates unobserved influences potentially correlated with the assignment variable, and the policy function $T: \mathbb{R} \mapsto \mathbb{R}$ is differentiable away from a kink location normalized to zero. The parameter of interest is the average partial effect of the policy variable at the kink
\begin{align*}
\tau_{RKD} = \E \left[ \dfrac{\partial Y(T,X,E)}{\partial T} \middle| X = 0 \right],
\end{align*}
\noindent
which inherits its causal interpretation from the definition of $Y$. Let $f_{E|X}(e|x)$ denote the conditional probability density function of $E$ given $X=x$. Under regularity conditions, we can write the first derivative of the CEF in 
this framework as
\begin{align*}
\mu^{1}(x) = \E \left[ \dfrac{\partial Y(T,X,E)}{\partial T} \middle| X = x \right] T^{1}(x) + \E \left[ \dfrac{\partial Y(T,X,E)}{\partial X} \middle| X = x \right] + \E \left[ Y \dfrac{\partial \ln f_{E|X}(e|X)}{\partial X} \middle| X = x \right].
\end{align*}

\noindent
Since $T$ is kinked at zero, this decomposition implies that, if the average partial effect of the assignment variable and the distribution of unobservables are continuous at zero, $\tau_{RKD}$ is identified by
\begin{align*}
\tau_{RKD}  =  \dfrac { \mu_{+}^{1}(0) - \mu_{-}^{1}(0) }{ T_{+}^{1}(0) - T_{-}^{1}(0)}.
\end{align*}
\noindent
\citet{card2015inference} characterize models for which this is the case and discuss conditions under which $\tau_{RKD}$ is equivalent to the "treatment on the treated" parameter in \citet{florens2008identification} or the "local average response" parameter in \citet{altonji2005cross}, respectively. In the next section, we focus on the sharp RKD, that is we assume that the denominator of $\tau_{RKD}$ is known, such that inference on $\tau_{RKD}$ is solely concerned with the jump in the first derivative of $\mu(x)$,   $\theta = \mu^{1}_{+}(0) -  \mu^{1}_{-}(0)$, which we refer to as the kink parameter. In Appendix \ref{appendix:F}, we extend our method to fuzzy designs by using a strategy recently proposed by \citet{noack2019bias}.

\section{Optimized Honest Confidence Intervals}

\paragraph{3.1 Problem Statement.} We seek to construct efficient confidence intervals of the form $\mathcal{I}_{\alpha} = [ \underline{\theta}, \bar{\theta} ]$ which cover the kink parameter $\theta$ with at least probability $1-\alpha$ for some prespecified level $\alpha>0$ in large samples. Furthermore, we strengthen this requirement by demanding our confidence intervals to be honest in the sense of \citet{li1989honest} with respect to the class $\mathcal{F}(L)$
\begin{align*}
\underset{n \rightarrow \infty}{\lim \inf} \text {  } \underset{\mu \in \mathcal{F}(L)}{\inf}  \Pr \left[ \mu^{1}_{+}(0) - \mu^{1}_{-}(0) \in \mathcal{I}_{\alpha} \right]  \geq 1 - \alpha.
\end{align*}
\noindent
The uniform requirement imposed by honesty disciplines our inference in the sense that it requires us to specify and take into account plausible adversarial distributions in our asymptotic approximation. In the present setting, this means to specify $L$ and guarantee coverage for the worst-case function in $\mathcal{F}(L)$. Honesty ensures that, for any tolerance level $\eta$, we can find a sample size $n_{\eta}$ such that for $n>n_{\eta}$ coverage of $\mathcal{I}_{\alpha}$ is above $1-\alpha -\eta$ for all $f \in \mathcal{F}(L)$. As discussed in \citet{armstrong2020simple}, the requirement to explicitly specify $L$ is not a disadvantage of uniform procedures, as methods that rely on pointwise guarantees of the type
\begin{align*}
\text{for every $f \in \mathcal{F}$, } \underset{n \rightarrow \infty}{\lim \inf} \Pr \left[ \mu^{1}_{+}(0) - \mu^{1}_{-}(0) \in \mathcal{I}_{\alpha} \right] \geq 1 - \alpha,
\end{align*}
\noindent
must implicitly restrict $L$ to justify that coverage is controlled over a given function class $\mathcal{F}$. This holds true, irrespective of the regularization problem that such procedures need to solve.
%\footnote{\citet{armstrong2020simple} point out that a undersmoothed bandwidth sequence implicitly defines a sequence of smoothness constants such that coverage is controlled and that robust bias correction can be considered a form of undersmoothing} 

\paragraph{3.2 General Approach.}

We center the confidence interval around a linear estimator of the form $\hat{\theta} = \sum_{i=1}^{n} w_{i}Y_i$, with weights $w_{N}$ optimized for a uniform criterion. The estimator is linear in the sense that the weights depend only on $X_N$ and a non-random tuning parameter $\kappa>0$, which we keep implicit in our notation. This choice is motivated by the relative minimax-efficiency result in \citet{donoho1994statistical} and the fact that the mean squared error of linear estimators, and related quantities governing the length of our confidence intervals, depend on the unknown function only through the bias. In order to ensure honesty of the enclosing interval, we compute the magnitude of the exact worst-case conditional bias $\bar{B}(w_{N}) = \sup_{ \mu  \in \mathcal{F}(L)} \E [ \hat{\theta} - \theta | X_{N}]$ of the estimator over $\mathcal{F}(L)$ during the optimization, and inflate the interval width accordingly. Following \citet{imbens2019optimized}, we obtain the estimate for the kink parameter by numerical convex optimization, solving a version of the linear minimax problem under known variance $\sigma_{N}^{2}$
\begin{align}
\min_{w_{N} \in \mathbb{R}^{n}}  \sum_{i=1}^{n} w_i ^2 \sigma_i^2 + \kappa \bar{B}(w_{N})^2  \qquad  \bar{B}(w_{N}) = \sup_{ \mu  \in \mathcal{F}(L)} \left\{  \sum_{i=1}^{n} w_{i} \mu(X_i) - (\mu^{1}_{+}(0) -  \mu^{1}_{-}(0)) \right\}. \label{qp1}
\end{align}
\noindent
Note that, since the class $\mathcal{F}(L)$ is symmetric with respect to zero, we do not need an absolute value inside the supremum, as the worst-case negative and positive biases over $\mathcal{F}(L)$ have the same magnitude. For $\kappa=1$, the objective function thus corresponds to the uniform conditional MSE of $\hat{\theta}$ over $\mathcal{F}(L)$. 
In general, $\kappa$ governs the worst-case conditional bias-variance tradeoff and is either chosen to minimize the uniform MSE or the interval length. Since the solution of (\ref{qp1}) depends on the data only through $X_{N}$, the confidence interval obtained by optimizing the interval length over $\kappa$ provides the same statistical guarantee as those obtained for a fixed $\kappa$.

\paragraph{3.3 Bias Characterization.} In order to translate (\ref{qp1}) into a tractable optimization problem we rely on the restrictions defining $\mathcal{F}(L)$. For any $\mu\in \mathcal{F}(L)$ and $X_{i} \in R_{X}$ we can write 
\begin{align*}
\mu_{+}(X_i) = \mu_{+}(0) + \mu^{1}_{+}(0) X_i + R_{+}(X_i)\qquad  \mu_{-}(X_i) = \mu_{-}(0) + \mu^{1}_{-}(0) X_i + R_{-}(X_i),
\end{align*}

\noindent %$ R(x) \leq 0.5 L \sign(x) x^{2}$ 
where $R_{+}$ and $R_{-}$ denote the remainders of the expansions. Note that by definition $R(0)= 0$, $R^{1}(0) = 0$, and $|R^{2}(x)| \leq L$ for all $\mu \in \mathcal{F}(L)$ and $x \in \mathcal{X}$. The worst-case conditional bias $\bar{B}(w_{N})$ of a general linear estimator of $\theta$ over $\mathcal{F}(L)$ is
\begin{align*}
\sup_{ \mu  \in \mathcal{F}(L)} \left\lbrace \sum_{i=1}^{n} w_{i,+} \left[ \mu_{+}(0) + \mu^{1}_{+}(0) X_i + R_{+}(X_i) \right] + \sum_{i=1}^{n} w_{i,-} \left[ \mu_{-}(0) + \mu^{1}_{-}(0) X_i + R_{-}(X_i) \right] - \theta \right\rbrace.
\end{align*}

\noindent
Since the assumption $\mu \in \mathcal{F}(L)$ does not impose any constraints on $(\mu^{0},\mu^{1})$ at zero, this expression is infinite unless the following discrete moment conditions are satisfied
\begin{align*}
\sum_{i=1}^{n} w_{i,+} = 0 \qquad \sum_{i=1}^{n} w_{i,+} X_i = 1 \qquad \sum_{i=1}^{n} w_{i,-}= 0 \qquad  \sum_{i=1}^{n} w_{i,-} X_i = -1 .
\end{align*}
\noindent
As a consequence, any solution to (\ref{qp1}) must satisfy these constraints, a fact that we utilize in the implementation of our estimator. We refer to a linear estimator and the corresponding set of weights with associated $X_{N}$ as "of the correct order" if these constraints are met, in which case the conditional bias of $\hat{\theta}$ is given by the weighted sum of approximation errors of the Taylor approximation to $\mu_{\pm}$ near zero. Since $\mu \in \mathcal{F}(L)$ implies that $\mu^{1}$ is absolutely continuous, an integration by parts argument allows further characterization of the conditional bias under the constraints, yielding
\begin{equation*}
\E [ \hat{\theta} - \theta | X_{N}] = \sum_{i=1}^{n} w_{i} R(X_{i}) = \sum_{i=1}^{n} w_{i,+} \int_{0}^{X_i} \mu^{2}(t) (X_{i} -t) dt
- \sum_{i=1}^{n} w_{i,-}(X_i) \int_{X_i}^{0} \mu^{2}(t) (X_{i} -t) dt.
\end{equation*}
\noindent
Applying an argument based on Fubini's Theorem then obtains
\begin{align*}
\E [ \hat{\theta} - \theta | X_{N}] &= \int_{0}^{\infty} \mu^{2}(t) \sum_{i: X_{i} \in [t,\infty) } w_{i,+} (X_i - t) dt - \int_{-\infty}^{0} \mu^{2}(t) \sum_{i: X_{i} \in (-\infty,t]}w_{i,-} (X_i - t)dt.
\end{align*}
\noindent
This representation of the conditional bias characterizes the choice of an an adversarial nature that needs to pick $\mu$ out of $\mathcal{F}(L)$ in response to $(w_{N},X_{N})$ for weights that satisfy the above constraints. Let $\bar{w}(t) = D(t) \sum_{i: X_{i} \geq t }w_{i,+} (X_i - t) - (1-D(t)) \sum_{i: X_{i} < t} w_{i,-} (X_i - t).$
In this notation, the conditional bias is $\int_{\mathbb{R}} \mu^{2}(t)  \bar{w}(t) dt$, which an adversarial nature maximizes by setting $\mu^{2}(t) = \sign (\bar{w}(t)) L$, yielding  
\begin{align}
\bar{B}(w_{N}) = L \int_{\mathbb{R}} | \bar{w}(t)| dt. \label{eq2}
\end{align}

\noindent
It follows that the worst-case conditional bias of a linear estimator, if it is finite, is proportional to $L$ and that the worst-case function in $\mathcal{F}(L)$ at which it is attained is a quadratic spline with piecewise constant second derivative of magnitude $L$. Another consequence of (\ref{eq2}) is that, for any given set of weights of the correct order and bound $L$, the computation of $\bar{B}(w_{N})$ amounts to finding the roots of $\bar{w}(t)$. This has practical value, as the sign of $\bar{w}(t)$ is known for local polynomial estimators under regularity conditions, a fact we utilize in our estimation strategy.
%Appendix A contains a formal derivation of this result.

\paragraph{3.4 Estimation via Dual Optimization.} Our approach to implementing the optimized linear confidence intervals relies on a renormalization justified by the result in (\ref{eq2}) and the convexity of (\ref{qp1}). From equation (\ref{eq2}) it follows that for linear estimators with finite worst-case bias, it holds that  $\sup_{ \mu \in \mathcal{F}(L)} \E [ \hat{\theta} - \theta | X_{N}] = L  \sup_{ R \in \bar{\mathcal{F}}(1)} \sum_{i=1}^{n} w_{i} R(X_i)$ with $\bar{\mathcal{F}}(1)$ defined as
\begin{align*}
\bar{\mathcal{F}}(1) = \lbrace f : f(0)=f^{1}(0)=0 \wedge | f_{\pm}^{1}(x) - f_{\pm}^{1}(x') | \leq |x-x'|, (x,x') \in \mathcal{X}_{\pm} \rbrace.
\end{align*}
The class $\bar{\mathcal{F}}(1)$ can be understood as the class of remainder functions corresponding to the conditional expectation functions in $\mathcal{F}(1)$, which is reflected by the additional constraints that correspond to the aforementioned properties of remainders. The renormalization allows us to equivalently state the optimization problem defining $\hat{\theta}$ as follows. For a fixed $\kappa$, we write the primal problem as

\begin{equation}
\begin{aligned}
\minimize_{w_{N},r} \quad & \sum_{i=1}^{n} w_{i}^2 \sigma_{i}^{2} + \kappa L^{2} r^{2} \\
\textrm{subject to} \quad & \sup_{ R \in \bar{\mathcal{F}}(1)}  \left[ \sum_{i=1}^{n} w_i R(X_i) \right] \leq r\\
        &\sum_{i=1}^{n} w_{i,+} = 0 \qquad \sum_{i=1}^{n} w_{i,-} = 0 \\
& \sum_{i=1}^{n} w_{i,+} X_i = 1 \qquad \sum_{i=1}^{n} w_{i,-} X_i = - 1. 
\end{aligned}
\label{qp3}
\end{equation}

\noindent
The weights solving problems (\ref{qp1}) and (\ref{qp3}) are equivalent since, at the optimal value of $r$, the objective functions are identical and any candidate solution to (\ref{qp1}) lies in the feasible set for $w_{N}$ of (\ref{qp3}). This reformulation is helpful, as we require the optimal weights as well as a sharp uniform upper bound on $\E [ \hat{\theta} - \theta | X_{N}]$ to construct our confidence interval and, more importantly, rely on the additional constraints of $\mathcal{\bar{F}}(1)$ in our implementation. The Lagrangian of (\ref{qp3}) is given by
\begin{align*}
L(w_{N},r,\nu,\lambda) &= \sup_{ R \in \bar{\mathcal{F}}(1)} \sum_{i=1}^{n} w_{i}^2 \sigma_{i}^{2} + \kappa L^{2} r^{2} +\nu \left( \sum_{i=1}^{n} w_i R(X_i) - r \right)
+ \lambda_1 \left( \sum_{i=1}^{n} w_{i,+} \right) \\
&+ \lambda_2 \left( \sum_{i=1}^{n} w_{i,-} \right)
+ \lambda_3 \left( \sum_{i=1}^{n} w_{i,+} X_i -1 \right) + \lambda_4 \left( \sum_{i=1}^{n} w_{i,-} X_i +1 \right).
\end{align*}

\noindent
In order to solve (\ref{qp3}) we rely on a second equivalence result. In Appendix \ref{appendix:B}, we show that the local polynomial weights with corresponding worst-case conditional bias lie in the feasible set of (\ref{qp3}), implying that a refined Slater's condition applies to (\ref{qp3}).  As a consequence, strong duality holds and any primal optimal point is also a minimizer of $L(w_{N},r,\nu^{*},\lambda^{*})$ where $(\nu^{*},\lambda^{*})$ is the solution to the dual problem
\begin{align*}
\maximize_{\nu,\lambda} q(\nu,\lambda) = \inf_{w_{N},r} L(w_{N},r,\nu,\lambda) \\
\textrm{subject to} \qquad \nu \in \mathbb{R}^{1}_{+}, \lambda \in \mathbb{R}^{4}. \nonumber
\end{align*}
\noindent
In Appendix \ref{appendix:C}, we show that we can interchange the order of the infimum and the supremum in the dual objective by applying a minimax theorem, yielding an inner convex quadratic minimization problem that is solved analytically. This results in closed-form expressions for the primal parameters as functions of the dual parameters and a remainder function $R \in \bar{\mathcal{F}}(1),$
\begin{flalign}
w_i = \dfrac{\lambda_1 D(X_i)  + \lambda_2 (1-D(X_i))  + \lambda_3 D(X_i) X_i + \lambda_4 (1-D(X_i)) X_i  + \nu R(X_i)}{ - 2\sigma_i^2} \phantom{iii} r = \dfrac{ \nu}{2 \kappa L^{2}} && \label{eq4}
\end{flalign}

\noindent
as well as a simplified expression for the dual objective 
\begin{align*}
q(\nu,\lambda) = \sup_{ R \in \bar{\mathcal{F}}(1)} & - \dfrac{1}{4} \sum_{i=1}^{n}  \dfrac{ \left[ \lambda_1 D(X_i)  + \lambda_2 (1-D(X_i))  + \lambda_3 D(X_i) X_i + \lambda_4 (1-D(X_i)) X_i + \nu R(X_i)\right]^{2} }{\sigma_i^2}\\
& - \dfrac{1}{4} \dfrac{ \nu^{2} }{\kappa L^{2}}  - \lambda_3 + \lambda_4. 
\end{align*}

\noindent % Since the supremum over $\bar{\mathcal{F}}(1)$ is attained,
Let $(w^{*},r^{*}) = (w(\nu^{*},\lambda^{*}),r(\nu^{*},\lambda^{*}))$ for the element of $\bar{\mathcal{F}}(1)$ that attains the supremum in $q(\nu,\lambda)$. It follows from strong duality and strict convexity of $L(w,r,\nu^{*},\lambda^{*})$ that $(w^{*},r^{*})$ is the solution of (\ref{qp3}). As a consequence, we can recover the weights solving (\ref{qp1}) as well as the associated worst-case conditional bias over $\mathcal{F}(L)$ by the solution of the simplified dual problem 
\begin{alignat}{2}
&\maximize_{\nu,\lambda, R} \phantom{12} && - \dfrac{1}{4} \sum_{i=1}^{n}  \dfrac{ \left[ \lambda_1 D(X_i)  + \lambda_2 (1-D(X_i))  + \lambda_3 D(X_i) X_i + \lambda_4 (1-D(X_i)) X_i + \nu R(X_i)\right]^{2} }{\sigma_i^2} \nonumber\\
& && - \dfrac{1}{4} \dfrac{ \nu^{2} }{\kappa^{2} L^{2}}  - \lambda_3 + \lambda_4.  \nonumber\\
& \textrm{subject to} && \nu \in \mathbb{R}_{+}^{1}, \lambda \in \mathbb{R}^{4}, R \in \bar{\mathcal{F}}(1),  \label{qp5}
\end{alignat}

\noindent
in conjunction with the mapping (\ref{eq4}) between primal and dual parameters and the result (\ref{eq2}). This translates the primal problem of $n+1$ parameters into a problem over the space $\bar{\mathcal{F}}(1)$ and five dual parameters, which we solve numerically by discretization as described in Appendix \ref{appendix:D}.

\vspace*{-0.45cm}
\vspace{\baselineskip}
\noindent
\textit{Remark 1.} Abstracting from the constraint $R \in \bar{\mathcal{F}}(1)$, the dual problem (\ref{qp5}) is a standard quadratic program and the remaining challenge is to find a suitable approximation strategy to the the functional constraint. In our implementation, we approximate the function on an equidistant grid to permit approximation of the second order constraint via finite central differences, e.g. $R^{2}(x) = [R(x+h)-2R(x) + R(x+h)]/h^2 + O(h^2)$ (see Appendix \ref{appendix:D} for details). This approach is formally justified by Proposition 2 of \citet{imbens2019optimized}, which states that, for assignment variables with compact and convex support, the optimal weights can be recovered with arbitrary small $L_{2}$- error under the proposed discretization strategy.

\vspace*{-0.45cm}
\vspace{\baselineskip}
\noindent
\textit{Remark 2.} Under the proposed strategy, the optimization approach to bias-aware inference allows us to incorporate shape constraints in a simple fashion.  As explained in more detail in Appendices \ref{appendix:D} and \ref{appendix:F}, any additional constraint on the CEF that can be approximated in terms of finite differences of Taylor remainders can be utilized by modifying the feasible set of (\ref{qp5}).

\paragraph{3.5 Interval Construction.}

\noindent
Given a solution $(\nu^{*},\lambda^{*},R^{*})$ to (\ref{qp5}) for a fixed value of $\kappa$, we recover the optimal weights $w_{N} \gets w^{*}$ and the corresponding worst-case bias magnitude $\bar{B}(w_{N}) \gets r^{*}L$ via the mapping (\ref{eq4}) and the result (\ref{eq2}) to construct the optimized linear interval. Intuitively, the construction relies on the following decomposition of $\hat{\theta} - \theta$
\begin{align*}
\hat{\theta} - \theta = \underbrace{ \sum_{i=1}^{n} w_{i}R(X_i)}_{= \E[\hat{\theta} - \theta|X_{N}]} + \underbrace{\sum_{i=1}^{n} w_{i} u_{i}}_{ = \hat{\theta} - \E[\hat{\theta}|X_{N}]}.
\end{align*}

\noindent
By definition, $\E[\hat{\theta} - \theta|X_{N}]$ is bounded in absolute value by $\bar{B}(w_{N})= \sup_{ \mu \in \mathcal{F}(L)}E [ \hat{\theta} - \theta | X_{N}]$ uniformly over $\mathcal{F}(L)$. Let $s_{n}^{2} = \sum_{i=1}^{n} w_{i}^{2} \sigma_{i}^{2}$ denote the conditional variance of $\hat{\theta}$ given $X_{N}$ and define the conditional bias to standard deviation ratio $t_{n} = s_{n}^{-1} \E[\hat{\theta} - \theta|X_{N}]$. The uniform bound implies that the t-statistic
\begin{align*}
\dfrac{ \hat{\theta} - \theta}{s_{n}} = \dfrac{\E[\hat{\theta} - \theta|X_{N}]}{s_{n}} + \dfrac{ \hat{\theta} - \E[\hat{\theta}|X_{N}]}{s_{n}}
\end{align*}
\noindent
is the sum of a term that is bounded in absolute value by $\bar{t}_{n} = \bar{B}(w_{N}) s_{n}^{-1}$ uniformly over $\mathcal{F}(L)$ and a term that, under conditions stated in the next section, converges to a standard normal distribution uniformly over $\mathcal{F}(L)$ by a suitable central limit theorem. Provided that this is the case, it follows that an honest $(1-\alpha)$ confidence interval for $\theta$ is given by
\begin{align}
\mathcal{I}_{\alpha} = \left[ \hat{\theta} \pm s_{n} \text{cv}_{1-\alpha}(\bar{t}_{n})   \right], \label{eq6}
\end{align}

\noindent
where $\text{cv}_{1-\alpha}$ denotes the $(1-\alpha)$ quantile of the folded normal distribution $|N(\bar{t}_{n},1)|$ with mean $\bar{t}_{n}$ and variance one, that is the distribution of the absolute value of a normal distribution with mean $\bar{t}_{n}$ and variance one. Intuitively, this construction works because the bias can not be negative and positive at the same time, which implies that a hypothetical interval that adds and substracts $\bar{B}(w_{N}) + z_{1-\alpha/2} s_{n}$ from $\hat{\theta}$ would be too conservative. For the sharp\footnote{In fuzzy discontinuity designs, the strategy underlying the construction of the interval (\ref{eq6}) can also be employed in principle. However, in this case the smoothing bias of the first stage estimator must be dealt with and additional problems arise. In Appendix \ref{appendix:F}, we discuss how our implementation utilizes the strategy proposed in \citet{noack2019bias} to extend the optimization approach to fuzzy discontinuity designs.}
regression kink design,  an honest $(1-\alpha)$ interval for $\tau_{RKD}$ is then immediately obtained by rescaling the upper and lower ends of $\mathcal{I}_{\alpha}$ by the inverse of the magnitude of the kink.

We construct two types of optimized linear confidence intervals according to (\ref{eq6}): Uniform MSE-optimal ($\kappa_{UMSE}=1$) and length-optimal ($\kappa_{LE}= \argmin_{\kappa>0} s_{n} \text{cv}_{1-\alpha}(\bar{t}_{n})$) intervals. However, the same construction principle can be applied to any uniform performance criterion that specifies a worst-case bias-variance trade-off, as the guarantees of (\ref{eq6}) hold for any fixed $\kappa>0$.\\

\noindent
\textit{Remark 3.} In order to obtain the length-optimal interval, we search for the optimal value $\kappa_{LE}$ using a combination of golden section search and successive parabolic interpolation as implemented in standard derivative free optimization libraries. This is feasible at high accuracy in practice as the runtime of a single optimization iteration as implemented is low (approximately $0.129$ seconds for a sample size of $6000$), leading to an average total runtime of $3.59$ seconds for the same sample size on a standard desktop computer (see Appendix \ref{appendix:G} for more details). This could in principle be further improved by restricting attention to values of $\kappa$ smaller than the uniform MSE-optimal choice $\kappa=1$. This is because the length-optimal weights will "oversmooth" relative to the uniform MSE-optimal weights, which \citet{armstrong2020simple} show for estimators in their regularity class (cf. Figure 1 therein).

%If the optimal linear estimator, for some implicit bandwidth parameter, generally falls in the class of estimators treated in \citet{armstrong2020simple}

\paragraph{3.6 Theoretical Properties.}

In order to discuss the statistical properties of confidence \linebreak intervals constructed according to (\ref{eq6}) we impose the following assumptions.\\

\vspace*{-0.45cm}

\noindent
\textbf{Assumption 1} Let $(C,\delta,\sigma_{\min}, \sigma_{\max}) \in \mathbb{R}_{+}^{4}$ be some fixed vector.
\begin{enumerate}[leftmargin=*, labelsep=*, align=left, itemsep=-0.2cm, font=\normalfont, label=(\roman*)]
\item  $\lbrace Y_{i}, X_{i} \rbrace_{i=1}^{n}$ is an i.i.d. random sample of size $n$ from a fixed population.
\item  $\mu \in \mathcal{F}(L)$ for some $L>0$. 
\item  $0 < \sigma_{\min}^{2} \leq E[(Y_i - \mu(X_i))^{2}|X_{i} = x] \leq \sigma_{\max}^{2}$ for all $x\in R_{X}$ and $\mu \in \mathcal{F}(L)$. 
\item  $\E[|Y_i - \mu(X_i)|^{2+\delta} | X_i = x] \leq C$ for all $x\in R_{X}$ and $\mu \in \mathcal{F}(L)$.
\item  The solution $w_{N}^{*}$ satisfies\\  
  $ \phantom{(a)} \qquad \dfrac{ \max_{i} w_{i}^{2} }{ \sum_{i=1}^{n} w_{i}^{2}} \overset{P}{\rightarrow} 0.$
\end{enumerate}

\noindent
Assumption 1 is sufficient for a central limit theorem to apply to $s_{n}^{-1} w_{N}^{T}u_{N} =s_{n}^{-1}\left(\hat{\theta} - \E[\hat{\theta}|X_{N}]\right)$ uniformly over $\mathcal{F}(L)$ and ensures the consistency of $\hat{\theta}$ in the identified setting. Part (i) is the standard model for survey data and provides that the kink parameter is a well defined quantity. Part (ii) and (iii) ensure that the quadratic program defining $\hat{\theta}$ is strictly convex, which guarantees that the optimal weights are uniquely recovered by the dual optimal parameters, provided that the data contains atleast two distinct points on either side of the threshold.

Assumptions (iii) and (iv) guarantee the existence of and establish bounds on the second and $(2+\delta)$-th absolute conditional moment functions of the CEF error uniformly over the support of the assignment variable and the class of permitted CEFs. The two assumptions restrict the class of permitted distributions beyond the CEF constraint (ii) in that they require uniformly bounded and non-zero conditional variances as well as the existence of a strictly finite higher order moment function. Assumptions 1 (i)-(v) are sufficient to establish that Lyapunov's condition applies to each element of $\mathcal{F}(L)$, implying convergence of $s_{n}^{-1} w_{N}^{T}u_{N}$ to a standard normal variable uniformly over $\mathcal{F}(L)$.

Assumptions (v) restricts the limit behavior of the set of optimal weights and are difficult to derive from higher-level conditions. This is because, to the best of our knowledge, a closed form solution to (1) is not known and a general characterization of $w_{N}^{*}$ beyond the spline property derived in Section 3 is difficult. While this is unattractive from a theoretical point of view, the good news is that we can verify that the condition is approximately met in any given application, and our implementation reports the finite sample counterpart to part (v). Assumption (v) together with (iii) is sufficient for $s_{n}^{2} = o_{p}(1)$ uniformly over $\mathcal{F}(L)$ and implies consistency of $\hat{\theta}$ in the identified setting.

Roughly speaking, Assumption 1 rules out distributions such that, for some $X_{i} \in R_{X}$ and $\mu \in \mathcal{F}(L)$, in the limit $w_i u_i$ is "too large", in the sense that it dominates the behavior of the sequence $s_{n}^{-1}  w_{N}^{T} u_{N}$. This rules out that only a "small" proportion of the data is driving the estimate under this function. Appendix \ref{appendix:E} contains a formal discussion of how the relevant components of Assumption 1 can be used to show that Lyapunov's condition holds  conditionally on $X_{N}$ uniformly over $\mathcal{F}(L)$, which is the key ingredient in ensuring that the optimized interval attains honesty. Once uniform convergence of $s_{n}^{-1}[\hat{\theta} - \E[\hat{\theta}|X_{N}] \overset{D}{\rightarrow} N(0,1)$ is established, it follows from the definitions of the worst-case bias to standard deviation ratio $\bar{t}_{n}$ and the critical value $\text{cv}_{1-\alpha}$ that, for a standard normal random variable $Z\sim N(0,1)$, it holds uniformly over $\mathcal{F}(L)$ that
\begin{align*}
\underset{n \rightarrow \infty}{\lim \inf} \text {  }  \Pr \left( |Z+ t_{n}| \leq \text{cv}_{1-\alpha}(\bar{t}_{n}) | X_{n} \right) \geq 1 - \alpha.
\end{align*}
\noindent
Taken together, the two results imply that (\ref{eq6}) is honest, which we record in Proposition 1.\\

\noindent
\textbf{Proposition 1} Suppose that Assumption 1 holds. Then uniformly over $\mathcal{F}(L)$
\begin{equation*}
s_{n}^{-1}[\hat{\theta} - \E[\hat{\theta}|X_{N}]  \overset{D}{\rightarrow} N(0,1) \qquad \text{and} \qquad  t_{n} \leq \bar{t}_{n} = s_{n}^{-1} \bar{B}(w_{N}),
\end{equation*}
\noindent
and the interval $\mathcal{I}_{\alpha} = \left[ \hat{\theta} \pm s_{n} \text{cv}_{1-\alpha}(\bar{t}_{n})   \right]$ satisfies
\begin{align*}
\underset{n \rightarrow \infty}{\lim \inf} \text {  } \underset{\mu \in \mathcal{F}(L)}{\inf}  \Pr \left[ \mu^{1}_{+}(0) - \mu^{1}_{-}(0) \in \mathcal{I}_{\alpha} \right]  \geq 1 - \alpha
\end{align*}
for $t_{n}$, $\bar{B}(w_{N})$ and $\text{cv}_{1-\alpha}$ as defined in Section 3.5.\\

% Honest vs default
\noindent
The relevant difference of the statistical guarantee given in Proposition 1 relative to those that pointwise approaches to nonparametric confidence interval construction rely on is as follows: It ensures that, for any tolerance level $\eta$, one can find a sample size $n_{\eta}$ such that for all $n>n_\eta$ coverage is at least $1-\alpha-\eta$ for \textit{all} functions in $\mathcal{F}(L)$. In contrast, pointwise procedures can not generally ensure the existence of such a sample size for any given non-trivial tolerance level without restricting the curvature or a higher order derivative, since the true CEF is unknown. Thus, their coverage properties can theoretically be poor even in large samples. Consequently, the uniform guarantee provided by honesty is required for reliably good finite sample performance. Once such a restriction is imposed, it follows from the definition of optimized intervals that they are the minimax optimal choice (under known variances) amongst all linear intervals.

% Partial
Another attractive property of (\ref{eq6}) and "bias-aware" intervals in general is that they remain valid, irrespective of whether the assignment variable has support arbitrarily close to the threshold or not, in the sense that the statistical guarantee of the interval remains the same. In settings in which this is not the case, the interval will have positive length in the limit, but not necessarily cover the whole identification interval, that is the interval of values for $\theta$ that are consistent with the distribution $(X,Y)$ and the restriction imposed by $\mu \in \mathcal{F}(L)$. Partial identification intervals of this type were proposed in \citet{imbens2004confidence} and are also useful in other non-standard situations, e.g. if one wants to exclude data for reasons such as data entry errors or other institutional characteristics that could justify such a choice.\\

\noindent
In order to discuss two potential threats to the quality of the approximation underlying the construction of (\ref{eq6}), we further introduce the following two assumptions.\\

\noindent
\textbf{Assumption 2} Let  $C_{1} \in \mathbb{R_{+}}$ be fixed and $\hat{s}_{n}$ denote an estimator at our disposal.
\begin{enumerate}[leftmargin=*, labelsep=*, align=left, itemsep=-0.2cm, font=\normalfont, label=(\roman*)]
\item Assumption 1 holds with $\delta=1$.
\item Assumption 1 holds and $\hat{s}_{n} - s_{n} = o_{p}(1)$  uniformly over $\mathcal{F}(L)$. 
\end{enumerate}

\noindent
Assumption 2 (i) allows us to clarify the role of the ratio $\bar{w}_{R} = \left[ \max_{i}|w_{i}|\right]  \left[\sum_{i=1}^{n} |w_{i}|\right]^{-1}$ addressed by Assumption 1 (v) with respect to the quality of the normal approximation underlying (\ref{eq6}).  It follows from the Berry-Essen Theorem (cf. Theorem 3 in \cite{deasymptotic}) that under Assumption 2 (i)
\begin{align*}
\underset{z \in \mathbb{R}^{1}\text{ } \mu \in \mathcal{F}(L)}{\sup} \left\lvert \Pr \left[ \left( \dfrac{\hat{\theta} - \E[\hat{\theta}| X_{N}] }{s_n}  \right) \leq z \left\vert\vphantom{\frac{1}{1}}\right.  X_{N} \right] - \Phi(z) \right\rvert
\leq  D \dfrac{C}{\sigma_{min}^{3}} \dfrac{ \max_{i}|w_{i}| }{ \sum_{i=1}^{n} |w_{i}|},
\end{align*}
\noindent
where $\Phi$ denotes the standard normal CDF and the constant $D$ lies in  $0.4097 < D \leq 0.56$. This illustrates the role of Assumption 1 (v) in ensuring the quality of the distributional approximation. In particular, under the maintained assumptions, one would expect that the finite sample coverage rate of (\ref{eq6}) at the worst-case function is close to nominal whenever the ratio $\bar{w}_{R}$ is small. We therefore report $\bar{w}_{R}^{2}$ as a diagnostic statistic in our implementation and recommend to verify that this is the case in practice. If the ratio is "large", in the sense that the weights concentrate on a small set of observations, it is recommended to modify $\kappa$. In doing so, one effectively trades the quality of the estimator resulting from the initial choice of $\kappa$ in terms of the respective performance criterion for an improvement in the quality of the distributional approximation.

Assumption 2 (ii) emphasizes that the honesty property of (\ref{eq6}) was derived under known variances and that, in principle, one needs an appropriate estimator for $\sigma_{N}$ in order to preserve honesty under estimated conditional variance. Assumption 2 (ii) provides that such a uniformly consistent estimator of $s_{n}$ is available. In this case, the feasible interval that replaces $s_{n}$ with $\hat{s}_{n}$ remains asymptotically uniformly valid. This is pointed out because commonly used estimators for the conditional variance have a leading bias that is proportional to $\mu^{1}$, which is unrestricted over $\mathcal{F}(L)$. \citet{noack2019bias} propose an estimator for the conditional variance based on a regression adjusted version of the nearest-neighbor estimator of \citet{abadie2014inference} that has leading bias proportional to $\mu^{2}$ and can thus preserve honesty under second order bounds. Our implementation contains their proposed estimator as well as standard estimators of $s_{n}$. 

Note that the estimator solving (\ref{qp1}) is the finite-sample minimax linear estimator of $\theta$ over $\mathcal{F}(L)$ only under known conditional variances $\sigma_{N}^{2}$. If the conditional variances need to be estimated, the estimator is no longer guaranteed to achieve the minimax risk in finite samples. However, in the case of homoskedasticity, it suffices to estimate $\sigma^{2}_N$ by an efficient estimator for the conditional variance to obtain honest and asymptotically minimax optimal intervals. Under heteroskedasticity, a uniformly consistent estimator for $\sigma^{2}_{N}$ is required to maintain honesty as discussed above, and the minimax properties of the estimator solving (\ref{qp1}) depend on this choice.

%% Practical implementation
\paragraph{3.7 Practical Implementation.}  So far we have assumed that the curvature bound $L$ and the conditional variance $\sigma_{N}^{2}$ are known. In practice, it needs to be specified how $\sigma_{N}$ should be estimated and $L$ chosen. While the previous discussion gives some guidance on how to estimate $\sigma_{N}$, the most important choice in implementing our proposed method is the choice of the tuning parameter $L$, which, without additional assumption, can not be determined from the data without undermining the honesty of (\ref{eq6}) (cf. \citet{armstrong2018optimal} and references therein). This is due to a result in \citet{low1997nonparametric}, who shows that when $\mathcal{F}$ is a derivative smoothness class, it is, without further assumptions, not possible to adapt to $\mathcal{F}$ while maintaining uniform coverage at the same time. 

 % Choice of curvature bound
\paragraph*{3.7.1. Choice of $L$.} As a consequence of Low's impossibility result, the curvature bound $L$ has to be chosen a priori and application-specific knowledge on what constitutes plausible amounts of curvature is required to obtain suitable values of $L$. We reiterate that this requirement is not unique to bias-aware approaches to inference, as confidence intervals based on pointwise procedures must implicitly restrict $L$ to be informative at any given tolerance level.

In the absence of reliable information on the magnitude of $L$, it is recommended to conduct a sensitivity analysis by considering a range of plausible bounds together with rule of thumb (ROT) estimates of $L$ based on modelling the CEF over the largest part of its domain that is plausibly informative. In our implementation, we consider three rules of thumb. The first was suggested by \citet{armstrong2020simple} and is based on fitting a global quartic polynomial on each side of the threshold. The ROT estimate of $L$ is then obtained by computing the global maximum of the absolute value of the second derivative of the polynomial implied by the estimated coefficients. The second rule of thumb we consider was proposed in \citet{imbens2019optimized}. It involves fitting a quadratic polynomial on each side of the cutoff and computing an estimate of $L$ by scaling the maximum second derivative magnitude by a factor of 2-4. Finally, we propose a third rule of thumb that is based on fitting a cubic smoothing spline with evenly spaced knots on each side of the threshold. The bound $\hat{L}_{ROT}$ is then estimated by the maximum magnitude of the implied second derivative. This approach is heuristically motivated by the fact that the worst-case function of our estimator is a quadratic spline. In our simulation study, we report results based on this approach.

While it is not possible to consistently recover $L$ from the data, we are aware of two methods that were proposed to guard against overly optimistic choices of $L$ and to gain intuition for what might constitute plausible degrees of curvature. The first method is due to \citet{kolesar2018inference}, who propose a method to estimate a lower bound on $L$ based on the observations that any function in $\mathcal{F}(L)$ can, between any two points that are $\Delta$ units apart, not depart from a straight line by more than $\dfrac{L \Delta^{2}}{8}$. The second method is due to \citet{noack2019bias}, who propose a graphical procedure based on the solution to a constrained least squares problem to visualize "extreme" elements of $\mathcal{F}(L)$. The idea is to plot this element while iteratively increasing the curvature bound until the resulting function become implausibly erratic.
We generally recommend to combine subject knowledge, ROT estimates and such heuristic devices to gain intuition in any given application.

% Estimation of sigma
\paragraph*{3.7.2. Estimation of $\sigma_{N}$.} The discussion in the previous section implies that the estimator derived under known variances underlying the optimized linear confidence intervals can be understood as motivated by a homoskedastic model.
In order to ensure that the inference based on (\ref{eq6}) is robust to heteroskedasticity, it is required to construct confidence intervals using an approriate estimator for the conditional variances $\sigma_{N}^2$, analogous to a regression analysis that uses ordinary least squares estimators but builds confidence intervals using Eicker–Huber–White standard errors. In our implementation, we initialize $\sigma_{N}$ by a naive homoskedastic estimate to obtain the weights, before building confidence intervals based on a function that implements different heteroskedasticity-robust estimators of the conditional variance, including estimators based on standard estimates of $\sigma_{i}$ relying on the residuals of linear regressions, nearest-neighbor estimates proposed and considered in \citet{abadie2006large} and \citet{abadie2014inference}, as well as the uniformly consistent modification proposed in \citet{noack2019bias}. The results reported in the simulation study in the next section are obtained using the nearest-neighbor approach of \citet{abadie2014inference} to estimate $\sigma_{N}$ based on 10 nearest-neighbor matches.

%%%% Section 4

\newpage

\section{Comparison with other Methods}

In this section, we compare the performance of optimized linear confidence intervals to a variety of procedures based on local linear regression in a simulation study.
In order to be precise about the comparison we briefly introduce the considered methods.

\paragraph{4.1 Local Linear Methods.} The local linear estimate $\hat{\theta}_{LL}$ of $\theta$ with bandwidth $h$ is the coefficient on $X_i D_i$ in a weighted OLS regression of $Y_{i}$ on the vector $(1,X_i,D_i,X_iD_i)$, using only observations $i$ such that $|X_{i}| \leq h$, with weights determined by a kernel function. Under regularity conditions, it holds that if the density of the assignment variable $f_{X}(x)$ is continuous and bounded away from zero in an open neighborhood around the threshold, the MSE of the local linear estimator with bandwidth sequence $h_{n} \rightarrow 0$ evaluated at a function $\mu$ is
\begin{align*}
\MSE(h_{n};\mu) = h_{n} \left[ B^{2} + o_{p}(1) \right] + \frac{1}{nh_{n}^{3}} \left[V + o_{p}(1) \right],
\end{align*}

\noindent
where $B \propto (\mu^{2}_{+}(0)-\mu^{2}_{-}(0))$ is the leading asymptotic bias and $V \propto (\sigma_{+}(0)^{2} + \sigma_{-}(0)^{2})f_{X}(0)^{-1}$ is the asymptotic variance. If $B \neq 0$, an asymptotic MSE-optimal bandwidth sequence is thus
\begin{align*}
h_{PMSE} = n^{-1/5} \left[ \dfrac{ 3 V}{2 B^{2}} \right]^{1/5}.
\end{align*}

\noindent
In our simulation study, we compare the performance of uniform procedures to methods that rely on plug-in estimates of this quantity which add a regularization term to the denominator that shrinks with the sample size \citep{imbens2012optimal}. We denote such estimates by $\hat{h}_{PMSE}$ and compute them using the plug-in estimators proposed by \citet{calonico2014robust}. The undersmoothing bandwidths are computed relative to the obtained pointwise asymptotic MSE-optimal estimate $\hat{h}_{US} = n^{-1/20} \hat{h}_{PMSE}$. The two bandwidth choices required for the RBC intervals are either both set pointwise optimal $b= \hat{b}_{PMSE}$, $h=\hat{h}_{PMSE}$, where $\hat{b}_{PMSE}$ refers to the pointwise asymptotic MSE-optimal plug-in estimate for the local quadratic bias estimator, or both set to the pointwise asymptotic MSE-optimal estimate $\hat{h}_{PMSE}$ for the local linear estimator. In addition, we consider RBC bandwidth choices $h_{CE}$, $b_{CE}$ that optimize the pointwise asymptotic coverage error \citep{calonico2018coverage}, which can be considered an intermediate form of undersmoothing and robust bias correction. 

Under regularity conditions and restrictions on the rate of the bandwidth sequence $h_{n} \rightarrow 0$
\begin{align*}
\sqrt{nh_{n}^{3}} \left[ \hat{\theta}_{LL}(h_{n}) - \theta - h_{n} B_{n} \right] \overset{d}{\rightarrow} N(0,V),
\end{align*}
\noindent
where $B_{n} \overset{P}{\rightarrow} B$. The methods that we consider differ in whether and how they take into account the incorrect centering induced by the smoothing bias. In our simulation, we consider the following four types of two-sided local linear intervals for the above bandwidth choices:
\begin{alignat*}{3}
&\text{\textit{Conventional}} 		\qquad &&\mathcal{I}_{Conv.} &&= \left[ \hat{\theta}_{LL}(\hat{h}_{PMSE})  \pm z_{1-\alpha /2} \sqrt{ \hat{V}_{Conv.} /n\hat{h}_{PMSE}^{3}} \right],\\
&\text{\textit{Undersmoothed}}		 \qquad &&\mathcal{I}_{US} &&= \left[ \hat{\theta}_{LL}(\hat{h}_{US})  \pm z_{1-\alpha /2} \sqrt{ \hat{V}_{US} /n\hat{h}_{US}^{3}} \right],\\
&\text{\textit{Robust Bias Correction}} \qquad &&\mathcal{I}_{RBC} &&= \left[ \hat{\theta}_{LL}(\hat{h}_{RBC}) - \hat{h}_{RBC} \hat{B}_{n}(\hat{b}_{RBC})  \pm z_{1-\alpha /2} \sqrt{ \hat{V}_{RBC} /n\hat{h}_{RBC}^{3}} \right],\\
&\text{\textit{Fixed Length}} 			\qquad &&\mathcal{I}_{FL} &&= \left[  \hat{\theta}_{LL}(\hat{h}_{FL}) \pm \text{cv}_{1-\alpha}(\bar{t}_{n}) \sqrt{ \hat{V}_{FL} /n \hat{h}_{FL}^{3}}  \right],
\end{alignat*}
%$z_{1-\alpha/2}$ denotes the $(1-\alpha/2)$ quantile of the standard normal distribution 
\noindent
where $z_{1-\alpha /2}$ denotes the $(1-\alpha /2)$ quantile of the standard normal distribution, $\hat{V}$ denotes an estimate of the respective asymptotic variance, and $\text{cv}_{1-\alpha}$ as well as $t_{n}$ are defined as in (\ref{qp5}) for the worst-case magnitude of the conditional bias of $\hat{\theta}_{LL}(\hat{h}_{FL})$.

The first two types, conventional and undersmoothed confidence intervals, essentialy assume the smoothing bias away. While the conventional method directly assumes that $h_{n} B_{n} s_{n}^{-1} \approx 0$, undersmoothed intervals rely on the asymptotic promise that $h_{US}/h_{PMSE} \underset{n \rightarrow \infty}{\rightarrow} 0$, implying that $\sqrt{nh_{n}^{3}} \left[ \hat{\theta}(h_{n}) - \theta - h_{n} B_{n} \right] = \sqrt{nh_{n}^{3}} \left[ \hat{\theta}(h_{n}) - \theta \right] + o_{p}(1) \overset{d}{\rightarrow} N(0,V)$, which is uninformative about the smoothing bias in a given sample.

The last two types explicitly address the smoothing bias. RBC intervals are centered around a bias-corrected point estimate, using a higher order local polynomial estimator to estimate the bias. They differ from traditional bias-corrected intervals, which are known to perform poorly in finite samples \citep{hall1992effect}, in that they do not require $h_{RBC}/b_{RBC} \underset{n \rightarrow \infty}{\rightarrow} 0$. As a consequence, the standardized bias-correction term is not negligible asymptotically, leading to a different asymptotic variance $V_{RBC}$ that captures the additional uncertainty introduced by bias estimation. In our simulation, all RBC bias estimates are based on local quadratic estimators. The fixed length intervals are constructed in the same spirit as the optimized intervals discussed in Section 3 with weights defined by the local linear estimator. They take into account the exact magnitude of the worst-case conditional bias by inflating the critical value according to the ratio $\bar{t}_{n}$, and are therefore valid for any bandwidth choice.  The bandwidths for the fixed length local linear intervals are obtained analogously to $\kappa_{UMSE}$ and $\kappa_{LE}$ by minimizing the finite sample uniform MSE $h_{UMSE}$ or the interval half-length $h_{HL}$. The variances for interval construction are estimated by the conditional variances of the estimators implied by the weights, using the nearest-neighbor approach of \citet{abadie2014inference} based on 10 nearest neighbor matches.

\newpage

\paragraph{4.2 Monte Carlo Setup.}

\begin{figure}[h]
\centering
\makebox[\textwidth][c]{%
\begin{tabular}{cc}
  \includegraphics[width=85mm]{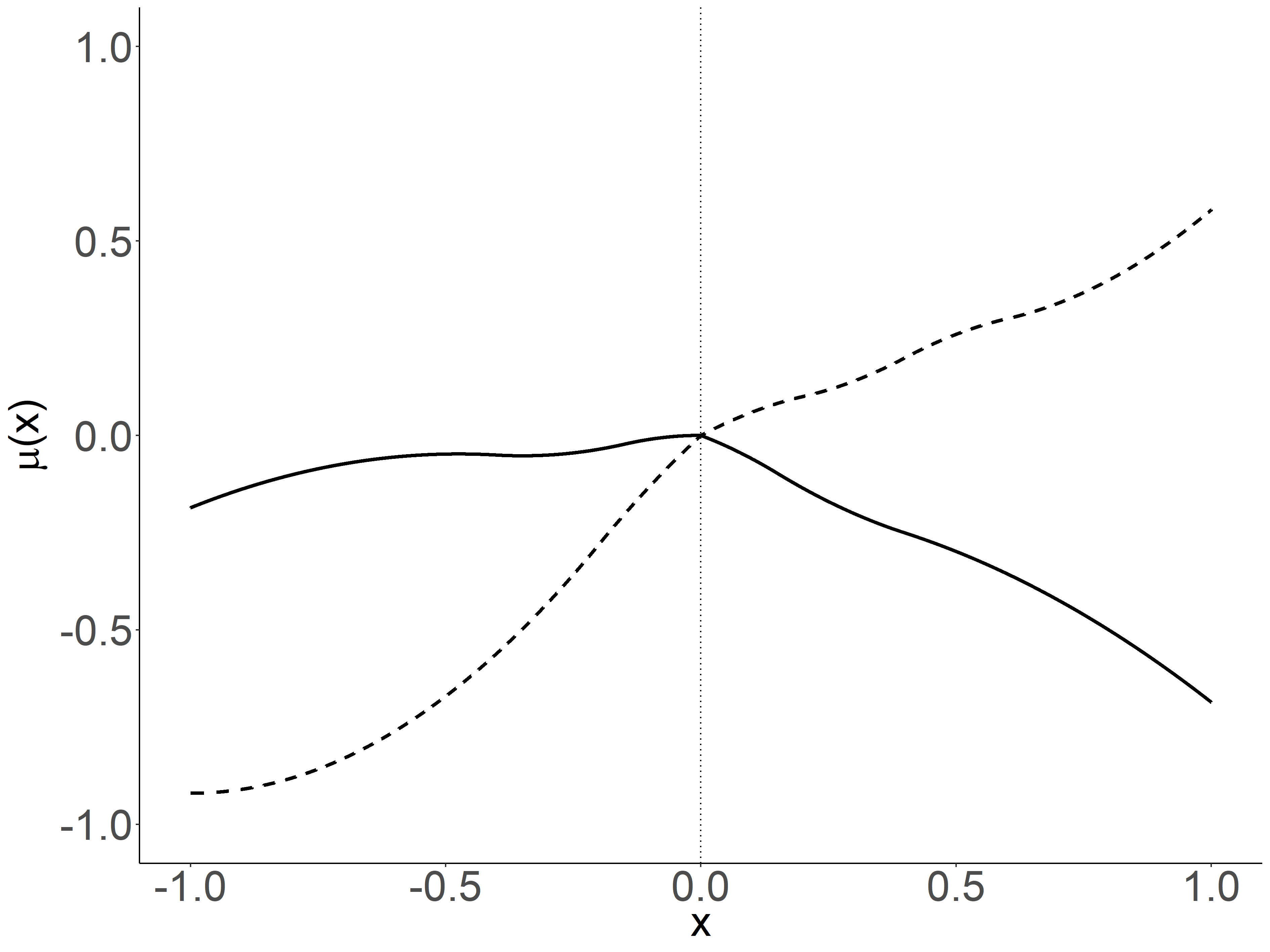} &   \includegraphics[width=85mm]{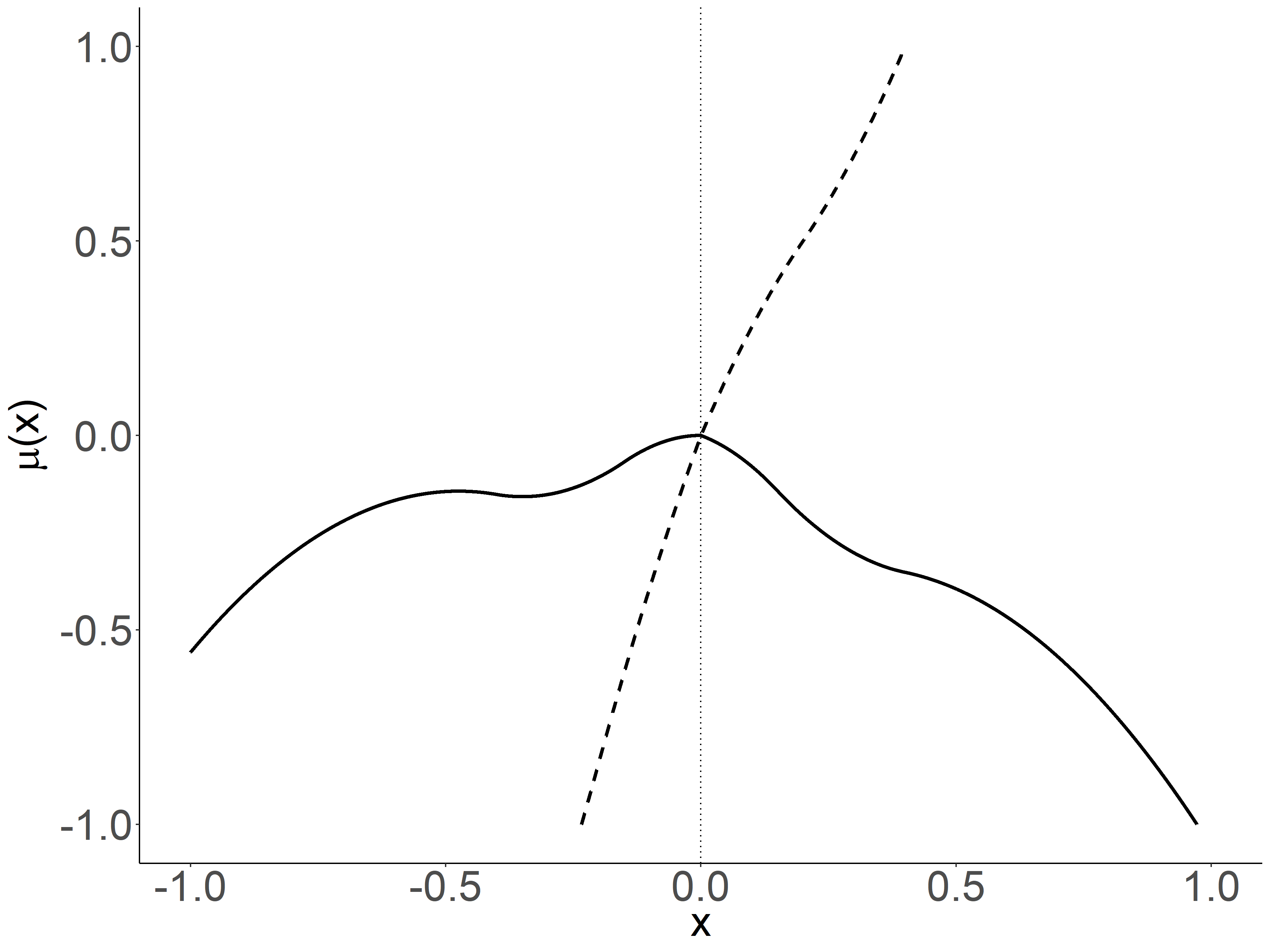} \\
 (a) $L=2$ & (b) $L=6$\\[6pt]
\end{tabular}}
\caption{Shape of $\mu_{1}$ (solid) and $\mu_{2}$ (dashed) for $\theta=-0.5$ and $L \in \lbrace{2,6 \rbrace}$ on $[-1,1]$.}
\label{Fig:one}
\end{figure}

In order to investigate the performance of the optimized linear confidence intervals relative to the methods introduced above we conduct a simulation study. The setups differ in the conditional mean function, its degree of curvature and the distribution of the assignment variable, yielding a total of 8 different settings. The assignment variable is drawn from an equidistant uniform distribution with support $\lbrace -1, -1 + \frac{2}{K}, \cdots, 1- \frac{2}{K}, 1 \rbrace$, where the parameter $K$ controls the number of support points and $K_\infty$ means the continuous uniform distribution with support $[-1,1]$. The outcome data is generated according to 
\begin{equation*}
Y_{i} = \mu_{j}(X_i) + \varepsilon_{i} \qquad j \in  \lbrace{ 1,2 \rbrace},
\end{equation*}
\noindent
where the CEF error $\varepsilon_{i}$ is drawn from a mean zero normal distribution with $\sigma=0.1$ and 
\begin{align*}
\mu_{1}(x) &= D(x) \theta x  +  \dfrac{L}{2} \left[ -x^{2} + 1.75 s_{+}^{2}(|x|-0.15) - 1.25 s_{+}^{2}(|x|-0.4)  \right]  \\
\mu_{2}(x) &= D(x) \theta x + \dfrac{L}{2}  \left[ (x+1)^{2} - 2 s_{+}^{2}(x+0.2) + 2 s_{+}^{2}(x-0.2) - 2s_{+}^{2}(x-0.4) + 2s_{+}^{2}(x-0.6) - 0.92 \right],
\end{align*}

\noindent
where $s_{+}^{2}(x) = D(x)x^{2}$ denotes square of the plus function. Note that both CEFs are second order splines with maximal second order derivative magnitude $L$ and thus elements of $\mathcal{F}(L)$. The function $\mu_{1}$ attains the second order bound only in $(-0.15,0.15)$ while $\mu_{2}$ attains the bound everywhere on $[-1,1]$ with alternating signs in each interval defined by the knots. In the simulation, we set $\theta=-0.5$ and consider bounds $L \in \lbrace{2,6 \rbrace}$. Figure \ref{Fig:one} displays the shape of the functions.\\

% FIGURE 1

\newpage

\paragraph{4.3 Monte Carlo Results.}

\noindent
Tables \ref{Tab:one} and \ref{Tab:two} show the results of 5000 Monte Carlo runs\footnote{In Appendix \ref{appendix:I}, we provide analogous results for 20.000 Monte Carlo runs that were conducted on a cluster due to the required computational resources.}  with sample size $n=2000$ for $\mu_{1}$ and $\mu_{2}$ respectively. The top panel in each table displays the results for the case when the assignment variable is drawn from the continunous uniform distribution, while the bottom panel shows the results for 80 equidistant support points. We use a triangular kernel for all local linear methods. In the case that a bandwidth selector chooses a bandwidth for which the respective estimator is not defined, we manually adjust the bandwidth such that it covers three support points on either side of the cutoff.\footnote{This occured only in the discrete setting.} The left panel in each table reports the results for the low curvature version of the respective CEF, while the right panel reports the results for the high curvature variant. Columns 1 and 2 indicate the method and the tuning target.  The curvature bound $L$ is either chosen by the rule of thumb $\hat{L}$ or fixed to 2 or 6, as indicated in the tuning subscript. We report the empirical coverage rate at nominal level 95\% (Cov.), the average length relative to the optimal linear interval with correct curvature bound (RL), as well as the average tuning parameter choice ($h/\kappa$) of each method.

Unsurprisingly, conventional and undersmoothing confidence intervals show below nominal coverage in all designs, with undercoverage of undersmoothed intervals becoming more severe in the high curvature regime. The performance of robust bias-corrected intervals varies with the tuning target. While both, the default RBC method\footnote{This refers to the default in the authors' R implementation of RBC CIs.} that picks both bandwidths using the respective asymptotic MSE-optimal estimate and the coverage error optimized RBC interval undercover severly, the RBC interval obtained by setting both bandwidths to the local linear pointwise MSE bandwidth or the UMSE bandwidth under the ROT estimate of $L$ show close to nominal coverage and are insensitive to the true curvature. However, for the latter this comes at a cost in terms of their length relative to fixed-length and optimized intervals. In all designs, RBC intervals with tuning that attains close to nominal coverage are at least approximately twice as long as the infeasible length-optimal interval and 40\% longer than feasible length optimized intervals under the ROT choice of $L$.
Fixed-length and optimized intervals show above or close to nominal coverage under the correct curvature, with the lowest empirical coverage at 94.7\% $(\mu_{1,2},K=80,L=6)$. As one would expect, they are conservative when the true curvature is lower than specified and undercover when the true curvature is higher than specified. The rule of thumb choice of $L$ tends to overestimate the curvature with the exception of $(\mu_{1},K_{\infty},L=6)$, leading to above nominal coverage of both uniform interval types in most

\newpage

\setcounter{table}{0}

% TBL 1

\newgeometry{top=0.75cm,bottom=1.25cm,footskip=0cm}

\begin{table}[h!] %\centering

\hspace*{-1.50cm}  \resizebox{1.20\textwidth}{!}{\begin{threeparttable}
  
\caption{Monte Carlo Results: Coverage and Relative Length I}
\ra{1.1}
\begin{tabular}{@{}lllcclcccl@{}}\toprule
$\mu_{1}(x)$ & & & \multicolumn{3}{c}{$L = 2$} & \phantom{abc}& \multicolumn{3}{c}{$L = 6$} \\
\cmidrule{4-6} \cmidrule{8-10}  
Method & Tuning & \phantom{abc} & Cov. & RL & $h / \kappa$ &&  Cov. & RL & $h / \kappa$ \\ \midrule
\multicolumn{2}{c}{\textit{Cont. Design}}\\
Conv. & $h_{PMSE}$ 				&&  43.0  &  0.566 &  0.226			&& 	18.3  &  0.398 &  0.183 \\
US  & $h_{US}$ 					&&  87.1  &  0.999 &  0.154 		&& 	64.2  &  0.702 &  0.125 \\
RBC & $b_{PMSE},h_{PMSE}$ 		&& 53.9   & 0.849 &  0.226 (0.462)	&& 	38.1  &  0.582 &  0.183 (0.388) \\
RBC & $h=b=h_{PMSE}$ 				&& 95.2  &  2.140 &  0.226		&& 	95.1  &  1.510 &  0.183 \\
RBC & $b_{CE},h_{CE}$ 			&& 84.4  &  1.160 &  0.154 (0.462) &&	68.9  &  0.807 &  0.125 (0.388) \\
RBC & $h=b=h_{UMSE,L=\hat{L}}$		&& 95.5  &  4.390 &  0.167		&& 	94.7  &  3.940 &  0.104\\
LL-FL & $h_{UMSE,L=\hat{L}}$ 	&& 95.0  &  1.710 &  0.167			&& 	95.2  &  1.530 &  0.104\\
LL-FL & $h_{HL,L=\hat{L}}$ 		&&  96.8 &   1.610 &  0.221			&& 	93.1  &  1.440 &  0.138\\
Opt & $\kappa_{UMSE,L=\hat{L}}$ 	&& 94.9  &  1.700 &  1.000		&& 	95.1  &  1.520 &  1.000\\
Opt & $\kappa_{HL,L=\hat{L}}$		&& 96.7  &  1.600 &  0.246		&& 	93.0  &  1.430 &  0.246\\
\midrule
LL-FL & $h_{UMSE,L=2}$ 		&& 95.1  &  1.060 &  0.187				&& 	21.1  &  0.550 &  0.187\\
LL-FL & $h_{HL,L=2}$ 		&& 97.0  &  1.000 &  0.247				&& 	 0.42  &  0.517 &  0.247\\
Opt & $\kappa_{UMSE,L=2}$ 	&& 95.1  &  1.060 &  1.000				&& 	21.0  &  0.549 &  1.000\\
Opt & $\kappa_{HL,L=2}$ 	&& 97.0  &  1.000 &  0.246				&&  0.36   & 0.517  & 0.246\\
\midrule
LL-FL & $h_{UMSE,L=6}$ 		&& 99.3  &  2.060  & 0.121				&& 	95.0  &  1.060 &  0.121\\
LL-FL & $h_{HL,L=6}$ 		&& 100.0 &   1.940 &  0.160				&& 	95.0  &  1.000 &  0.160\\
Opt & $\kappa_{UMSE,L=6}$ 	&& 99.4  &  2.050  & 1.000				&& 	95.1  &  1.060 &  1.000\\
Opt & $\kappa_{HL,L=6}$ 	&& 100.0 &   1.930 &  0.245				&& 	95.0  &  1.000 &  0.245\\
\midrule
\multicolumn{2}{c}{\textit{Disc. Design }}\\
Conv. & $h_{PMSE}$ 					&& 44.8  &  0.578 &  0.225		&& 	17.8  &  0.403 &  0.183 \\
US  & $h_{US}$ 						&& 87.8  &  1.040 &  0.154 		&& 	65.3  &  0.723 &  0.126 \\
RBC & $b_{PMSE},h_{PMSE}$ 			&& 54.4  &  0.869 &  0.225 (0.463)		&& 	37.2  &  0.593 &  0.183 (0.389) \\
RBC & $h=b=h_{PMSE}$ 				&& 95.4  &  2.320 &  0.225		&& 	95.2  &  1.670 &  0.183 \\
RBC & $b_{CE},h_{CE}$ 				&& 84.8  &  1.200 &  0.154 (0.463)	&&	69.6  &  0.832 &  0.126 (0.389)\\
RBC & $h=b=h_{UMSE,L=\hat{L}}$		&& 95.0  &  4.550 &  0.170		&& 	94.3  &  4.380 &  0.108\\
LL-FL & $h_{UMSE,L=\hat{L}}$ 		&& 95.4  &  1.900 &  0.170		&& 	95.4  &  1.760 &  0.108\\
LL-FL & $h_{HL,L=\hat{L}}$ 			&& 97.3  &  1.810 &  0.224		&& 	93.4  &  1.670 &  0.142\\
Opt & $\kappa_{UMSE,L=\hat{L}}$ 	&& 95.4  &  1.860 &  1.000 		&& 	95.0  &  1.710 &  1.000\\
Opt & $\kappa_{HL,L=\hat{L}}$		&& 97.2  &  1.760 &  0.246		&& 	93.2  &  1.620 &  0.243\\
\midrule
LL-FL & $h_{UMSE,L=2}$ 				&& 95.0  &  1.070 &  0.189		&& 	19.7  &  0.548  & 0.189\\
LL-FL & $h_{HL,L=2}$ 				&& 97.1 &   1.000 &  0.251		&& 	0.30   & 0.515  & 0.251\\
Opt & $\kappa_{UMSE,L=2}$ 			&& 95.4  &  1.060 &  1.000		&& 	20.9  &  0.546 &  1.000\\
Opt & $\kappa_{HL,L=2}$ 			&& 97.2  &  1.000 &  0.242		&& 	0.32   & 0.513 &  0.242\\
\midrule
LL-FL & $h_{UMSE,L=6}$ 		&&  99.5  &  2.100  & 0.125				&& 	95.0  &  1.080 &  0.125\\
LL-FL & $h_{HL,L=6}$ 		&& 100.0  &  1.960  & 0.163				&& 	94.9  &  1.010 &  0.163\\
Opt & $\kappa_{UMSE,L=6}$ 	&& 99.4   &   2.080 &  1.000			&& 	94.9  &  1.070 &  1.000\\
Opt & $\kappa_{HL,L=6}$ 	&& 100.0  &  1.950  & 0.237				&& 	94.7  &  1.000 &  0.237\\
\bottomrule

\end{tabular}
\label{Tab:one}

\begin{tablenotes}
\small
   \item \textit{Note:} Empirical coverage rate (Cov.), average length relative to optimized interval with true smoothness bound (RL) and tuning parameter choice ($h / \kappa$) of conventional (Conv.), undersmoothed (US), robust bias-corrected (RBC), fixed length (LL-FL) and optimized linear (Opt) 95\% confidence intervals over 5.000 Monte Carlo draws. The pilot RBC bandwidh is reported in parentheses if it is selected separately from $h$.
\end{tablenotes}
\end{threeparttable}}
\end{table}

\restoregeometry

%%%% TBL 2

\newpage
\newgeometry{top=0.75cm,bottom=1.25cm,footskip=0cm}

\begin{table*}[h!]\centering
\hspace*{-1.50cm}  \resizebox{1.20\textwidth}{!}{\begin{threeparttable}
  
\caption{Monte Carlo Results: Coverage and Relative Length II}
\ra{1.1}
\begin{tabular}{@{}lllcclcccl@{}}\toprule
$\mu_{2}(x)$ & & & \multicolumn{3}{c}{$L = 2$} & \phantom{abc}& \multicolumn{3}{c}{$L = 6$} \\
\cmidrule{4-6} \cmidrule{8-10}  
Method & Tuning & \phantom{abc} & Cov. & RL & $h / \kappa$ &&  Cov. & RL & $h / \kappa$ \\ \midrule
\multicolumn{2}{c}{\textit{Cont. Design}}\\
Conv. & $h_{PMSE}$ 					&&  42.7  &  0.599 &  0.221 		&& 	55.4  &  0.593 & 0.143 \\
US  & $h_{US}$ 						&&  84.6  &  1.060 &  0.151			&& 	79.7  &  1.050 & 0.098 \\
RBC & $b_{PMSE},h_{PMSE}$ 			&& 62.2  &  0.908 &  0.221 (0.439)			&& 	82.8  &  0.778 & 0.143 (0.336)\\
RBC & $h=b=h_{PMSE}$  				&& 95.4  &  2.260 &  0.221			&& 	93.9  &  2.250 & 0.143\\
RBC & $b_{CE},h_{CE}$ 				&& 87.0  &  1.230 &  0.151 (0.439)			&&	86.8  &  1.150 & 0.098 (0.336)\\
RBC & $h=b=h_{UMSE,L=\hat{L}}$		&& 95.1  &  5.510 &  0.134			&& 	94.3  &  4.870 & 0.088\\
LL-FL & $h_{UMSE,L=\hat{L}}$ 		&& 98.1  &  2.140 &  0.134			&& 	98.7  &  1.890 & 0.088\\
LL-FL & $h_{HL,L=\hat{L}}$ 			&& 99.3  &  2.010 &  0.178			&& 	99.6  &  1.780 & 0.116\\
Opt & $\kappa_{UMSE,L=\hat{L}}$ 	&& 98.1  &  2.130 &  1.000			&& 	98.7  &  1.880 & 1.000\\
Opt & $\kappa_{HL,L=\hat{L}}$		&& 99.3  &  2.010 &  0.246			&& 	99.6  &  1.770 & 0.246\\
\midrule
LL-FL & $h_{UMSE,L=2}$ 		&& 95.0  &  1.060 &  0.187				&& 	20.6  &  0.550 & 0.187\\
LL-FL & $h_{HL,L=2}$ 		&& 95.4  &  1.000 &  0.247				&& 	 0.02  &  0.517 & 0.247\\
Opt & $\kappa_{UMSE,L=2}$ 	&& 95.1  &  1.060 &  1.000				&& 	20.5 &   0.549 & 1.000\\
Opt & $\kappa_{HL,L=2}$ 	&& 95.3  &  1.000 &  0.246				&& 	0.02  &  0.517  & 0.246\\
\midrule
LL-FL & $h_{UMSE,L=6}$ 		&& 99.3  &  2.060 &  0.121				&& 	95.0 &   1.060 & 0.121\\
LL-FL & $h_{HL,L=6}$ 		&& 100.0 &   1.940 &  0.160				&& 	95.0 &   1.000 & 0.160\\
Opt & $\kappa_{UMSE,L=6}$ 	&& 99.4  &  2.050 &  1.000				&& 	95.1 &   1.060 & 1.000\\
Opt & $\kappa_{HL,L=6}$ 	&& 100.0 &   1.930 &  0.245				&& 	95.0 &   1.000 & 0.245\\
\midrule
\multicolumn{2}{c}{\textit{Disc. Design }}\\
Conv. & $h_{PMSE}$ 					&& 45.0  &  0.621 &  0.218			&& 	54.2 &   0.607 & 0.143 \\
US  & $h_{US}$ 						&& 84.4  &  1.120 &  0.149 			&& 	80.2 &   1.000 & 0.106\\
RBC & $b_{PMSE},h_{PMSE}$ 			&& 62.1  &  0.939 &  0.218 (0.436)			&& 	83.2  &  0.801 & 0.143 (0.334) \\
RBC & $h=b=h_{PMSE}$ 				&& 95.5  &  2.520 &  0.218 			&& 	94.1 &   2.680 & 0.143 \\
RBC & $b_{CE},h_{CE}$ 				&& 85.8  &  1.300 &  0.149 (0.436)		&&	87.5 &   1.110 & 0.106 (0.334)\\
RBC & $h=b=h_{UMSE,L=\hat{L}}$		&& 95.0  &  5.990 &  0.137			&& 	94.1 &   5.730 & 0.092\\
LL-FL & $h_{UMSE,L=\hat{L}}$ 		&& 98.3  &  2.490 &  0.137			&& 	98.9 &   2.060 & 0.092\\
LL-FL & $h_{HL,L=\hat{L}}$ 			&& 99.5  &  2.370 &  0.181			&& 	99.8 &   1.940 & 0.121\\
Opt & $\kappa_{UMSE,L=\hat{L}}$ 	&& 98.4  &  2.420 &  1.000			&& 	98.8 &   2.000 & 1.000\\
Opt & $\kappa_{HL,L=\hat{L}}$		&& 99.4  &  2.300 &  0.246			&& 	99.8 &   1.890 & 0.239 \\
\midrule
LL-FL & $h_{UMSE,L=2}$ 				&& 95.0  &  1.070 &  0.189			&& 	18.9 &   0.548 & 0.189\\
LL-FL & $h_{HL,L=2}$ 				&& 95.6  &  1.000 &  0.251			&& 	0.04  &  0.515 & 0.251\\
Opt & $\kappa_{UMSE,L=2}$ 			&& 95.2  &  1.060 &  1.000			&& 	20.3 &   0.546 & 1.000\\
Opt & $\kappa_{HL,L=2}$ 			&& 95.3  &  1.000 &  0.242		&& 	0.06 &   0.513 & 0.242\\
\midrule
LL-FL & $h_{UMSE,L=6}$ 		&& 99.5  &  2.100  & 0.125				&& 	95.0  &  1.080 &  0.125\\
LL-FL & $h_{HL,L=6}$ 		&& 100.0 &   1.960 &  0.163				&& 	94.9  &  1.010 & 0.163\\
Opt & $\kappa_{UMSE,L=6}$ 	&& 99.4  &  2.080  & 1.000			&& 	94.9 &   1.070  & 1.000\\
Opt & $\kappa_{HL,L=6}$ 	&& 100.0 &   1.950 &  0.237			&& 	94.7 &   1.000  & 0.237\\
\bottomrule

\end{tabular}
\label{Tab:two}

\begin{tablenotes}
\small
   \item \textit{Note:} Empirical coverage rate (Cov.), average length relative to optimized interval with true smoothness bound (RL) and tuning parameter choice ($h / \kappa$) of conventional (Conv.), undersmoothed (US), robust bias-corrected (RBC), fixed length (LL-FL) and optimized linear (Opt) 95\% confidence intervals over 5.000 Monte Carlo draws. The pilot RBC bandwidh is reported in parentheses if it is selected separately from $h$.
\end{tablenotes}
\end{threeparttable}}
\end{table*}

\restoregeometry

\newpage

\noindent
designs with lowest empirical coverage at 93.0\% $(\mu_{1},K_{\infty},L=6)$.  The resulting intervals thus tend to be conservative but are still substantially shorter than their pointwise counterparts under valid tuning. Interestingly, the length-optimal bias-aware intervals show higher coverage rates than their uniform MSE-optimal counterparts on average. Moreover, the length-optimal weights of both, fixed length and optimized estimators "oversmooth" relative to the uniform MSE-optimal choice of the respective tuning parameter. In all designs, the optimally tuned fixed-length intervals demonstrate performance on par with their optimized counterparts, indicating that the high minimax efficiency of local linear estimators under second order bounds demonstrated in \citet{armstrong2020simple} for estimating the value of the CEF at a point also holds for the estimation of first derivatives.

\paragraph{4.4 Gains from Optimization.} The popularity of local linear estimators in empirical practice is motivated by their intuitive appeal and a range of attractive theoretical properties of local polynomials, in particular their asymptotic minimax efficiency over the Taylor class of functions  (\cite{fan1993local}, \cite{fan1997local}, \cite{cheng1997automatic}). \citet{armstrong2020simple} show that local polynomial estimators can also attain high minimax efficiency in the Hölder class of functions defined by derivative bounds among a large class of estimators to which a central limit theorem applies and that have worst-case bias and standard deviation that scale as powers of a bandwidth parameter.  This result relies on their observation that, for relevant performance criteria, the asymptotic minimax performance of two estimators in this class does not depend on the criterion but is solely governed by their worst-case biases, their standard deviations and their rate exponents $r = \gamma_{b}/(\gamma_{b}-\gamma_{s})$, where $\gamma_{b}$ and $\gamma_{s}$ denote the scaling exponents of the worst-case bias and the standard deviation respectively. Moreover, they show that the optimal worst-case bias to standard deviation ratio depends only on the criterion and $r$ (cf. Theorem 2.1 in \cite{armstrong2020simple}).  

Their analytic results provide us with guidance on what to expect with respect to the asymptotic efficiency gains of optimized linear confidence intervals and allow us to state a lower bound for the efficiency gain of our method relative to uniform MSE-optimal fixed length intervals. For the local linear estimator of the kink parameter, $r=0.4$ and their calculations imply an efficiency gain of approximately 6\% at $\alpha= 0.05$ for moving from the uniform MSE-optimal to the length-optimal local linear interval (cf. Figure 3 in \cite{armstrong2020simple}), which is consistent with our Monte-Carlo results. Thus, a lower bound for the efficiency gain of the optimized interval relative to the UMSE-optimal fixed length interval is 6\%. This

\newpage

%For their leading example, inference on the value of the CEF at a point,
%Using this result, they in particular show that under second order bounds, the local linear estimator with triangular kernel and uniform MSE optimal bandwidth is 99.99\% efficient in this class, implying that the fixed-length interval centered around this estimate is close to optimal.
%%% FIGURE 2

\begin{figure}[h!]
\centering
\makebox[\linewidth]{%
\begin{tabular}{cc}
  \includegraphics[width=80mm]{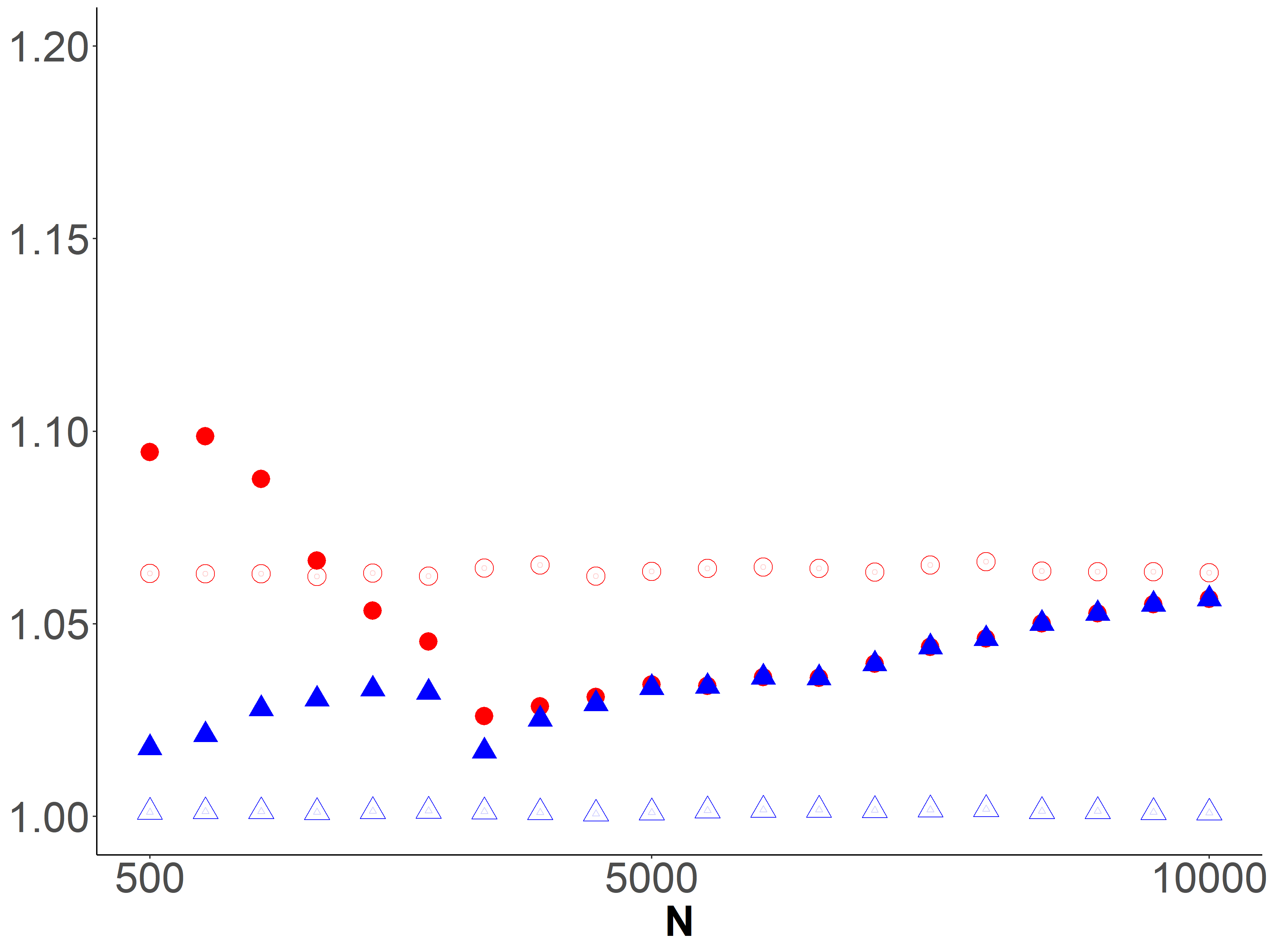} &   \includegraphics[width=80mm]{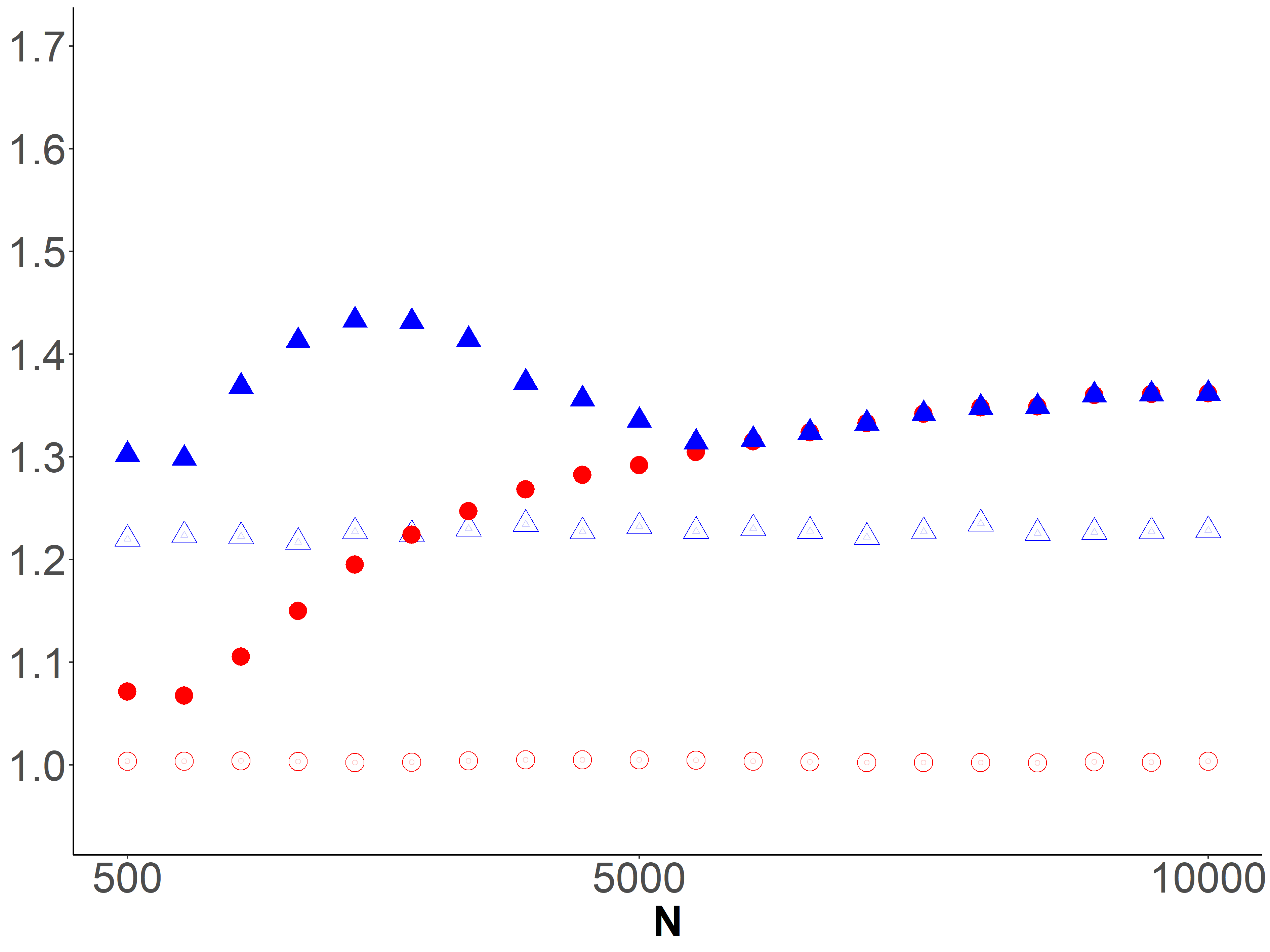} \\
(a) Excess Length & (b) Excess UMSE\\[6pt]
\end{tabular}}
\caption{Excess length (a) and excess UMSE (b) of the length-optimal (triangle) and UMSE-optimal (circle) two-sided 95\% local linear fixed-length interval and estimator relative to the respective optimized linear interval/estimator for continuous (blank) and discrete (filled) designs across different sample sizes.}
\label{Fig:two}
\end{figure}

%%%% FIGURE 3

\begin{figure}[h!]
\centering
\makebox[\linewidth]{%
\begin{tabular}{ccc}
\raisebox{85pt}{\parbox[b]{.05\textwidth}{\begin{turn}{-90}$K=50$\end{turn}}}%
  \includegraphics[width=55mm]{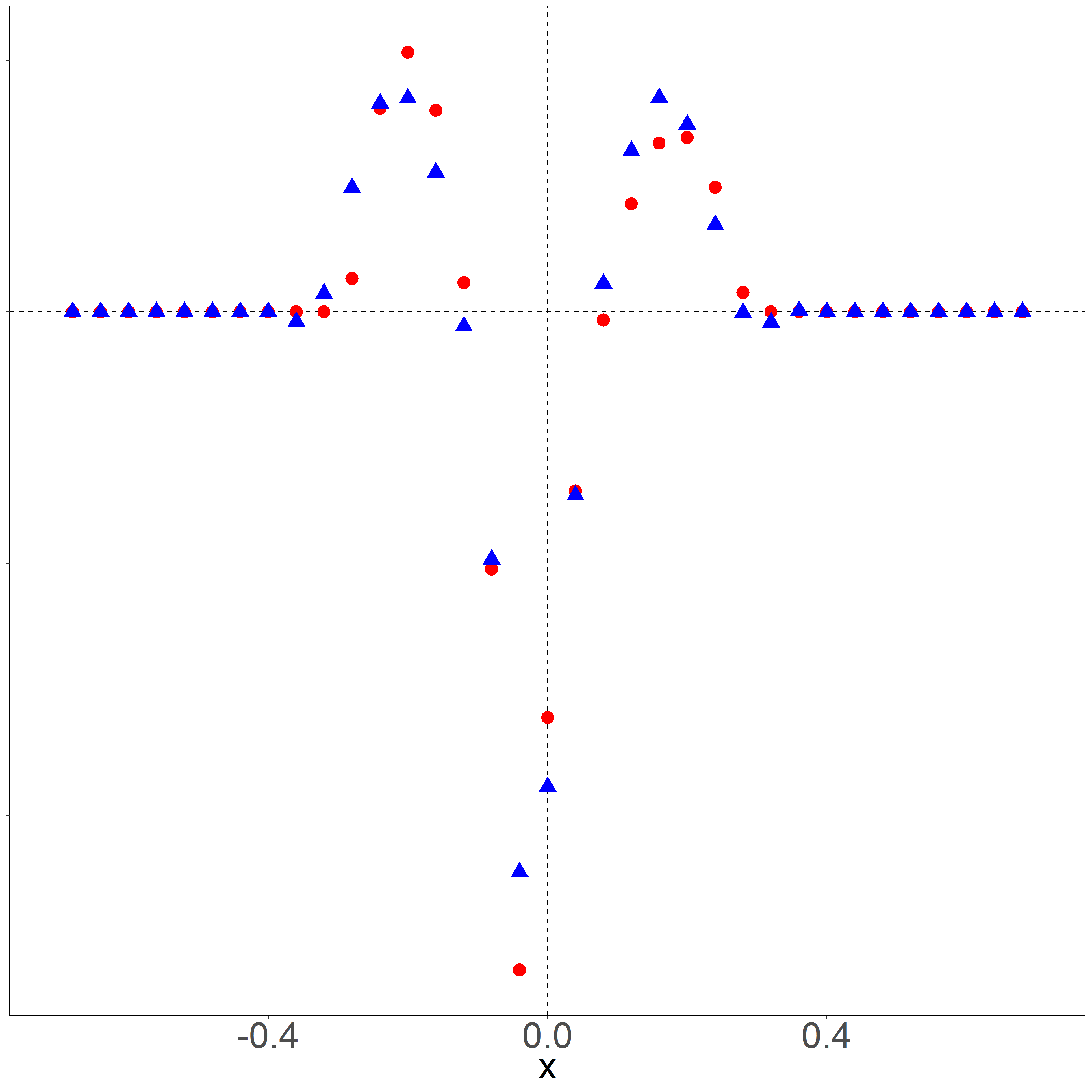} &   \includegraphics[width=55mm]{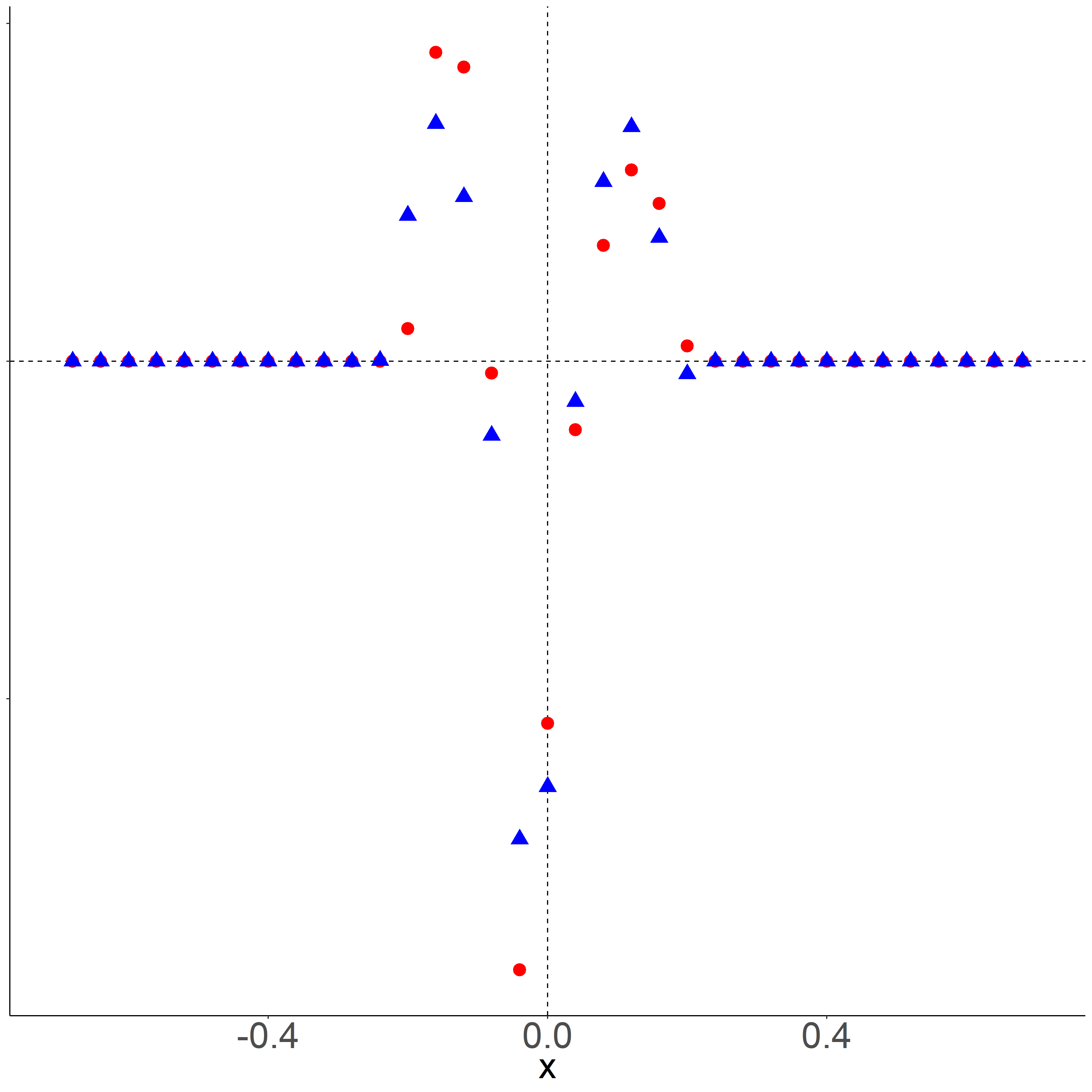} & \includegraphics[width=55mm]{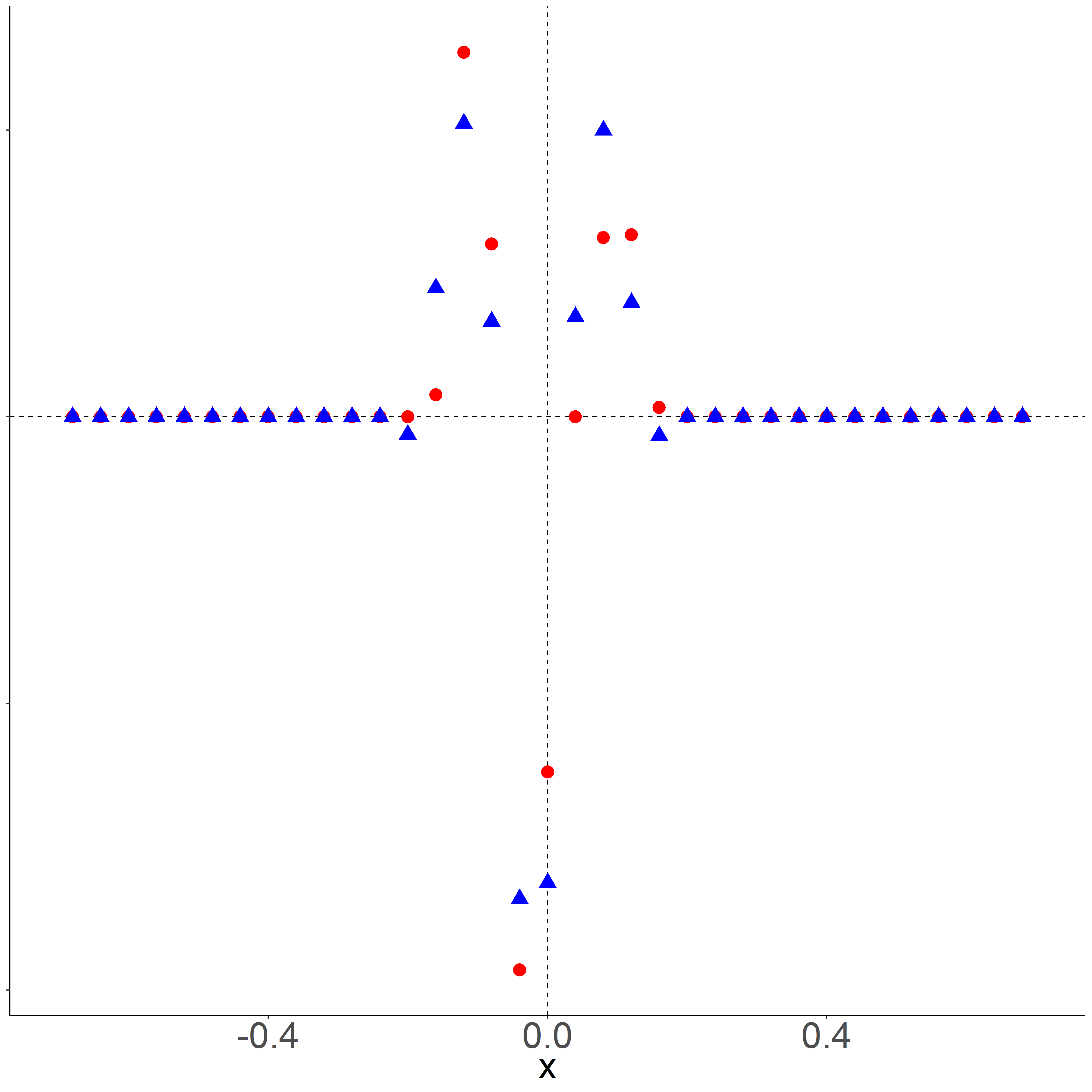} \\
  \raisebox{85pt}{\parbox[b]{.05\textwidth}{\begin{turn}{-90}$K=2000$\end{turn}}}%
  \includegraphics[width=55mm]{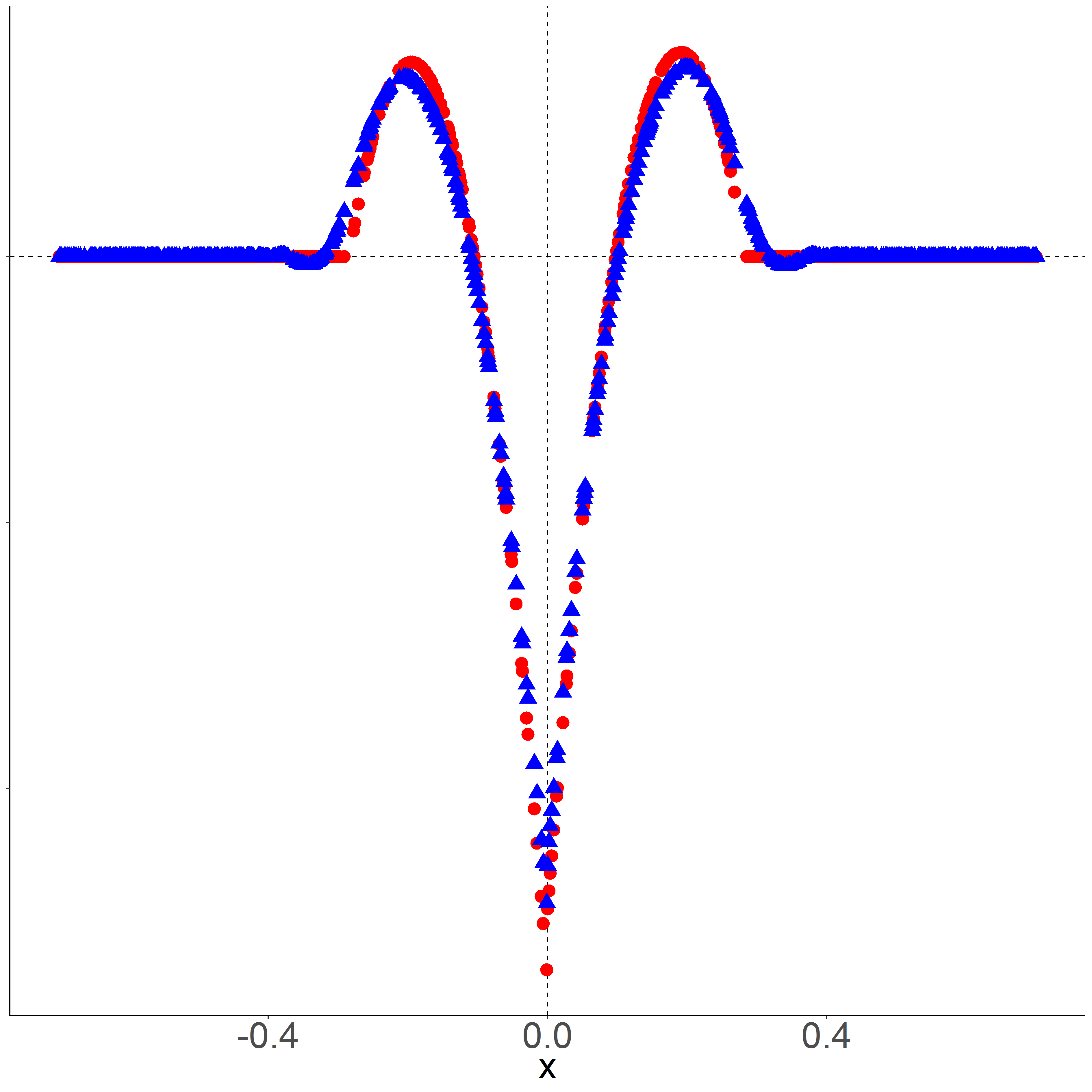} &   \includegraphics[width=55mm]{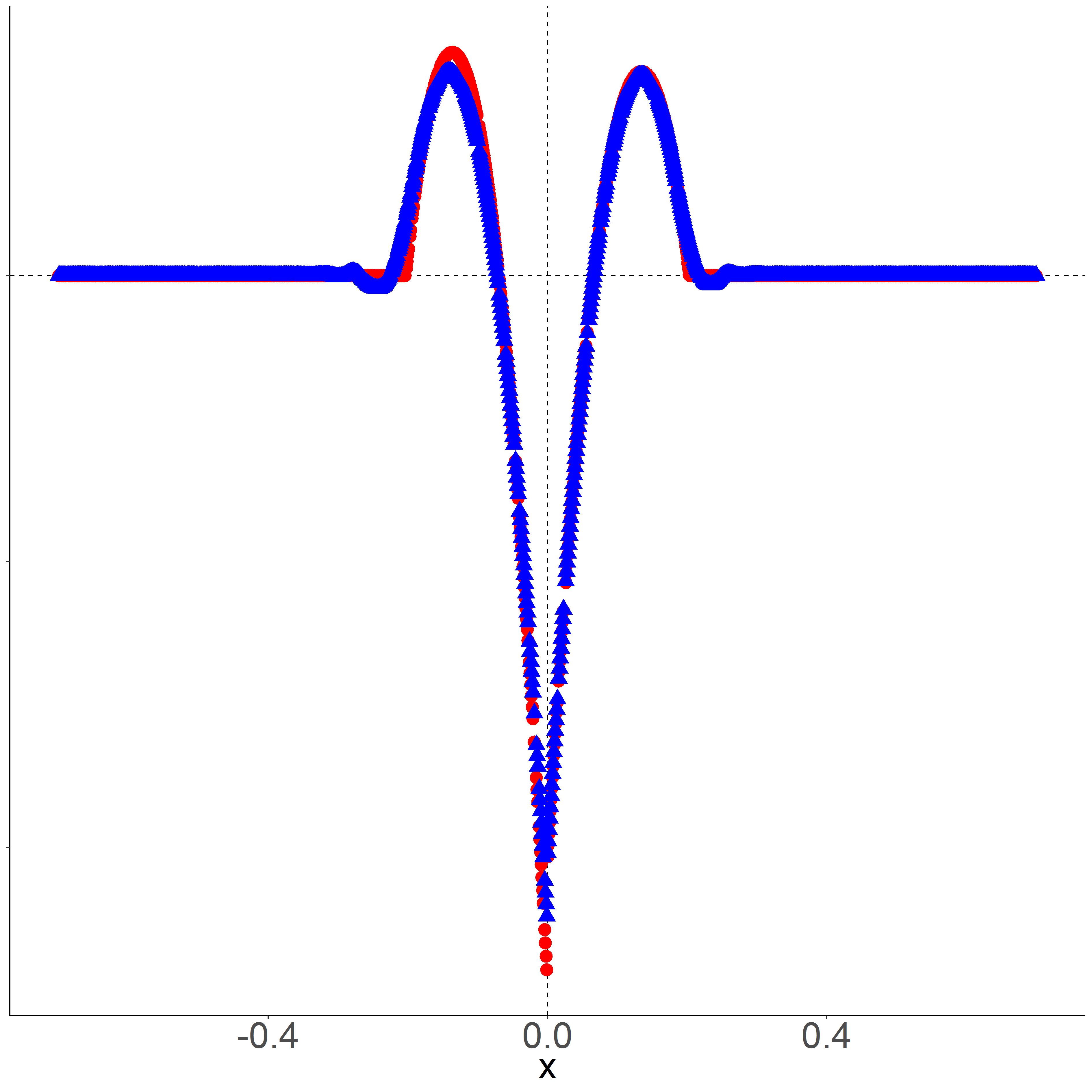} & \includegraphics[width=55mm]{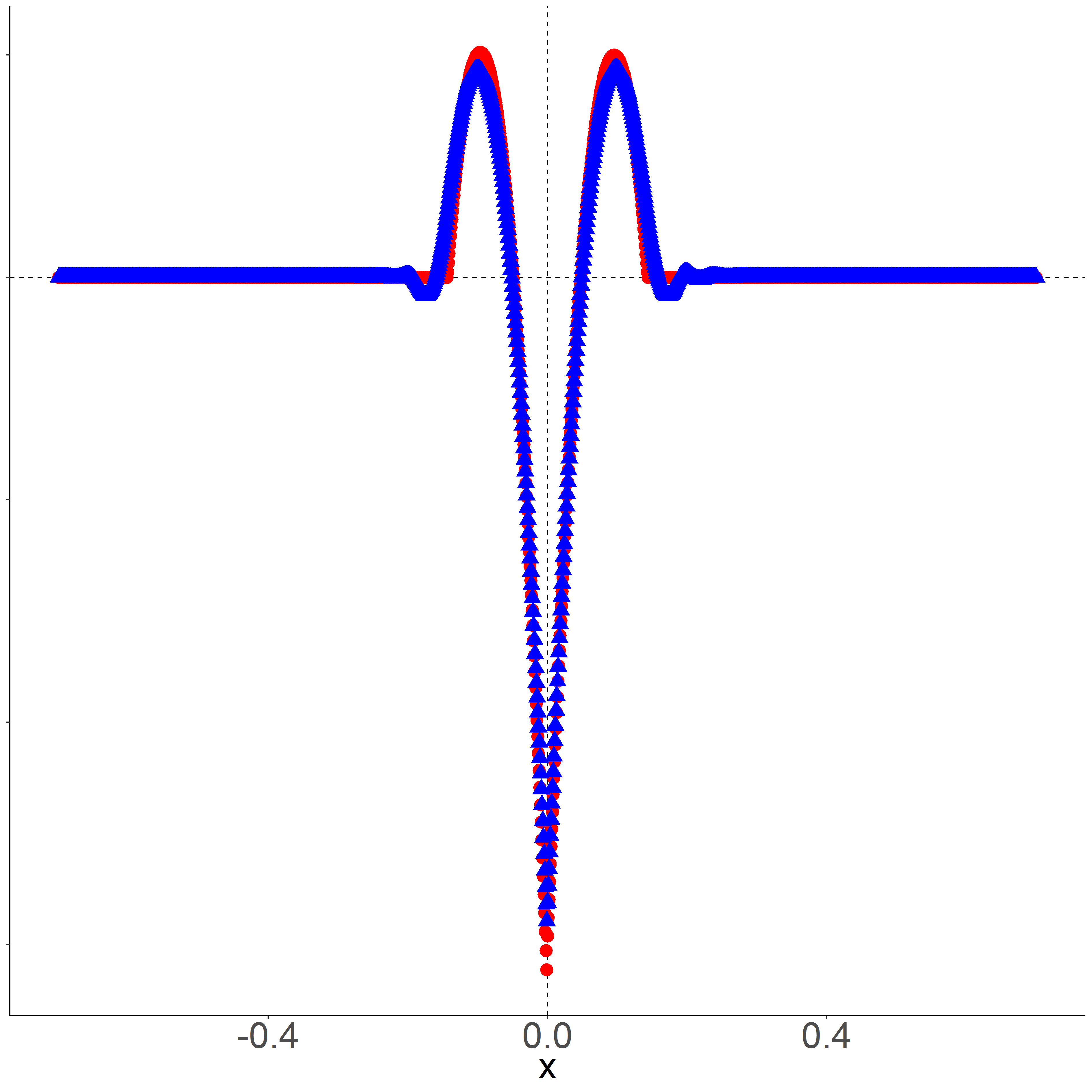} \\
  %\raisebox{85pt}{\parbox[b]{.05\textwidth}{\begin{turn}{-90}$r=200$\end{turn}}}%
  %\includegraphics[width=50mm]{Code_Results/plots/N_1000_points_200.png} &   \includegraphics[width=50mm]{Code_Results/plots/N_5000_points_200.png} & \includegraphics[width=50mm]{Code_Results/plots/N_25000_points_200.png} \\
(a) $n=1000$ & (b) $n=5000$ & (c) $n=25000$\\[6pt]
\end{tabular}}
\caption{Uniform MSE-optimal optimized linear (triangle) and fixed-length local linear (circle) weights for a coarse ($K=50$) and a dense ($K=2000$) uniform design on $[0,1]$ for different sample sizes, $L=2$ and $\sigma=0.2$. The plotted weights are scaled by $n^{4/5}$ to facilitate comparison across sample sizes.}
\label{Fig:three}
\end{figure}

\clearpage
\newpage
\noindent

\noindent
implies a larger lower bound for the efficiency gain relative to undersmoothed and RBC intervals based on valid bandwidth choices. Figure \ref{Fig:two} shows the relative risk of fixed 
length length-optimal and uniform MSE-optimal estimators in terms of interval length (a) and uniform MSE (b) relative to their optimized linear counterparts for the respective criterion for the continuous and discrete ($K=40$) uniform design for different sample sizes. It illustrates that the efficiency gains from optimization increase as the discreteness of the assignment variable becomes more severe.\footnote{This point was previously made in \citet{imbens2019optimized}. The convergence of the risk observed in Figure \ref{Fig:two} is due to manual bandwidth adjustments to ensure that the estimators are well defined.} This is because the shape of the optimized weighting function increasingly deviates from the local linear weights as the coarseness of the data becomes more severe, as shown in Figure \ref{Fig:three}, which depicts uniform MSE-optimal local linear and optimized weights for different degrees of discreteness.
In continuous designs fixed-length and the optimal linear kernels are nearly identical and the only advantage of the optimization based approach is that it can be easily modified to sharpen inference via shape constraints as discussed in Appendix \ref{appendix:F}.

\section{Empirical Illustration}

We apply our method to the data of \citet{landais2015assessing}, who estimates the effect of unemployment benefits on the duration of unemployment in a regression kink design. The paper exploits kinks in the schedule of unemployment benefits arising from a hard cap at a maximum benefit amount $b_{max}$. In the US, the weekly benefit amount $b$ received by an eligible unemployed is a fixed fraction $\gamma$ of a function of previous quarterly earnings $hqw$ in a base period up to the cap.
\begin{align*}
    b = \begin{cases} \begin{array}{lr}
        \gamma hqw, & \text{if } \gamma hqw \leq b_{max}\\
        b_{max}, & \text{if } \gamma hqw > b_{max}
        \end{array} 
        \end{cases}
\end{align*}

\begin{figure}[h!]
\centering
\makebox[\linewidth]{%
\includegraphics[scale=0.33]{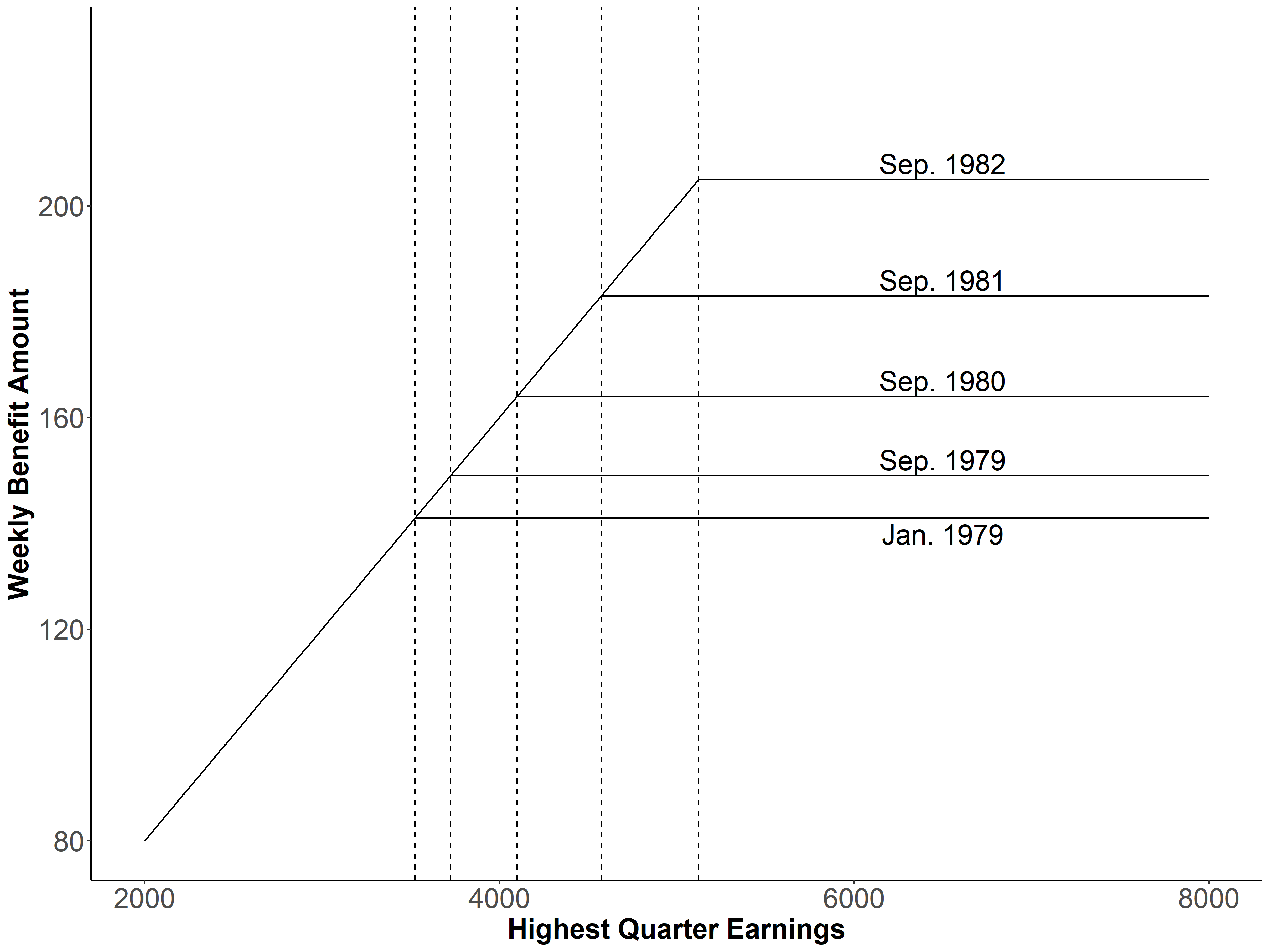}}
\caption{Weekly UI Benefit Schedule (USD), Louisiana 1979-1984 (\cite{landais2015assessing}).}
\label{Fig:four}
\end{figure}

\noindent
\citet{landais2015assessing} reports estimates of $\tau_{RKD}$ for five US states: Louisiana, Idaho, Missouri, New Mexico and Washington. For the sake of exposition, we focus on the results for Louisiana, which serves as the leading example in the paper. Figure \ref{Fig:four} displays the benefit schedule for Louisiana for the time period covered by the data. Due to adjustments, the maximum benefit level changed over time, resulting in five distinct kinks. For the period under consideration the weekly benefit rate was fixed at $\gamma=0.04$, which corresponds to a constant replacement rate of \\

%\makebox[\textwidth][c]{
%as reported in \citep{landais2015assessing}

\noindent
 52\% up to the respective kink, from where onwards the replacement rate decreases.

\begin{figure}[h!]
\centering
\makebox[\textwidth][c]{%
\begin{tabular}{cc}
    \includegraphics[width=80mm]{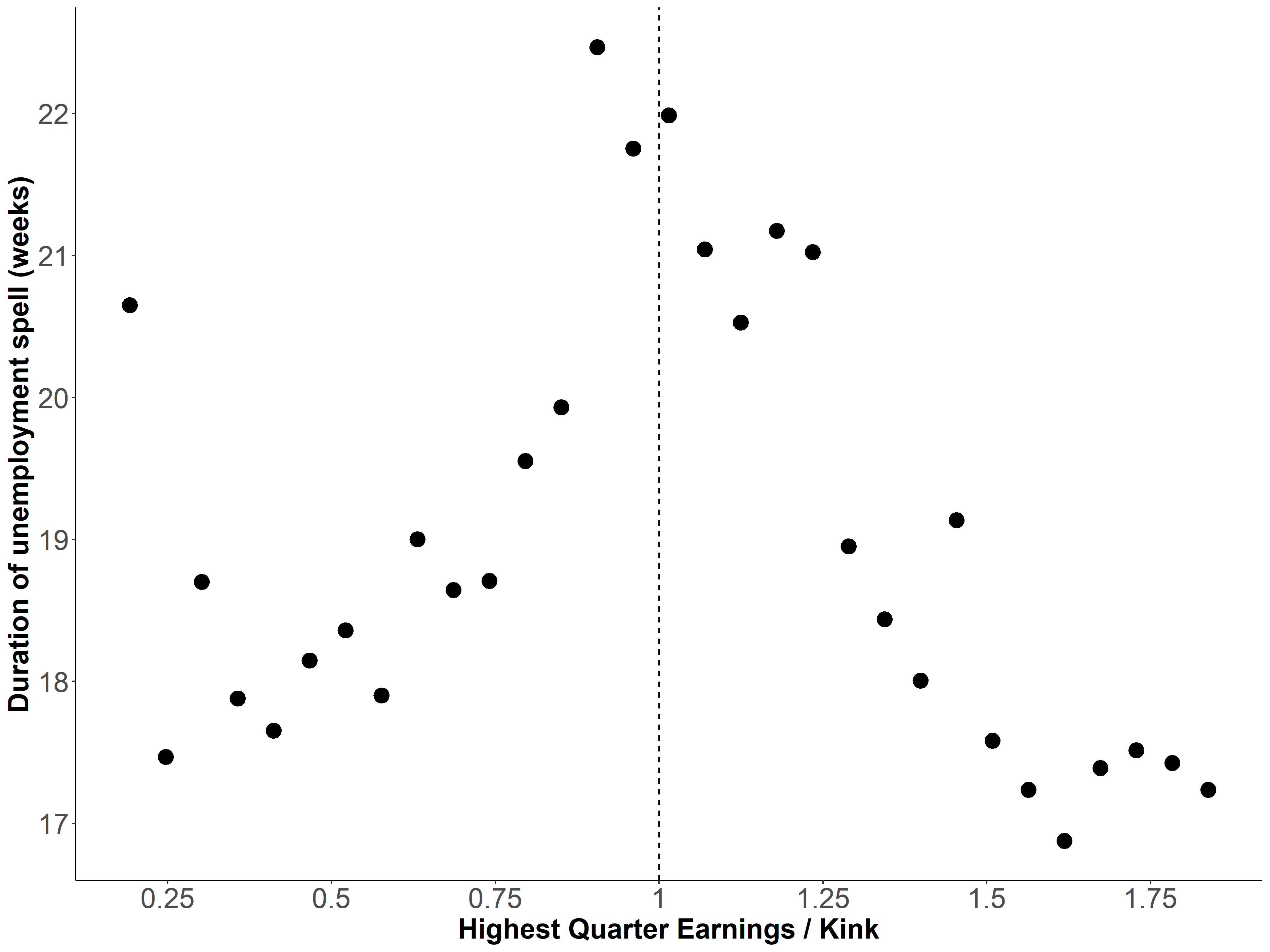} & \includegraphics[width=80mm]{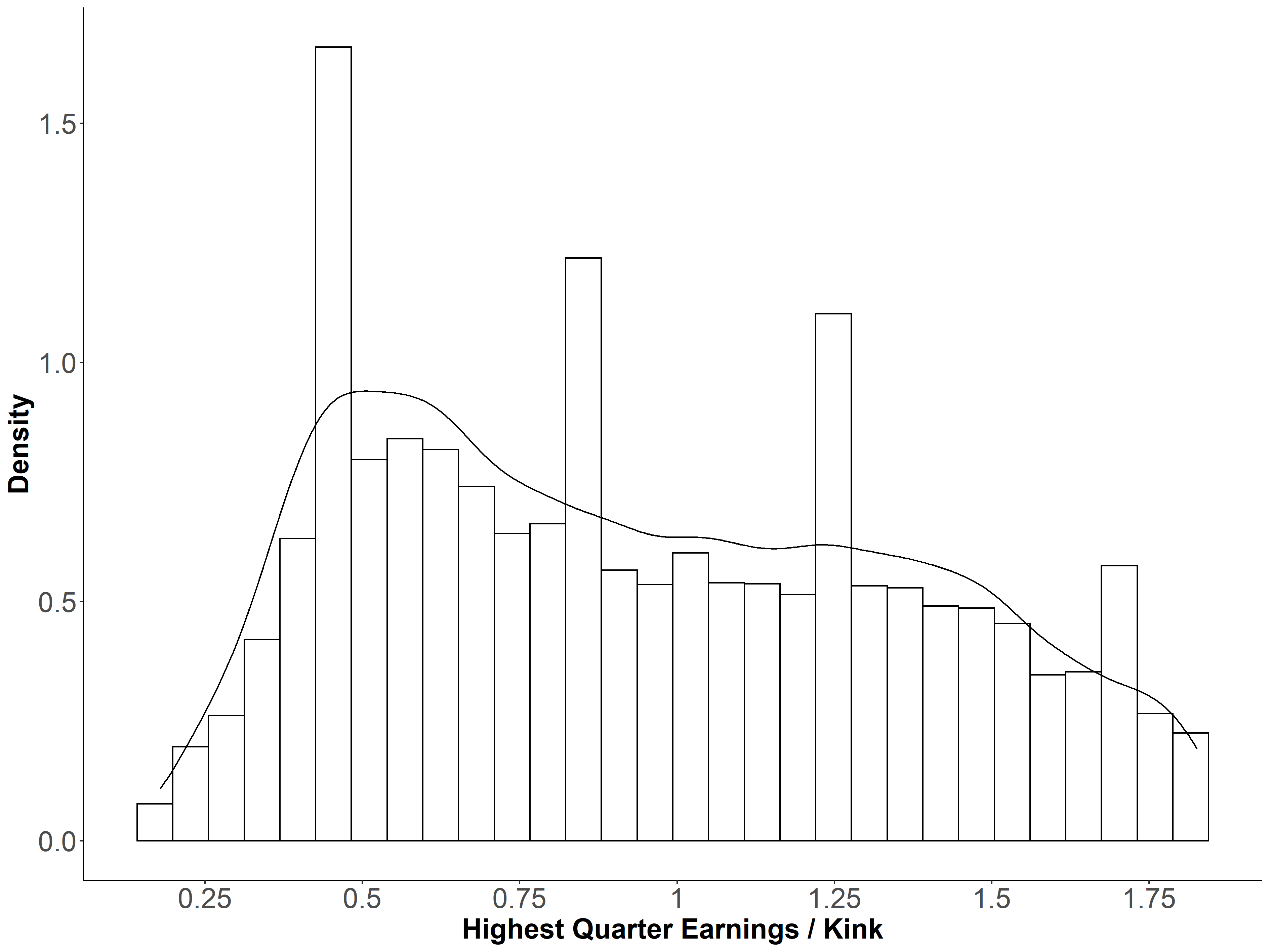} \\
(a) Average duration & (b) Density \\[6pt]
\end{tabular}}
\caption{Average duration (a) and density (b) in 30 equal width bins over [-0.15,1.85].}
\label{Fig:five}
\end{figure}

The paper utilizes data from the Continuous Wage and Benefit History (CWBH), a publicly available administrative UI data set for the US that contains  the universe of unemployment spells and wage records for the five US states from the late 1970s to 1984, with different states starting the recording at different points in time. For Louisiana, the dataset contains $n=44702$ unemployment spells for the whole time period. See \citet{landais2015assessing} Section II.A for a detailed description of the data. Figure \ref{Fig:five} plots the pooled data for all five time periods, where we have normalized the assignment variable (highest quarterly earnings) by the location of the respective kink and included only data within an 85\% interval of the kink  $[0.15,1.85]$. This corresponds to bandwidths in the range of 3000-4300 USD. In order to reduce noise, it shows the average unemployment durations, measured in weeks, in 30 equally wide bins. The paper reports RKD estimates of the effect of the benefit level on unemployment duration separately for each time period, with point estimates and standard errors rescaled to the 2010 USD price level. In the main specification, \citet{landais2015assessing} estimates $\tau_{RKD}$ using local linear regression with a fixed bandwidth $h=2500$\footnote{In the paper's online Appendix, the author reports robustness checks for the pooled sample consisting of the last two periods using bandwidths $1500$ and $4500$.} and reports conventional 95\% confidence intervals for $\tau_{RKD}$ based on Eicker-Huber-White standard errors. Table \ref{Tab:three} replicates\footnote{The point estimates in the top-left panel for period 3 and 4 deviate by $0.001$ from the results reported in the paper. We attribute this difference to rounding.} the results reported in Table 2 of the paper and additionally presents robust bias-corrected ($h=b=\hat{h}_{PMSE}$) and optimized linear $(L=\hat{L}_{ROT})$ point estimates and 95\% confidence intervals computed on the same data.

\vspace*{-1cm}
\noindent 
The point estimates reported in Table \ref{Tab:three} correspond to the estimated effects of a 1 USD increase in weekly benefits on the duration of paid unemployment in weeks. For example, the point estimate for period 4 reported in \citet{landais2015assessing} (top-left panel) suggest that a 1 dollar increase in weekly unemployment benefits leads to a $0.043$ weeks increase in the duration of unemployment at the kink. The estimates in the top-left panel correspond to (local) elasticities in the range of .2 and .7, suggesting that a 10\% increase in the average weekly benefit amount\\

\begin{table}[h!]\centering
  \begin{threeparttable}
  
\caption{RKD Estimates: Effect of the Benefit Level on Unemployment Duration}
\ra{1.05}

%\centering
%  \resizebox{\textwidth}{!}{\begin{minipage}{\textwidth}

\begin{tabular}{@{}lcccccccccccc@{}}\toprule
 & & \multicolumn{3}{c}{Landais (2015)} & &  \multicolumn{3}{c}{RBC}  \\
\cmidrule{3-5} \cmidrule{7-9}  
 Period & n &    Estimate & Interval & $h$ && Estimate & Interval & $h$  \\ \midrule
1 & 2493 & 0.006 & $[-0.005,0.018]$ & 2500    && 	 -0.012 	& $[-0.462,0.438]$ & 598 		\\
2 & 4580 & 0.018 &  $[0.007,0.028]$ & 2500	  && 	 -0.095 	& $[-0.293,0.103]$ & 1058  \\
3 & 4019 & 0.019 &  $[0.007,0.030]$ & 2500	  &&	 0.045 	& $[-0.250,0.339]$ & 886  \\
4 & 5577 & 0.043 &  $[0.025,0.061]$ & 2500	  &&	 -0.190 	& $[-0.517,0.137]$ & 1031  \\
5 & 10721 & 0.047 &  $[0.035,0.060]$ & 2500	  &&	 0.191 	&  $[0.043,0.339]$ & 1403  \\
\midrule
 & & \multicolumn{3}{c}{Optimized} & &  \multicolumn{3}{c}{Optimized}  \\
\cmidrule{3-5} \cmidrule{7-9}  

Period & n &    Estimate & Interval & $\kappa$ && Estimate & Interval & $\kappa$  \\ \midrule
1 & 2493 & 0.006 & $[-0.005,0.016]$ &0.638  	&& 	0.006 & $[-0.005,0.016]$  &  1	 		\\
2 & 4580 & 0.017 & $[-0.003,0.038]$ & 0.247  	&&	0.016 & $[-0.006,0.038]$  &  1	 		\\
3 & 4019 & -0.059 & $[-0.240,0.123]$ & 0.196 	&& 	-0.033 & $[-0.222,0.157]$ &   1 		 \\
4 & 5577 & 0.063 &  $[-0.042,0.168]$ & 0.210  	&& 	0.067 & $[-0.045,0.180]$ &   1  		 \\
5 & 10721 & 0.063 &  $[0.018,0.109]$ & 0.232  	&& 	0.075 &  $[0.027,0.123]$ &   1		 	 \\
\bottomrule

\end{tabular}
\label{Tab:three}

\begin{tablenotes}
\small
   \item \textit{Note:} Ad-hoc conventional local linear, robust bias-correction and optimized RKD point estimates and 95 \% confidence intervals for $\tau_{RKD}$ in the CWBH data for the periods Jan-Sep 1979 (Period 1), Sep 1979-Sep 1980 (Period 2), Sep 1980-Sep 1981 (Period 3), Sep 1981-Sep 1982 (Period 4) and Sep 1982-Dec 1983 (Period 5).
\end{tablenotes}
\end{threeparttable}
\end{table}

\newpage

\noindent
increases unemployment duration by 2 to 7\% on average at the kink. As can be seen from the confidence intervals reported in the remaining panels, this finding is sensitive to potential smoothing biases for most of the time periods under consideration, with both RBC and optimized intervals covering zero at the 95\% level for periods 1-4. However, while the RBC point estimates differ substantially from those reported in the top-left panel, with confidence intervals that are mostly uninformative for the question at hand, our procedure yields point estimates relatively close to those reported in \citet{landais2015assessing} and lower bounds of 95\% confidence intervals that are marginally below zero for most periods. This is because the estimated curvature is rather low relative to the estimators standard deviation in all data sets except for period 3, with $\hat{L}_{ROT} \times 10^5 = (0.0137, 0.058, 1.99, 0.477, 0.145)$. For the same reason, the difference between length-optimal and UMSE-optimal intervals is rather small in this data set as shown in the lower panel of Table \ref{Tab:four}. Overall, these results indicate that the uncertainty associated with the estimated effect of unemployment benefits on unemployment duration is higher than suggested by the top-left panel unless one is certain about the linearity of the CEF in the domain specified by the bandwidth.

As mentioned earlier, the optimization approach to bias-aware RKD inference allows us to readily sharpen inference by imposing shape constraints on the CEF. This is often useful, as the plateaus in the schedules that are typically exploited in regression kink designs often give rise to empirically plausible concavity and convexity restrictions. In Appendices \ref{appendix:F} and \ref{appendix:D}, we discuss this further and explain how shape constraints are implemented in practise. Table \ref{Tab:four} presents the counterparts of the optimized intervals reported in the lower panel of Table \ref{Tab:three} under the restriction that $\mu$ is a concave function, demonstrating that such constraints can substantially shrink our confidence set.\\

\begin{table}[h!]\centering
  \begin{threeparttable}
\caption{Optimized Intervals under Shape Constraints: Concavity}
\ra{1.1}
%\centering
%  \resizebox{\textwidth}{!}{\begin{minipage}{\textwidth}
\begin{tabular}{@{}lccccccccccc@{}}\toprule
 & &  \multicolumn{3}{c}{Length-optimal} & &  \multicolumn{3}{c}{UMSE-optimal}  \\
\cmidrule{3-5} \cmidrule{7-9}  
Period & $n$ &    Estimate & Interval & $\kappa$ && Estimate & Interval & $\kappa$  \\ \midrule
1 & 2493 & 0.001 & $[-0.006,0.009]$ & 1.220 && 0.001 & $[-0.006,0.009]$ &    1 \\
2 & 4580 & 0.005 & $[-0.002,0.011]$ & 1.154 && 0.005 & $[-0.002,0.011]$ & 1 \\
3 & 4019 & 0.002 & $[-0.004,0.008]$ & 1.173 && 0.002 & $[-0.004,0.008]$ &   1 \\
4 & 5577 & 0.034 & $[0.026,0.042]$ & 1.416 && 0.034 & $[0.026,0.043]$  & 1 \\
5 & 10721 & 0.023 & $[0.018,0.028]$ & 1.319 && 0.023 & $[0.018,0.028]$  &  1 \\
\bottomrule

\end{tabular}
\label{Tab:four}

\begin{tablenotes}
\small
   \item \textit{Note:} Shape constrained (concavity of $\mu$) optimized linear confidence intervals for $\tau_{RKD}$ in the CWBH data for the periods as in Table \ref{Tab:three}.
\end{tablenotes}
\end{threeparttable}
\end{table}

\section{Conclusion}

Motivated by the finite sample coverage problems of pointwise approaches to nonparametric inference, this paper proposes a robust and efficient alternative method for the construction of nonparametric confidence intervals in regression kink designs. Given a curvature bound, the method is fully data driven, easy to implement, and has excellent finite sample coverage and length properties due to its minimax construction that explicitly takes into account the worst-case smoothing bias in a given data set.\\

\newpage

\appendix
\section*{Appendices}
\addcontentsline{toc}{section}{Appendices}
\renewcommand{\thesubsection}{\Alph{subsection}}

\setcounter{section}{0}
\setcounter{figure}{0}
\setcounter{table}{0}

\renewcommand\thefigure{\thesubsection.\arabic{figure}}    
\renewcommand\thetable{\thesubsection.\arabic{table}}

%%%% APPENDIX A

\subsection{Bias Derivation}
\label{appendix:A}

The proposition that the weights solving problem (\ref{qp3}) also solve the program (\ref{qp1}) relied on two arguments. First, we argued that the weights solving (\ref{qp1}) must lie in the feasible set of (\ref{qp3}) by means of simple Taylor expansions. Second, we argued that by the result in (\ref{eq2}), the worst-case conditional bias of $\hat{\theta}$ over $\mathcal{F}(L)$ is proportional to the smoothness bound $L$ and, by definition of the remainder, the same as the worst-case conditional bias over the normalized class $\bar{\mathcal{F}}(L)$. The second argument relied on an explicit formula for the remainder of Taylor expansions near the threshold obtained via integration by parts and the Fubini-Tonelli Theorem.\\

%with $\int_{0}^{x} |\mu^{2}(t)|dt \leq \infty$ for each $x \in \mathcal{X}$
\noindent
\textit{Proof.} Since $\mu \in \mathcal{F}(L)$, $\mu^{1}$ is Lipschitz by definition and hence absolutely continous, and has second derivative almost everywhere with $\underset{x \in \mathcal{X}}{\sup}|\mu^{2}(x)| \leq L$
. By the Fundamental Theorem of Calculus we have that for each $X_i \in \mathcal{X}_{+}$ and $\mu \in \mathcal{F}(L)$
\begin{equation*}
\mu_{+}(X_i) = \mu_{+}(0) + \int_{0}^{X_i} \mu_{+}^{1}(t) dt.
\end{equation*}

\noindent
Set $u(t) = \mu_{+}^{1}(t)$ and $v(t) = X_i - t$.  Integration by parts then yields
\begin{align*}
\mu_{+}(X_i) &= \mu_{+}(0) -  \int_{0}^{X_i} u(t) v^{1}(t) dt\\
&= \mu_{+}(0) - \left[ \mu_{+}^{1}(t) (X_i - t) \right]_{0}^{X_i} + \int_{0}^{X_i} \mu_{+}^{2}(t) (X_i - t) dt\\
&= \mu_{+}(0) + \mu_{+}^{1}(0) X_i + \int_{0}^{X_i} \mu_{+}^{2}(t) (X_i -t) dt.
\end{align*}

\noindent
Analogously, for each $X_i \in \mathcal{X}_{-}$ and $\mu \in \mathcal{F}(L)$, $\mu_{-}(X_i) = \mu_{-}(0) + \mu_{-}^{1}(0) X_i - \int_{X_i}^{0} \mu_{-}^{2}(t) (X_i -t) dt$. The integral form of the remainder allows us to write the conditional mean as
\begin{align*}
\E [ \hat{\theta} | X_{N}]&=  \sum_{i=1}^{N} w_{+} \left[ \mu_{+}(0) + \mu_{+}^{1}(0) X_i + \int_{0}^{X_i} \mu_{+}^{2}(t) (X_i -t) dt \right]\\
& + \sum_{i=1}^{N} w_{-} \left[ \mu_{-}(0) + \mu_{-}^{1}(0) X_i - \int_{X_i}^{0} \mu_{-}^{2}(t) (X_i -t) dt \right],
\end{align*}

\noindent
which we used to confirm the first statement. The constraints
\begin{align*}
\sum_{i=1}^{n} w_{i,+} = 0 \qquad \sum_{i=1}^{n} w_{i,-} = 0 \qquad \sum_{i=1}^{n} w_{i,+} X_i = 1 \qquad \sum_{i=1}^{n} w_{i,-} X_i = -1 
\end{align*}
are in fact necessary conditions for the worst-case bias over $\mathcal{F}(L)$ to be finite, as, if they are not all satisfied, we can choose ($\mu_{+},\mu_{-},\mu_{+}^{1},\mu_{-}^{1})$ to induce an arbitrary conditional bias. As a consequence, we have that, if the worst-case conditional bias is finite, it is given by 
\begin{align*}
\sup_{ \mu \in \mathcal{F}(L)} \E [ \hat{\theta} - \theta | X_{N}] &= \sup_{ \mu \in \mathcal{F}(L)} \left[ \sum_{i=1}^{n} w_{i,+} \int_{0}^{X_i} \mu_{+}^{2}(t) (X_i -t) dt
- \sum_{i=1}^{n} w_{i,-}  \int_{X_i}^{0} \mu_{-}^{2}(t) (X_i -t) dt \right].
\end{align*}

\noindent
In order to derive the result (\ref{eq2}) we define the two measure spaces $(\mathbb{R},\mathscr{L},\upsilon_{1})$ and $(\mathbb{N},\mathscr{P}(\mathbb{N}),\upsilon_{2})$, where $\mathscr{L}$ denotes the Lebesgue $\sigma$-algebra, $\upsilon_{1}$ the Lebesgue measure, $\mathscr{P}(\mathbb{N})$ the power set of the natural numbers and $\upsilon_{2}$ the counting measure. Note that the measures are $\sigma$-finite on the real numbers and natural numbers respectively. Let $( \mathbb{N} \times \mathbb{R}, \mathscr{A}, \upsilon)$ denote the product space, where $\mathscr{A}$ is the generated product $\sigma$-algebra and $\upsilon$ is the unique product measure. Moreover, let $g: \mathbb{N} \times \mathbb{R} \mapsto \mathbb{R}$, $g(i,t) = \tilde{w}(i) \mu^{2}(t) ( \tilde{t}(i)-t)$, where $\tilde{w}(i) = w_{i} \cdot \mathbbm{1}_{[1 \leq i \leq n]}$ and $\tilde{t}(i) = X_{i} \cdot \mathbbm{1}_{[1 \leq i \leq n]}$ and define the sets $A_{+} = \lbrace (i,t) \in \mathbb{N} \times \mathbb{R} \phantom{i} \vert \phantom{i} 0 \leq t \leq \tilde{t}(i) \rbrace$ and $A_{-} = \lbrace (i,t) \in \mathbb{N} \times \mathbb{R} \phantom{i} \vert \phantom{i} \tilde{t}(i) \leq t < 0 \rbrace$. Note that by definition, $\sum_{i=1}^{n} w_{i,+} \int_{0}^{X_i} \mu_{+}^{2}(t) (X_i -t) dt = \int_{\mathbb{N}} \int_{0}^{\tilde{t}(i)} g(i,t) d \upsilon_{1}(t) d \upsilon_{2}(i) \equiv T_{1}$ and $\sum_{i=1}^{n} w_{i,-}  \int_{X_i}^{0} \mu_{-}^{2}(t) (X_i -t) dt =  \int_{\mathbb{N}} \int_{\tilde{t}(i)}^{0} g(i,t) d \upsilon_{1}(t) d \upsilon_{2}(i) \equiv T_{2}$. By the Fubini–Tonelli Theorem it holds\footnote{Note that if $\max_{1\leq i \leq n} \lbrace |w_{i}| \rbrace < \infty$ boundedness of $\mathcal{X}$ and $\mu^{2}$ implies $ \int_{A_{\pm}} |g(i,t)| d \upsilon  < \infty$.} that
%\sum_{i=1}^{n} w_{i,+} \int_{0}^{X_i} \mu_{+}^{2}(t) (X_i -t)dt =
\begin{align*}
T_{1} = \int_{A_{+}} g(i,t) d\upsilon &= \int_{0}^{\infty} \int_{\tilde{t}(i)\geq t} g(i,t) d \upsilon_{2}(i) d \upsilon_{1}(t) = \int_{0}^{\infty} \mu_{+}^{2}(t) \sum_{i: X_{i} \in [t,\infty) } w_{i,+}(X_i - t)dt,\\
T_{2} = \int_{A_{-}} g(i,t) d\upsilon &= \int_{-\infty}^{0} \int_{\tilde{t}(i) < t} g(i,t) d \upsilon_{2}(i) d \upsilon_{1}(t) = \int_{-\infty}^{0} \mu_{-}^{2}(t) \sum_{i: X_{i} \in (-\infty,t]} w_{i,-}(X_i - t)dt,
\end{align*}
\noindent
and it follows that under the constraints,
\begin{align*}
\sup_{ \mu \in \mathcal{F}(L)} \E [ \hat{\theta} - \theta | X_{N}] &= \sup_{ \mu \in \mathcal{F}(L)} \left[ \int_{0}^{\infty} \mu_{+}^{2}(t) \sum_{i: X_{i} \in [t,\infty) } w_{i,+}(X_i - t) dt - \int_{-\infty}^{0} \mu_{-}^{2}(t) \sum_{i: X_{i} \in (-\infty,t]} w_{i,-}(X_i - t)dt \right].
\end{align*}
\noindent
This yields (\ref{eq2}) and, by the definition of the remainder\footnote{Note that $|R^{1}(x)-R^{1}(x')| = |\mu^{1}(x) - \mu^{1}(x')| \leq |x-x'|$ for any $x,x' \in \mathcal{X}$ and $\mu \in \mathcal{F}(L)$.}, an immediate consequence is 
\begin{align*}
\bar{B}(w_{N}) = \sup_{ \mu \in \mathcal{F}(L)} \E [ \hat{\theta} - \theta | X_{N}] = L  \sup_{ \mu \in \mathcal{F}(1)} \E [ \hat{\theta} - \theta | X_{N}] = L \sup_{ R \in \bar{\mathcal{F}}(1)}  \sum_{i=1}^{n} w_{i} R(X_{i}).
\end{align*}
\noindent
with $\bar{\mathcal{F}}(1) = \left\lbrace f: f(0)=f^{1}(0)=0 \wedge | f^{1}(x) - f^{1}(x') | \leq |x-x'|, x,x' \in \mathcal{X} \right\rbrace$. This shows the equivalence of problems (\ref{qp1}) and (\ref{qp3}) in terms of optimal weights and implies that, by symmetry of $\mathcal{F}(L)$ with respect to zero, $r^{*}L$ is a sharp bound on the magnitude of $\bar{B}(w_{N})$.

\noindent
\textit{Remark.} Provided that  $\sum_{i=1}^{n} w_{i,+} X_{i}^j = - \sum_{i=1}^{n} w_{i,-} X_{i}^j =  \delta_{\varv,p}$ for $j=1,\dots,p$, iteratively integrating by parts and applying the same logic yields a general formula for the conditional bias of a linear estimator of $\theta^{\varv}$ as a function of the weights and $\mu^{p+1}$:
\begin{align*}
\frac{1}{p!} \left[ \int_{0}^{\infty} \mu_{+}^{p+1}(t) \sum_{i: X_{i} \in [t,\infty) } w_{i,+} (X_i - t)^{p} dt - \int_{-\infty}^{0} \mu_{-}^{p+1}(t) \sum_{i: X_{i} \in (-\infty,t]} w_{i,-} (X_i - t)^{p} dt \right].
\end{align*}

%\E [ \hat{\theta} - \theta | X_{N}] =

%%%%%%%%%%%%% APPENDIX B

\subsection{Constraint Qualification}
\label{appendix:B}

In Section 3.4, we argued that, by strong duality of (\ref{qp3}), we can recover the solution of (\ref{qp3}), and thus the weights solving (\ref{qp1}) as well as the associated worst-case bias over $\mathcal{F}(L)$, by solving the dual problem underlying (\ref{qp5}). Because (\ref{qp3}) is a strictly convex quadratic problem and our constraints are all linear equalities and inequalities, we can verify strong duality by showing that a refined Slater's condition (cf. \cite{boyd2004convex}) applies. This constraint qualification states that strong duality holds whenever there exists a feasible point for (\ref{qp3}), which we demonstrate by showing that the weights of the local polynomial estimator of order $p\geq 1$ together with $r$ corresponding to its worst-case conditional bias over $\mathcal{F}(1)$ satisfy all constraints of (\ref{qp3}).\\

\noindent
\textit{Proof.} Let $K_h(t)= K(t/h)$ for a bounded kernel function $K$ with support $[-1,1]$ indexed by a bandwidth $h>0$ and let $e_j$ denote the unit column vector of length $2p+2$ with a one at position $j$. The local polynomial estimator of order $p\geq \varv$ for a jump in the $\varv$-th derivative is $\hat{\theta}_{LP(\varv,p)}= \varv ! w_{LP(\varv,p)}^{T} Y_{N}$ with weights given by $w_{LP(\varv,p)} = e_{p+2+ \varv}^{T} \left( X_{0}^{T}W_0 X_0 \right)^{-1} X_{0}^{T} W_0$,
%\begin{align*}
%w_{LP(v,p)} = e_{p+2+v}^{T} \left( X_{0}^{T}W_0 X_0 \right)^{-1} X_{0}^{T} W_0,
%\end{align*}
\noindent
where the matrices $X_{0}$ $(n \times 2p+2)$ and $W_{0}$ $(n \times n)$ are defined as $W_0 = \operatorname{diag} \left( \begin{matrix} K_h(X_1) & \cdots & K_h(X_n) \end{matrix} \right)$ and $X_{0} = [\chi_{1},\dots, \chi_{n}]^{T}$ with $\chi_{i} = \left( 1, X_{i}, \dots, X_{i}^p,  D(X_i),  X_{i} D(X_{i}), \dots ,  X_{i}^p D(X_{i}) \right)$.
%\begin{align*}
%X_0 = \left[ \begin{matrix}
%1 & X_1 &\cdots & X_1^p & D(X_1) & X_1 D(X_1)& \cdots &  X_1^p D(X_1)\\
% 1 & X_2 & \cdots & X_2^p &  D(X_2) & X_2 D(X_2)& \cdots &  X_2^p D(X_2) \\
%\vdots & \vdots & \cdots & \vdots & \vdots & \vdots & \cdots & \vdots\\
%1 & X_n &\cdots & X_n^p & D(X_n) & X_n D(X_n)& \cdots &  X_n^p D(X_n)\\
%\end{matrix} \right]
%W_0 = \operatorname{diag} \left( \begin{matrix} K_h(X_1) & \cdots & K_h(X_n)
%\end{matrix} \right).
%\end{align*}
\noindent
From this definition, it immediately follows that a generalization of the equality constraints of (\ref{qp3}) hold, since the local polynomial estimator is unbiased for polynomials of order $j \leq p$
\begin{align*}
& \sum_{i=1}^{n} w_{i,+} X_i^{\varv} =  e_{p+2+\varv}^{T} \left( X_{0}^{T} W_0 X_0 \right)^{-1} X_{0}^{T} W_0 X_0 e_{p+2+\varv} = 1\\
& \sum_{i=1}^{n} w_{i} X_i^{\varv} = e_{p+2+\varv}^{T} \left( X_{0}^{T} W_0 X_0 \right)^{-1} X_{0}^{T} W_0 X_0 e_{1+\varv} = 0\\
& \sum_{i=1}^{n} w_{i,+} X_{i}^{j} = e_{p+2+\varv}^{T} \left( X_{0}^{T} W_0 X_0 \right)^{-1} X_{0}^{T} W_0 X_0 e_{p+2+j} = 0\\
& \sum_{i=1}^{n} w_{i} X_{i}^{j} = e_{p+2+\varv}^{T} \left( X_{0}^{T} W_0 X_0 \right)^{-1} X_{0}^{T} W_0 X_0 e_{1+j} = 0 \qquad \text{for } p \geq j \neq v.
\end{align*}
%By (\ref{eq2}), we have $\bar{B}_{n}(w_{LL}) = L \int_{-\infty}^{\infty} | \bar{w}(t)| dt$ with $\bar{w}(t)$ defined as in Section 3.3.

\noindent
Moreover, the integral form of the remainder immediately provides us with a feasible bound $r$ as $ \underset{ R \in \mathcal{\bar{F}}(1)}{\sup}  \E \left[ \hat{\theta}_{LP(\varv,p)} - \theta^{\varv}|X_{N} \right] = \underset{ R \in \mathcal{\bar{F}}(1)}{\sup}  \E \left[ \varv! \sum_{i=1}^{n} w_i R(X_i) | X_{N} \right] \leq  \sum_{i=1}^{n} \big \lvert w_{i} X_{i}^{p+1} \frac{\varv!}{(p+1)!} \big \rvert$. Thus, the problem (\ref{qp3}) is strictly feasible, and, by Slater's condition, strong duality holds.

It turns out that the simple bound obtained via this approach is sharp for general local polynomial estimators. To show this explicitly, assume that $\hat{\theta}_{LP(\varv,p)}$ is well defined and note that
\begin{align*}
\bar{w}_{LP(\varv,p),+}(t) &= \sum_{X_{i} \geq t} w_{i,LP(\varv,p)} (X_{i}-t)^{p} = e_{p+2+\varv}^{T} \left( X_{0}^{T} W_0 X_0 \right)^{-1} X_{0}^{T} W_{0} \left( \begin{matrix}
(X_{1}-t)^{p} D(X_{1}-t) \\
\vdots \\
(X_{n}-t)^{p} D(X_{n}-t)
\end{matrix} \right).
\end{align*}
%&= \sum_{i=1}^{n} w_{LP(v,p),i,+} (X_{i}-t)^{p-1} D(X_{i}-t)

\noindent
It follows from regression principles that $\bar{w}_{LP(\varv,p),+}(t)$ and $\bar{w}_{LP(\varv,p),-}(t)$ can be understood as the $(\varv+1)$-th coefficients from a weighted polynomial regression of  $\Delta_{i,+}(t) = (X_{i}-t)^{p} D(X_{i}-t)$ and $\Delta_{i,-}(t) = (X_{i}-t)^{p} (1-D(X_{i}-t))$ on $M_{i} = (1 , X_i , \cdots , X_i^p)$ based on subsets of the data with $X_{i} \geq 0$ and $X_{i} < 0$ respectively. The corresponding regressions in turn can be understood as Tikhonov regularized least-squares problems, which allows us to show that the regression coefficients $\bar{w}_{LP(\varv,p),+}(t)$ and $\bar{w}_{LP(\varv,p),-}(t)$ are attenuated versions of coefficients with deterministic sign. To show this formally, we define the following objects
\begin{alignat*}{2}
\mathbf{X}_{+}(t) &= [M_{i}^{T}, \dots ]: &&\text{ Matrix of $M_{i}$'s for units with $0\leq t \leq X_{i}$,}\\
\Gamma_{+}(t) &= [M_{i}^{T}, \dots]: &&\text{ Matrix of $M_{i}$'s for units with $0\leq X_{i}  < t$,}\\
W_{\mathbf{X}_{+}}(t) &= \operatorname{diag} (K_{h}(X_{i}), \dots): &&\text{ Matrix of weights for units with $0\leq t  < X_{i}$,}\\
W_{\Gamma_{+}}(t) &= \operatorname{diag} (K_{h}(X_{i}), \dots): &&\text{ Matrix of weights for units with $0\leq X_{i}  < t$,}\\
\Delta_{X,+}(t) &= [\Delta_{i,+}(t), \dots]: &&\text{ Vector of $\Delta_{i,+}(t)$'s for units with $X_{i} \geq 0$.}
\end{alignat*}
\noindent
Analogously,
\begin{alignat*}{2}
\mathbf{X}_{-}(t) &= [M_{i}^{T}, \dots]: &&\text{ Matrix of $M_{i}$'s for units with $X_{i} \leq t \leq 0$},\\
\Gamma_{-}(t) &= [M_{i}^{T}, \dots]: &&\text{ Matrix of $M_{i}$'s for units with $t\leq X_{i}  < 0$},\\
W_{\mathbf{X}_{-}}(t) &= \operatorname{diag} (K_{h}(X_{i}), \dots): &&\text{ Matrix of weights for units with $X_{i} \leq t \leq 0$},\\
W_{\Gamma_{-}}(t) &= \operatorname{diag} (K_{h}(X_{i}), \dots): &&\text{ Matrix of weights for units with $t\leq X_{i}  < 0$},\\
\Delta_{X,-}(t) &= [\Delta_{i,-}(t), \dots]: &&\text{ Vector of $\Delta_{i,+}(t)$'s for units with $X_{i} < 0$}.
\end{alignat*}

\noindent
Let $\norm{\cdot}_{2}$ denote the Euclidean norm. In this notation, $\bar{w}_{LP(\varv,p),+}(t)$ and $\bar{w}_{LP(\varv,p),-}(t)$ are the $(\varv+1)$-th elements of the solutions to the "regularized" least-squares problems 
\begin{align*}
\min_{\gamma \in \mathbb{R}^{p+1}} \norm{ W_{\mathbf{X}_{\pm}}(t)^{\frac{1}{2}} \left[ \mathbf{X}_{\pm}(t) \gamma - \Delta_{X,\pm}(t)\right] }_{2}^{2} +\norm{ W_{\Gamma_{\pm}}(t)^{\frac{1}{2}} \Gamma_{\pm}(t) \gamma }_{2}^{2}.
\end{align*}

\noindent
Assume without loss of generality that $\Delta_{X,+}(t)$ and $\Delta_{X,-}(t)$  are sorted in ascending and descending order respectively. By least squares algebra, it then follows that $\bar{w}_{LP(\varv,p),\pm}(t)$ is 
\begin{align*}
\hat{\gamma}_{\varv} = e_{\varv+1}^{T} \left( \mathbf{X}_{\pm}(t)^{T} W_{\mathbf{X}_{\pm}}(t) \mathbf{X}_{\pm}(t) + \Gamma_{\pm}(t)^{T} W_{\Gamma_{\pm}}(t) \Gamma_{\pm}(t) \right)^{-1} \mathbf{X}_{\pm}(t)^{T} 
\left[ \begin{matrix}
W_{\Gamma_{\pm}}(t) & \mathbf{0} \\
\mathbf{0} & W_{\mathbf{X}_{\pm}}(t)
\end{matrix} \right]
\Delta_{X,\pm}(t).
\end{align*}
\noindent
Note that, for any $t$, the "unregularized" subproblem $\min_{\gamma \in \mathbb{R}^{p+1}} \norm{ W_{\mathbf{X}_{\pm}}(t)^{\frac{1}{2}} \left[ \mathbf{X}_{\pm}(t) \gamma - \Delta_{X,\pm}(t)\right] }_{2}^{2}$ corresponds to the simple weighted least squares regressions of $(X_{i}-t)^{p}$ on $(1,X_{i},...,X_{i}^{p})$ on the data sets defined by $X_{i} \geq t$ and $X_{i} < t$ respectively. By the binomial theorem, it holds for any $p\geq 0$
\begin{align*}
(x-t)^{p} = \sum_{k=0}^{p}  {{p}\choose{k}} x^{p-k} (-t)^{k} = {{p}\choose{0}} x^{p} (-t)^{0} + {{p}\choose{1}} x^{p-1} (-t)^{1} + \cdots + {{p}\choose{p}} x^{0} (-t)^{p},
\end{align*}
\noindent
which implies that there always exists a perfect solution (in the mean squared error sense) to the "unregularized" subproblem, with coefficients that do not depend on the weights $W_{\mathbf{X}_{\pm}}(t)$ or the data $\mathbf{X}_{\pm}(t)$. More importantly, the sign of the coefficients does only depend on the sign of $t$ and the orders $p$ and $\varv$. Note that, for $t>0$, the coefficients $\lbrace {{p}\choose{k}} (-t)^{k} \rbrace_{k=0}^{p}$ in the polynomial expansion alternate in sign, starting with a positive coefficient on $x^{p}$. For $t<0$ all coefficients are positive.

It follows that for $t\geq0$, $\bar{w}_{LP(p,\varv)}(t)$ is positive if $p-\varv$ is even, and negative if $p-\varv$ is odd, while $\bar{w}_{LP(p,\varv)}(t)$ is positive for all $t<0$. As a consequence, the conditional bias $\int_{\mathbb{R}} \mu^{p+1}(t) \bar{w}(t) dt$ is maximized by setting $\mu^{p+1}(t) = L$ if $p-\varv$ is even and $\mu^{p+1}(t) = -L \sign(t)$ if $p-\varv$ is odd. This yields the following general formula of the worst-case bias of a local polynomial estimator under $p+1$ order bounds %$\psi(p,v) = - \mathbbm{1}_{[p=v]} + \mathbbm{1}_{[p>v]} (-1)^{p-v+1}$
\begin{align*}
 \bar{B}_{n}(w_{LP(\varv,p)}) &=  \dfrac{\varv! L}{(p+1)!} \left[ (-1)^{p-\varv} \sum_{i = 1}^{n} w_{i,+} X_{i}^{p+1}  + \sum_{i=1}^{n} w_{i,-} X_{i}^{p+1}  \right].
\end{align*}

%%%% APPENDIX C

\subsection{Dual Optimization}
\label{appendix:C}

Appendix \ref{appendix:A} justifies the equivalence of (\ref{qp1}) and (\ref{qp3}) by establishing (\ref{eq2}). Appendix \ref{appendix:B} shows strong duality of (\ref{qp3}) and the original dual underlying (\ref{qp5}), implying that the solution to (\ref{qp1}) \linebreak can be obtained by solving the dual problem. To derive the objective in (\ref{qp5}) and to obtain (\ref{eq4}), which we used to recover the solution of (\ref{qp3}), we relied on a reformulation of the dual objective that gave rise to a quadratic problem nested in the dual objective $q(\nu,\lambda)$. The problem was obtained by interchanging the order of the supremum and infimum  in the original dual objective
\begin{align*}
q(\nu,\lambda) &= \inf_{(w_{N},r)  } L(w,r,\nu,\lambda) \\
&= \inf_{(w_{N},r)} \sup_{ R \in \bar{\mathcal{F}}(1)} \sum_{i=1}^{n} w_{i}^2 \sigma_{i}^{2} + \kappa L^{2} r^{2} +\nu \left( \sum_{i=1}^{n} w_i R(X_i) - r \right)
+ \lambda_1 \left( \sum_{i=1}^{n} w_{i,+} \right) \\
&+ \lambda_2 \left( \sum_{i=1}^{n} w_{i,-} \right)
+ \lambda_3 \left( \sum_{i=1}^{n} w_{i,+} X_i -1 \right) + \lambda_4 \left( \sum_{i=1}^{n} w_{i,-} X_i +1 \right).
\end{align*}
\noindent
\textit{Proof.} To see that this is admissible, first note that the set of functions $\bar{\mathcal{F}}(1)$ is convex. Let $z \in [0,1]$ and consider any two $f_1,f_2 \in \bar{\mathcal{F}}(1)$, with $(x,x') \in \mathcal{X}_{\pm}$. By definition of $\bar{\mathcal{F}}(1)$,
\begin{align*}
& z f_1(0) + (1-z) f_2(0) = zf_1^{1}(0) + (1-z) f_2^{1}(0) =0\\
& z |f^{1}_{1}(x) - f^{1}_{1}(x')| +  (1-z)| f_{2}^{1}(x) - f_{2}^{1}(x')| \leq |x - x'|
\end{align*}
\noindent
which implies that $zf_{1}(x) + (1-z)f_{2}(x) \in \bar{\mathcal{F}}(1)$. Moreover, any sequence of functions $f_{n}(x)$ in $\bar{\mathcal{F}}(1)$ is uniformly bounded and uniformly equicontinuous. The first statement follows from arguments analogous to those in Appendix A since $|f_{n}(x)| \leq \lvert \int_{0}^{x} (x-t)dt \rvert = 0.5  x^2$ for all $f_{n} \in \bar{\mathcal{F}}(1)$. Uniform equicontinuity holds since for any $(x,y)$ in $[\underline{x},\bar{x}]$ and sequence $f_{n}$ in $\bar{\mathcal{F}}(1)$
\begin{align*}
|f_{n}(x) - f_{n}(y)| &\leq  \Big \lvert \int_{0}^{x} (x-t)dt - \int_{0}^{y} (y-t)dt \Big \rvert = |0.5 (x^{2}-y^{2})| \leq  \max \lbrace |\bar{x}|, |\underline{x}| \rbrace |x-y|,
\end{align*}
\noindent
so that for $\delta=\dfrac{\varepsilon}{\max \lbrace |\bar{x}|, |\underline{x}| \rbrace}$ it holds that $|f_{n}(x) - f_{n}(y)| \leq \varepsilon$ whenever $|x-y| < \delta$. It follows from the Arzelà-Ascoli Theorem that the set $\bar{\mathcal{F}}(1)$ is relatively compact in $C([\underline{x},\bar{x}])$, the space of continuous real-valued functions on $[\underline{x},\bar{x}]$.  To obtain the result, define $g: \mathbb{R}^{n+1} \times \bar{\mathcal{F}}(1) \mapsto \mathbb{R}^{1}$
\begin{align*}
g(\mathbf{x},R|\nu,\lambda) &= \sum_{i=1}^{n} w_{i}^2 \sigma_{i}^{2} + \kappa L^{2} r^{2} + \nu \left( \sum_{i=1}^{n} w_i R(X_i) - r \right) + \lambda_1 \left( \sum_{i=1}^{n} w_{i,+} \right)  \\
&+ \lambda_2 \left( \sum_{i=1}^{n} w_{i,-} \right) + \lambda_3 \left( \sum_{i=1}^{n} w_{i,+} X_i -1 \right) + \lambda_4 \left( \sum_{i=1}^{n} w_{i,-} X_i +1 \right),
\end{align*}

\noindent
where $\mathbf{x} = (w_1, \cdots, w_n,r)^{T}$. Clearly, $\mathbb{R}^{n+1}$ is convex and $g$ is continuous in both arguments. Moreover, the function is, for any given $R \in \bar{\mathcal{F}}(1)$, convex over $\mathbb{R}^{n+1}$ and, for any given $\mathbf{x} \in \mathbb{R}^{n+1}$, linear (concave) over $\bar{\mathcal{F}}(1)$. It then follows from Sion's minimax theorem \citep{sion1958general} that
\begin{equation*}
\inf_{\mathbf{x} \in \mathbb{R}^{n+1} } \sup_{ R \in \bar{\mathcal{F}}(1)} g(\mathbf{x},R|\nu,\lambda) = \sup_{ R \in \bar{\mathcal{F}}(1)} \inf_{ \mathbf{x} \in \mathbb{R}^{n+1}}  g(\mathbf{x},R|\nu,\lambda).
\end{equation*}

\noindent
Interchanging the infimum  and supremum in the dual objective substantially simplifies the minimax problem and allows us to obtain the result in (\ref{eq4}). To make this explicit, define $\mathbf{Q}= \operatorname{diag}(2\sigma_1^2, \cdots,2\sigma_n^2,2\kappa L^2)$, let $f(\mathbf{x}) = \frac{1}{2}\mathbf{x^{T} Qx}$ and denote by $f^{*}(\mathbf{y}) = \sup_{\mathbf{x}} \left[ \mathbf{x^{T} y} - f(\mathbf{x}) \right]$ the conjugate function of $f(\mathbf{x})$. Moreover, let  

\begin{align*}
\mathbf{C} = \left[ \begin{matrix}
D(X_1) & \cdots & D(X_n) & 0\\
1-D(X_1)  & \cdots & 1-D(X_n)  & 0 \\
D(X_1)X_1 & \cdots & D(X_n)X_n & 0\\
(1-D(X_1))X_1 & \cdots & (1-D(X_n))X_n & 0\\
\end{matrix} \right] 
\qquad
\mathbf{d} = \left[ \begin{matrix}
\phantom{\matminus}0\phantom{\matminus}\\
\phantom{\matminus}0\phantom{\matminus}\\
\phantom{\matminus}1\phantom{\matminus}\\
\matminus1\phantom{\matminus}\\ 
\end{matrix} \right]  
\qquad 
\mathbf{A}^{T} = \left[ \begin{matrix}
R(X_1)\\
\vdots\\
R(X_n)\\
-1 \end{matrix}
 \right]  
\qquad
\mathbf{b} = 0.
\end{align*}

\noindent
In this notation, the dual objective after the exchange of supremum and infimum is given by
\begin{align*}
q(\nu,\lambda) &= \sup_{ R \in \bar{\mathcal{F}}(1)} \inf_{\mathbf{x}} \left[ f( \mathbf{x}) + \nu \left( \mathbf{ Ax -b} \right) + \lambda^{T} \left( \mathbf{Cx-d} \right)  \right]\\
& = \sup_{ R \in \bar{\mathcal{F}}(1)}  - \lambda^{T} \mathbf{d}  + \inf_{\mathbf{x}} \left(  f( \mathbf{x}) + \left( \nu \mathbf{A} + \mathbf{\lambda^{T} C} \right) \mathbf{x} \right) \\
& = \sup_{ R \in \bar{\mathcal{F}}(1)} - \lambda^{T} \mathbf{d} - f^{*} \left( - \left( \mathbf{A}^{T} \mathbf{\nu + C'\lambda} \right)  \right)\\
& = \sup_{ R \in \bar{\mathcal{F}}(1)} - \lambda^{T} \mathbf{d} - \frac{1}{2}  \left(  \mathbf{A}^{T} \mathbf{\nu + C'\lambda}  \right)^{T} \mathbf{Q}^{-1} \left(  \mathbf{A}^{T} \mathbf{\nu + C^{T} \lambda}  \right)\\
& = \sup_{ R \in \bar{\mathcal{F}}(1)} - \dfrac{1}{4} \sum_{i=1}^{n}  \dfrac{ \left[ \lambda_1 D(X_i)  + \lambda_2 (1-D(X_i))  + \lambda_3 D(X_i) X_i + \lambda_4  (1-D(X_i))X_i + \nu R(X_i) \right]^{2} }{\sigma_i^2} \\
& \qquad \phantom{iiij} - \dfrac{1}{4} \dfrac{ \nu^{2} }{\kappa L^{2}}  - \lambda_3 + \lambda_4,
\end{align*}

\noindent
Note that for $\sigma_i >0, L > 0, \kappa>0$, the objective function of the inner minimization problem in the dual objective is a quadratic form with positive definite Hessian. As a consequence, we obtain the analytical expressions (\ref{eq4}) by $(w_{N},r) = - \mathbf{Q}^{-1} \left(  \mathbf{A}^{T} \mathbf{\nu + C^{T} \lambda}  \right)$ and can write $q(\nu,\lambda)$ using its conjugate function. Finally, since by (\ref{eq2}) the maximum exists, we can write the dual problem as a maximization problem over the space $\bar{\mathcal{F}}(1)$ which yields  (\ref{qp5}).

\noindent
\textit{Remark.} In order to pass this problem to a numeric solver, it is useful to turn it into a minimization problem and to reparameterize it to avoid products of optimization parameters, $ \tilde{R}(X_i) \leftarrow \nu R(X_i)$. This yields the baseline problem that we solve via discrete approximation.
\begin{alignat}{2}
& \minimize_{\nu,\lambda, \tilde{R}} \qquad &&   \sum_{i=1}^{n}   \dfrac{ \left[ \lambda_1 D(X_i)  + \lambda_2 (1-D(X_i))  + \lambda_3 D(X_i) X_i + \lambda_4 (1-D(X_i))X_i + \tilde{R}(X_i) \right]^{2} }{4\sigma_i^2} \nonumber \\
& &&   + \dfrac{\nu^2}{4\kappa L^{2}} + \lambda_3 -  \lambda_4 \label{qp7} \\
& \textrm{subject to} && \nu \in \mathbb{R}_{+}, \lambda \in \mathbb{R}^{4}, \tilde{R} \in \bar{\mathcal{F}}(\nu) \nonumber.
\end{alignat}

\subsection{Implementation}
\label{appendix:D}

Our implementation solves the QP (\ref{qp7}) by approximating the continuous argument $\tilde{R}$ on a discrete equidistant grid with distance $\Delta=h$ between any two adjacent grid points. Let $\underline{x}$ and $\bar{x}$ denote the minimal and maximal elements of $\lbrace{X_{i} - c \rbrace}_{i=1}^{n}$. In a first step, we divide the interval $I=\left( \underline{x} - \frac{h}{2}, \bar{x} + \frac{h}{2} \right]$ in $J=(\bar{x}-\underline{x})/h$ disjoint intervals $I_{j} = \left( \underline{x} + (j-\frac{3}{2})h, \underline{x} + (j-\frac{1}{2})h \right]$. Let $x_{j}$ denote the center of interval $j$, so that $x_{1} = \underline{x}$ and $x_{J} = \bar{x}$. In a second step, we assign each data point $X_{i}$ to its closest grid point. For example, if $X_{i}$ falls into $I_{j}$ we assign $j \gets i$ and store this mapping. Moreover, we keep track of the two grid points ($x_{c,-},x_{c,+})$ that are closest to the normalized cutoff zero (ties are possible only if zero itself is a grid point). 
\begin{figure}[h!]
\centering
\hspace*{-1.00cm}
\begin{tikzpicture}[>={Bracket[width=6mm,line width=3pt,length=1.5mm]},scale=0.75]
% Die Grundlinie:
\draw(0,0)--(24,0);
% Striche und Beschriftung in Abständen 0, 2, 4, 6, ...

\foreach \x/\xtext in {1.5/$\phantom{h} \underline{x} \phantom{h}$,4.5/$\underline{x}+h$,7.5/$\cdots$,10.5/$0$,13.5/$\cdots$,16.5/$\phantom{h}x_j\phantom{h}$,19.5/$\cdots$,22.5/$\phantom{h}\bar{x}\phantom{h}$}
    \draw(\x,2pt)--(\x,-2pt) node[below] {\xtext};
 
% potential centers at 1.5 (first) ,4.5 (second),7.5 (...) ,10.5 (0) ,13.5 (...) ,16.5 (xj),19.5 (cdots) ,22.5 (last)
 
% xc- and xc+

\fill[black] (9.25,0) circle (0.75mm) node[below=2mm] {$\color{black}{x_{c,-}}$};
\fill[black] (11.75,0) circle (0.75mm) node[below=2mm] {$\color{black}{x_{c,+}}$};
% Width
\draw[decorate, decoration={brace}, yshift=4ex]  (3,0) -- node[above=0.4ex] {$\Delta=h$}  (6,0);
% Intervals

% Connecting Intervals
\draw[{Parenthesis[width=6mm]}-{Bracket[width=6mm]},  thick] 
    (6,0) node[below=2mm] {$\phantom{a}$} -- ( 15,0) node[below=2mm] {$\phantom{a}$};
\draw[{Parenthesis[width=6mm]}-{Bracket[width=6mm]},  thick] 
    (18,0) node[below=2mm] {$\phantom{a}$} -- ( 21,0) node[below=2mm] {$\phantom{a}$};
    
 % First Interval (xmin), 1 point
%\fill[black] (1.5,0) circle (0.75mm) node[below=2mm] {$\phantom{a}$};
%\fill[red] (1.5,0) circle {0.75} -- (180:1ex) arc (180:0:1ex) -- cycle; % Fill a half circle filled with second colour (arg#1), if specified
\path [draw=red,fill=black] (1.5,0) circle (0.75mm) node[below=2mm] {$\phantom{a}$};

%\fill[orange] (1.5,0) circle (1ex);
%\clip (-1ex,-1ex) rectangle (1ex,0);
%\fill[green] (1.5,0) circle (1ex);

\draw[{Parenthesis[width=6mm]}-{Bracket[width=6mm]},  thick] 
    (0,0) node[below=2mm] {$\underline{x} - \frac{h}{2}$} -- ( 3,0) node[below=2mm] {$\phantom{a}$};
    
\fill[red] (2.5,0) circle (0.75mm) node[below=2mm] {$\phantom{a}$};

\begin{scope}[shift={(0,0)}]
    \path[draw my path={->,>=stealth}] (2.5,0.1) to[bend right] node[my node] {} (1.5,0.1);
 \end{scope}

% Second interval   (xmin +h) , 2 points 
\draw[{Parenthesis[width=6mm]}-{Bracket[width=6mm]},  thick] 
    (3.01,0) node[below=2mm] {$\phantom{a}$} -- ( 6,0) node[below=2mm] {$\phantom{a}$};

\fill[black] (4.5,0) circle (0.75mm) node[below=2mm] {$\phantom{a}$};
\fill[red] (3.25,0) circle (0.75mm) node[below=2mm] {$\phantom{a}$};
\fill[red] (5,0) circle (0.75mm) node[below=2mm] {$\phantom{a}$};

\begin{scope}[shift={(0,0)}]
    \path[draw my path={->,>=stealth}] (3.25,0.1) to[bend left] node[my node] {} (4.45,0.1);
 \end{scope}  
\begin{scope}[shift={(0,0)}]
    \path[draw my path={->,>=stealth}] (5,0.1) to[bend right] node[my node] {} (4.5,0.1);
 \end{scope}

% Third interval (xj), 2 points
\draw[{Parenthesis[width=6mm]}-{Bracket[width=6mm]},  thick] 
    (15.01,0) node[below=2mm] {$\phantom{a}$} -- ( 18,0) node[below=2mm] {$\phantom{a}$};
    
\fill[black] (16.5,0) circle (0.75mm) node[below=2mm] {$\phantom{a}$};
\fill[red] (15.5,0) circle (0.75mm) node[below=2mm] {$\phantom{a}$};
\fill[red] (17.0,0) circle (0.75mm) node[below=2mm] {$\phantom{a}$};
 
\begin{scope}[shift={(0,0)}]
    \path[draw my path={->,>=stealth}] (17,0.1) to[bend right] node[my node] {} (16.5,0.1);
\end{scope} 
\begin{scope}[shift={(0,0)}]
    \path[draw my path={->,>=stealth}] (15.5,0.1) to[bend left] node[my node] {} (16.45,0.1);
\end{scope}
 
% Last Interval (xmax), 1 point
  
 \draw[{Parenthesis[width=6mm]}-{Bracket[width=6mm]},  thick] 
    (21.01,0) node[below=2mm] {$\phantom{a}$} -- ( 24,0) node[below=2mm] {$\bar{x} + \frac{h}{2}$};
 
\path [draw=red,fill=black] (22.5,0) circle (0.75mm) node[below=2mm] {$\phantom{a}$};
 
%\fill[black] (22.5,0) circle (0.75mm) node[below=2mm] {$\phantom{a}$};
\fill[red] (21.7,0) circle (0.75mm) node[below=2mm] {$\phantom{a}$};

\begin{scope}[shift={(0,0)}]
    \path[draw my path={->,>=stealth}] (21.7,0.1) to[bend left] node[my node] {} (22.5,0.1);
 \end{scope} 
 
\end{tikzpicture}
\caption{Illustration of the discretization strategy. Red (black) dots indicate data (grid) points.}
\end{figure}
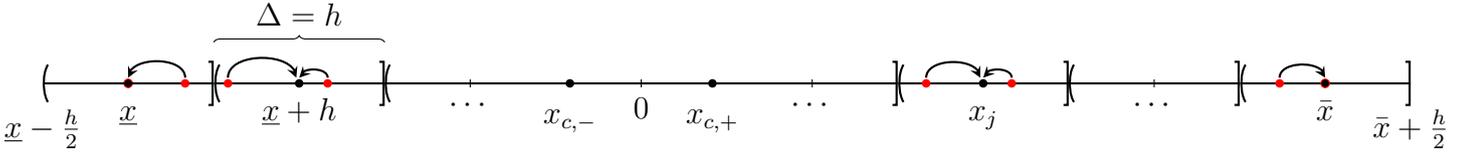

\noindent
Note that by construction $0 \geq x_{c,-} \underset{J \rightarrow \infty}{\rightarrow} 0$ and  $0 \leq x_{c,+} \underset{J \rightarrow \infty}{\rightarrow} 0$ since $h \underset{J \rightarrow \infty}{\rightarrow} 0$. Since $\mu \in \mathcal{F}(L)$,
\begin{alignat*}{2}
\tilde{R}(0) -  \tilde{R}(x_{c,-}) &= O(|x_{c,-}|) \text{  as } x_{c,-} \rightarrow 0  \phantom{as}  \qquad \tilde{R}^{1}(0) - \dfrac{ \tilde{R}(0) - \tilde{R}(x_{c,-})}{|x_{c,-}|}  &&= O(|x_{c,-}|) \text{  as }  x_{c,-} \rightarrow 0\\
\tilde{R}(0) -  \tilde{R}(x_{c,+}) &= O(|x_{c,+}|) \text{  as } x_{c,+} \rightarrow 0  \phantom{as} \qquad \tilde{R}^{1}(0) - \dfrac{ \tilde{R}(x_{c,+}) - \tilde{R}(0)}{|x_{c,+}|} &&=  O(|x_{c,+}|) \text{  as } x_{c,+} \rightarrow 0
\end{alignat*}
\noindent
and we take into account the constraint $\tilde{R}(0)=\tilde{R}^{1}(0)=0$ by enforcing $\tilde{R}(x_{c,-})= \tilde{R}(x_{c,+})=0$. In order to take into account the second order derivative constraint $|\tilde{R}^{2}(x)| \leq L$, we utilize the equidistance of grid points to impose the  constraints via second order central difference approximations $\tilde{R}^{2}(x) - [\tilde{R}(x + h ) - 2\tilde{R}(x) + \tilde{R}(x-h)]h^{-2} = O(h^{2})$ as $h \rightarrow 0$. This yields 2 linear equality and $2J$ linear inequality constraints\footnote{Note that we need to enforce the second order constraints only at grid points that have data points in their respective interval. We can thus increase precision without linearly growing the number of constraints.} that we enforce during the optimization
\begin{equation*}
\begin{bmatrix}
 0  & 0 \\
 \vdots   & \vdots      \\
 1   &  0      \\
 0   &  1       \\
 \vdots   &  \vdots        \\
  0   &  0      \end{bmatrix}^{T}
  \begin{bmatrix}
 \tilde{R}(\underline{x}) \\
 \vdots \\
 \tilde{R}(x_{c,-}) \\
 \tilde{R}(x_{c,+}) \\
    \vdots \\
 \tilde{R}(\bar{x})
    \end{bmatrix} = \mathbf{0}_{2} \qquad
      	\begin{bmatrix}
 	1 & -2	& 1  & 0  &  \cdots    &  \cdots         &         0             \\
  	0 &  1 	& -2 & 1  &  \cdots    &  \cdots         &         0             \\
   \vdots & \vdots & \ddots    &  \ddots     & \ddots    & \cdots   & \vdots        \\
    -1 & 2	& -1  & 0  &  \cdots    &  \cdots         &         0             \\
  	0 &  -1 & 2 & -1  &  \cdots    &  \cdots         &         0             \\
   \vdots & \vdots & \ddots    &  \ddots     & \ddots    & \cdots   & 0        \\
    \end{bmatrix}
   \begin{bmatrix}
 \tilde{R}(\underline{x}) \\
 \tilde{R}(\underline{x}+h) \\
    \vdots \\
 \tilde{R}(\bar{x})
    \end{bmatrix} \leq \mathbf{1}_{2J} \nu h^{2}.
\end{equation*}

\noindent
Let $\mathbf{A}_{1}$ and $\mathbf{A}_{2}$ denote the two matrices above. In order to write the quadratic program (\ref{qp7}) in standard form, we define $\tilde{\mathbf{x}} = (\tilde{w}_{J}, \nu, \lambda, \tilde{R}_{J})^{T}$, where $\tilde{w}_{j} = - 2 \sigma(x_{j})^{2} w_{j}$ and $\tilde{R}_{j} = \nu R(x_{j})$ are the reparameterized weights and remainders evaluated at the approximation points. The quadratic and linear part of the objective are 
$\tilde{\mathbf{H}} = \operatorname{diag}(0.5\sigma(\underline{x})^{-2}, \cdots, 0.5\sigma(\bar{x})^{-2},0.5\kappa^{-1} L^{-2},\mathbf{0}_{J+4}^{T})$ and $\tilde{d} = (\mathbf{0}_{J}^{T},0,0,0,1,-1,\mathbf{0}_{J}^{T})^{T}$. In this notation, (\ref{qp7}) under discretization of $\tilde{R} \in \bar{\mathcal{F}}(\nu)$ is
\begin{alignat}{2}
& \minimize_{\tilde{\mathbf{x}}} \qquad &&  \dfrac{1}{2} \tilde{\mathbf{x}}^{T} \tilde{\mathbf{H}} \tilde{\mathbf{x}} + \tilde{d}^{T} \tilde{\mathbf{x}} \nonumber \\
& \text{subject to} && [ -\mathbf{I}_{J}, \mathbf{0}_{J \times 1}, \tilde{\mathbf{C}}^{T} \lambda, \mathbf{I}_{J}] \tilde{\mathbf{x}} = \mathbf{0}_{J} \nonumber \\
& && [\mathbf{0}_{2 \times J+5}, \mathbf{A}_{1}] \tilde{\mathbf{x}} = \mathbf{0}_{2} \label{qp8} \\
& && [\mathbf{0}_{2J \times J+5}, \mathbf{A}_{2}] \tilde{\mathbf{x}} \leq \mathbf{1}_{2J} \nu h^{2}, \nu \in \mathbb{R}_{+}, \nonumber 
\end{alignat}

\noindent
where $\tilde{\mathbf{C}}$ is defined in analogy to the matrix $\mathbf{C}$ in Appendix \ref{appendix:C}, with the approximation points $x_{j}$ ($j=1\dots,J$) playing the role of the data points $X_{i}$ ($i=1,\dots,n$). In the implementation, we reformulate (\ref{qp8}) in terms of the equivalent Second Order Cone Program to avoid issues of standard solvers associated with the semidefiniteness of $\tilde{\mathbf{H}}$ induced by the $J+4$ zero rows. To this end, let $q$ be an additional parameter and note that since $\tilde{\mathbf{H}}$ is positive semidefinite, we can write $\tilde{\mathbf{x}}^{T}\tilde{\mathbf{H}}\tilde{\mathbf{x}} = \tilde{\mathbf{x}}^{T}\mathbf{R^{T}R}\tilde{\mathbf{x}} = \norm{\mathbf{R}\tilde{\mathbf{x}}}^{2}_{2}$, allowing us to equivalently write the problem with an objective that is linear in $\ddot{\mathbf{x}} = (\tilde{\mathbf{x}},q)^{T}$
\begin{alignat*}{2}
& \minimize_{\ddot{\mathbf{x}}} \qquad &&  e_{2J+6}^{T} \ddot{\mathbf{x}} + [\tilde{d},0]^{T} \ddot{\mathbf{x}} \nonumber \\
& \text{subject to} && q+1 \geq \sqrt{ \norm{\mathbf{R}\tilde{\mathbf{x}}}^{2}_{2} + q ^{2}} \\
& && \qquad \vdots
\end{alignat*}

\noindent
Given a solution to (\ref{qp8}), we assign weights to data points according to the mapping between the index sets of $X_{i}$ and $x_{j}$, and rescale each weight to satisfy the moment conditions exactly.

\subsection{Proof of Proposition 1}
\label{appendix:E}

\noindent
The honesty property of the interval (\ref{eq6}) for any given bound $L$ relies on an (asymptotic) upper bound for the worst-case smoothing bias and uniform convergence of $ s_{n}^{-1} [\hat{\theta} - \E[\hat{\theta} | X_{n}]] \overset{D}{\rightarrow} N(0, 1)$ over $\mathcal{F}(L)$. It then follows from standard arguments that a confidence interval based on the approximation $(\hat{\theta} - \theta)/s_n  \mathrel{\dot\sim} |N(\bar{t}_n, 1)|$ yields asymptotically uniformly valid confidence intervals.\\

\noindent
\textit{Proof.} We first show that under Assumption 1, conditionally on $X_{N}$, Lyapunovs condition applies uniformly over all permitted CEFs. By Lyapunov's Central Limit Theorem and assumptions A1 (i)-(ii) this requires us to show that for all $\mu \in \mathcal{F}(L)$
\begin{equation*}
 \dfrac{1}{s_{n}^{2+\delta}} \sum_{i=1}^{n} \E \left[ | w_i(Y_i-\mu(X_i)) - \E\left[ w_i (Y_i - \mu(X_i))| X_{N} \right] |^{2+\delta} | X_{N} \right] = o_{p}(1).\\
\end{equation*}

\noindent
By the definition of $\mu$,  $\E\left[ w_i (Y_i - \mu(X_i))| X_{n} \right]=0$, and the Lyapunov condition holds since
\begin{alignat*}{2}
 \dfrac{\sum_{i=1}^{n} \E \left[ | w_i(Y_i-\mu(X_i)) |^{2+\delta} | X_{N} \right]}{ \left[ \sqrt{\sum_{i=1}^{n} \sigma_i^{2} w_i^{2}} \right]^{2+\delta}}  &\leq   \dfrac{\sum_{i=1}^{n} |w_i|^{2+\delta} \E \left[ | (Y_i-\mu(X_i)) |^{2+\delta} | X_{N} \right]}{ \left[ \sqrt{\sum_{i=1}^{n} \sigma_i^{2} w_i^{2}} \right]^{2+\delta}} &&  \\
& \leq  C \dfrac{ \sum_{i=1}^{n}  |w_i|^{2+\delta} }{ \left[ \sqrt{\sum_{i=1}^{n} w_i^{2} \sigma_i^{2}} \right]^{2+\delta}}   && \text{by A1 (iv)}   \\
& \leq  C \dfrac{ \max_i \lbrace w_i \rbrace^{2} \sum_{i=1}^{n}  |w_i|^{\delta}  }{ \left[ \sigma_{min} \sqrt{\sum_{i=1}^{n} w_i^{2} } \right]^{2+\delta}}  && \text{by A1 (iii)}    \\
& =   \dfrac{C  }{ \sigma_{min}^{2+\delta}  }  \dfrac{\max_i \lbrace w_i \rbrace^{2}}{  \sum_{i=1}^{n} w_i^{2}} = o_{p}(1)  && \text{by A1 (v)}
\end{alignat*}
\noindent
where the second and third line utilize the uniform moment bounds and the last equation relies on the limit behavior of the ratio $\bar{w}_{R}$. It then follows from standard arguments that a uniform asymptotic upper bound on undercoverage of (\ref{eq6}) is given by
\begin{align*}
 \mathclap{\liminf_{n \rightarrow \infty} \left( \underset{\mu \in \mathcal{F}(L)}{\inf} \Pr \left[  \left \lvert s_{n}^{-1}[\hat{\theta} - \E[\hat{\theta} | X_{N}]] + t_{n} \right \rvert \leq \text{cv}_{1-\alpha}(\bar{t}_{n}) \mid X_{N} \right] -
 \underset{\mu \in \mathcal{F}(L)}{\inf} \Pr \Big[ \left \lvert Z + t_{n} \right \rvert \leq \text{cv}_{1-\alpha}(\bar{t}_{n}) \mid X_{N} \Big] \right) = 0,}
\end{align*}
where $Z$ denotes a standard normal variable. Then by the definition of $\text{cv}_{1-\alpha}$ and $\bar{t}_{n}$ it holds
\vspace*{-0.1cm}
\begin{align*}
\underset{n \rightarrow \infty}{\lim \inf} \text {  } \underset{\mu \in \mathcal{F}(L)}{\inf}  \Pr \left[ \mu^{1}_{+}(0) - \mu^{1}_{-}(0) \in \mathcal{I}_{\alpha} \right]  \geq 1 - \alpha.
\end{align*}

\subsection{Extensions}
\label{appendix:F}

\paragraph{F.1 Fuzzy Discontinuity Design.}

The inference problem discussed in the main body of the paper is a special case of the general problem of inference on the ratio of jumps in the $\varv-$th derivative of two CEFs at a point (normalized to zero). Let 
$\mu_{Y}(x) = \E[Y_{i} | X_{i} = x]$ and $\mu_{T}(x) = \E[T_{i} | X_{i} = x]$. The general parameter of interest in discontinuity designs is
\begin{align}
\tau_{RD}  =  \dfrac { \mu_{Y,+}^{\varv}(0) - \mu_{Y,-}^{\varv}(0) }{ \mu_{T,+}^{\varv}(0) - \mu_{T,-}^{\varv}(0)} = \dfrac{\theta^{\varv}_{Y}}{\theta^{\varv}_{T}}, \label{eq9}
\end{align}
\noindent
under the assumption that $\theta^{\varv}_{T} \neq 0$, with $\varv=0$ corresponding to RDDs and $\varv=1$ to RKDs. First note that, provided a bound on the magnitude of the respective derivatives of order $p > \varv$, we can, irrespective of the order $\varv$, employ the optimization approach discussed in Section 3 seperately to $\theta^{\varv}_{Y}$ and $\theta^{\varv}_{T}$ by restricting the feasible set via $\sum_{i=1}^{n} w_{i,+} X_{i}^{k} = - \sum_{i=1}^{n} w_{i,+} X_{i}^{k} = \delta_{\varv,k}$ as in Appendix \ref{appendix:A}. This is because, by the arguments in Appendix \ref{appendix:B}, we can always derive a suitable constraint qualification to employ the logic of Section 3. 

In the sharp case, that is if the institutional rule $T$ governing treatment assignment is deterministic $\Var[T(X_{i})|X_{i}] = 0$ and fully implemented $\Pr \lbrace T_i = T(X_i) \rbrace = 1$, it is sufficient to do so for just $\theta^{\varv}_{Y}$ and to conduct inference on  $\tau_{RD}$ in complete analogy to Section 3. However, in the fuzzy setting, that is if the rule is probablistic, or if there are deviations from the assignment rule, the denominator needs to be estimated and inference on $\tau_{RD}$ requires a method that adresses the nonlinearity induced by the ratio. As a consequence, additional complications for inference arise and our optimization approach can not be directly employed. Fuzzy designs can arise for various reasons such as noncompliance, multivariate rules with unobserved assignment variables or measurement errors, and occur frequently in applications.

The standard approach to inference in fuzzy settings is to estimate $\theta^{\varv}_{Y}$ and $\theta^{\varv}_{T}$ seperately and rely on a linearization of the estimator ratio to build confidence intervals based on a delta method argument. 
The delta method approach to inference in discontinuity designs has three important limitations. The first two concern the validity of the distributional approximation in settings with discrete assignment variables or weak identification, that is if the denominator $\theta_{T}$ is close to zero. In both cases, the validity of the approximation breaks down, as otherwise asymptotically negligible terms in the expansion have non-zero probability limits. A third problem is that, due to the nonlinearity of the estimator, the bias-aware approach can not control the exact worst-case smoothing bias.

Motivated by these shortcomings of delta method based inference, \citet{noack2019bias} propose a method to conduct inference based on an Anderson-Rubin type inversion argument \citep{anderson1949estimation}. This approach is known to allow for valid pointwise inference under weak identification and importantly also allows them to apply bias-aware methods to construct honest confidence sets, as the method avoids linearization. In order to extend our optimization based approach to fuzzy designs, we employ their strategy.

The basic idea is to consider the parameter $\theta^{\varv}(\tau) = \theta^{\varv}_{Y} - \tau \theta^{\varv}_{T}$ and to construct honest $(1-\alpha)$ confidence intervals for $\theta^{\varv}(\tau)$ for different values of $\tau$. Note that by definition of $\theta^{\varv}_{Y}$ and $\theta^{\varv}_{T}$, this corresponds to constructing CIs for the jump in the $\varv$-th derivative of $\mu_{\tau}(x) = \E[Y_i - \tau T_i|X_{i}=x]$ at the discontinuity point for any given value of $\tau$. A confidence set for $\tau_{RD}$ is then obtained by collecting all values of $\tau$ for which the auxiliary interval covers zero, and adjusting this set depending on whether an honest confidence interval for $\theta^{\varv}_{T}$ contains zero or not. The procedure thus reduces the fuzzy inference problem to repeated sharp problems and allows us to immediately employ optimized intervals. The only particularity that arises under this strategy relates to the shape of the confidence sets, which varies depending on whether the auxiliary confidence interval for $\theta^{\varv}_{T}$ covers zero or not. Let $\mathcal{I}_{T}=[\underline{\theta}^{\varv}_{T},\bar{\theta}^{\varv}_{T}]$ denote an optimized interval for $\theta^{\varv}_{T}$ and let $\mathcal{I}_{\tau_{RD}}=[\underline{\tau}_{RD},\bar{\tau}_{RD}]$ denote the confidence interval\footnote{$\mathcal{I}_{\tau}$ has this shape by continuity of $w^{*}$, $\bar{B}(w_{N})$ and $s_{n}$ in $\tau$, and the folded normal quantile function in $\bar{t}_{n}$.} obtained by employing the above method. The following five cases are possible for some $a<b$.

\begin{table*}[h!]\centering
  \begin{threeparttable}
  
\caption*{Table F.1 Shape of Anderson-Rubin Confidence Sets for $\tau_{RD}$}
\ra{1.1}

\begin{tabular}{llcccc}\toprule
& \multicolumn{2}{c}{Case} &   & Shape of Confidence Set \\
\cmidrule{2-3} \cmidrule{5-5}  
\phantom{aaaa} & i)   & $ 0 \not \in \mathcal{I}_{T}$  &  & $[a,b]$ & \phantom{aaaa} \\
& ii) & $ 0  \in \mathcal{I}_{T}, 0 \not \in \mathcal{I}_{\tau}$ && $(-\infty, a] \cup [b,\infty).$ & \\
& iii) & $ 0 \in \mathcal{I}_{T},  0  \in \mathcal{I}_{\tau}$ && $(-\infty,\infty)$ & \\
& iv)  & $ \underline{\theta}_{T}=0,  \underline{\tau}_{RD}>0$  && $[a,\infty)$& \\
& v)   & $ \bar{\theta}_{T} = 0 , \bar{\tau}_{RD} <0$  && $(-\infty, b]$ &  \\
\bottomrule\\
\end{tabular}
%\end{minipage}}
%\begin{tablenotes}
%\small
%   \item \textit{Note:} Ad-hoc conventional local linear, robust bias-correction and optimized RKD point estimates and 95 \% confidence intervals for $\tau_{RKD}$ in the CWBH data for the periods Jan-Sep 1979 (Period 1), Sep 1979-Sep 1980 (Period 2), Sep 1980-Sep 1981 (Period 3), Sep 1981-Sep 1982 (Period 4) and Sep 1982-Dec 1983 (Period 5).
%\end{tablenotes}
\end{threeparttable}
\end{table*}

\noindent
See \citet{noack2019bias} for a detailled discussion of the construction and properties of bias-aware Anderson-Rubin Confidence Sets based on local polynomial estimators. As discussed in their paper, the cases iv) and v) are empirically irrelevant, as the events $\underline{\theta}^{\varv}_{T}=0$ and $\bar{\theta}^{\varv}_{T} = 0$ occur with probability zero. Adapting their findings to our setting, we impose the following modification of Assumption 1 to ensure that the Anderson-Rubin Confidence Sets obtained via optimization are well defined and honest.\\

\noindent
\textbf{Assumption F.1} Let $(C,\delta,\sigma_{\min}, \sigma_{\max}) \in \mathbb{R}_{+}^{4}$ be some fixed vector.
\begin{enumerate}[leftmargin=*, labelsep=*, align=left, itemsep=-0.2cm, font=\normalfont, label=(\roman*)]
\item  $\lbrace Y_{i}, X_{i}, T_{i} \rbrace_{i=1}^{n}$ is an i.i.d. random sample of size $n$ from a fixed population.
\item  $(\mu_{Y} , \mu_{T}) \in \mathcal{F}(L) \times \mathcal{F}_{T}(L_{T})$ for some $L \geq 0$ and $L_{T} \geq 0$ with \\
 $\qquad \qquad \qquad \mathcal{F}_{p}(L) = \lbrace f : | f_{\pm}^{p-1}(x) - f_{\pm}^{p-1}(x') | \leq L |x-x'|, (x,x') \in \mathcal{X}_{\pm} \rbrace$\\, 
 $\qquad \qquad \phantom{iiii} \mathcal{F}_{T}(L) = \lbrace f : f \in \mathcal{F}(L_{T}) \wedge  | f_{+}^{\varv}(0) - f_{-}^{\varv}(0) | > 0  \rbrace.$
\item  For all $x\in R_{X} $, $(\mu_{Y} , \mu_{T}) \in \mathcal{F}(L) \times \mathcal{F}_{T}(L_{T})$ and $\tau \in \mathbb{R}$,\\
	 $\qquad \qquad \qquad 0 < \sigma_{\min}^{2} \leq E[(Y_i - \tau T_{i}  - \mu_{\tau}(X_i))^{2}|X_{i} = x] \leq \sigma_{\max}^{2}$.  
\item  For all $x\in R_{X}$, $(\mu_{Y} , \mu_{T}) \in \mathcal{F}(L) \times \mathcal{F}_{T}(L_{T})$ and $\tau \in \mathbb{R}$,\\
  $ \qquad \qquad \qquad  \E[|Y_i - \tau T_{i} - \mu_{\tau}(X_i)|^{2+\delta} | X_i = x] \leq C$. 
\item For each $\tau \in \mathbb{R}$, the solution $w^{*}$ satisfies \\
 $  \qquad \qquad \phantom{(a)} \qquad \dfrac{ \max_{i}w_{i}^{2} }{ \sum_{i=1}^{n} w_{i}^{2}} \overset{P}{\rightarrow} 0.$\\
\end{enumerate}

\noindent
The adjustments in Assumption F.1 ensure that (\ref{eq9}) is well defined and that we can construct honest CIs for $\theta^{\varv}_{T}$ and $\theta^{\varv}_{\tau_{RD}}$ (for each value of $\tau$), by utilizing the bound on the $p$-th derivatives encoded in $\mathcal{F}_{p}(L)$ and $\mathcal{F}_{T}(L_{T})$. Note that by the logic discussed in Section 3, the worst-case bias of our estimator over $\mathcal{F}(L) \times \mathcal{F}_{T}(L_{T})$ is proportional to $L+|\tau| L_{T}$. In order to compute the confidence sets, it is thus required to construct optimized intervals for different values of $\tau$, taking into account the relevant functional constraint.

In order to implement this procedure, we first compute a length-optimized interval for $\theta^{\varv}_{T}$. In a second step, we compute UMSE optimized intervals on a prespecified grid of values of $\tau$ by calling our method on data with modified outcome variable $\tilde{Y}_{i} = Y_{i} - \tau T_{i}$, adjusting the smoothness bound as needed. If the grid contained a sufficiently large range of values of $\tau$, this yields approximate values for $\underline{\tau}_{RD}$ and $\bar{\tau}_{RD}$. We then compute length-optimized intervals starting at the two approximate bounds until we found the roots of $\hat{\theta}^{\varv}(\tau) \pm s_{n}(\tau) \text{cv}_{1-\alpha}(\bar{t}_{n}(\tau))$. Finally, we report confidence sets according to Table F.1.

The method of \citet{noack2019bias} thus allows us to extend the optimization approach to fuzzy RKDs and RDDs. The latter is is a special case of the multivariate RDD problem considered in \citet{imbens2019optimized}, which we extend to the (univariate) fuzzy setting. Given the results in \citet{imbens2019optimized} and our simulations for RKDs one would expect these intervals to be particularly useful in fuzzy settings with discrete running variables.

\newpage

\paragraph{F.2 Shape Constraints.}
An attractive feature of the honest optimization approach to RD inference in sharp designs is that it is rather simple to utilize additional structural information on the CEF under consideration. In our view, this is a noteworthy feature of combining the bias-aware approach with optimization techniques, as utilizing such information for inference is typically rather difficult. This is because in general the distribution of a restricted estimator depends in a non-trivial fashion on whether and where the shape constraints are binding, which is typically not known a priori (\cite{freyberger2018inference}).

Under the bias-aware approach, this dependence operates through the worst-case bias and is rather simple in structure. In principle, any additional constraint that can be approximated in terms of finite differences of Taylor remainders and that does not break the convexity and compactness of $\bar{\mathcal{F}}(1)$ can be directly utilized to sharpen inference. Such information can easily be incorporated by adding suitable constraints to (\ref{qp8}). For example, concavity of $\mu$ could be utilized by the following modification to the feasible set of (\ref{qp8})
\begin{align*}
\begin{bmatrix}
 	1 & -2	& 1  & 0  &  \cdots    &  \cdots         &         0             \\
  	0 &  1 	& -2 & 1  &  \cdots    &  \cdots         &         0             \\
   \vdots & \vdots & \ddots    &  \ddots     & \ddots    & \cdots   & \vdots        \\
    -1 & 2	& -1  & 0  &  \cdots    &  \cdots         &         0             \\
  	0 &  -1 & 2 & -1  &  \cdots    &  \cdots         &         0             \\
   \vdots & \vdots & \ddots    &  \ddots     & \ddots    & \cdots   & 0        \\
    \end{bmatrix}
   \begin{bmatrix}
 \tilde{R}(\underline{x}) \\
 \tilde{R}(\underline{x}+h) \\
    \vdots \\
 \tilde{R}(\bar{x})
    \end{bmatrix} \leq \left( \begin{matrix} \mathbf{0}_{J} \\ 
    \mathbf{1}_{J} \nu h^{2} \end{matrix} \right).
\end{align*}

\noindent
Analogously, monotonicity of $\mu$ can be imposed via the constraint
\begin{align*}
\begin{bmatrix}
 	1 & 0	& -1  & 0  &  \cdots    &  \cdots         &         0             \\
  	0 &  1 	& 0 & -1  &  \cdots    &  \cdots         &         0             \\
  	 0 &  0 	& 1 & 0  &  -1    &  \cdots         &         0             \\
   \vdots & \vdots & \ddots    &  \ddots     & \ddots    & \ddots    & \vdots        \\
    \end{bmatrix}
   \begin{bmatrix}
 \tilde{R}(\underline{x}) \\
 \tilde{R}(\underline{x}+h) \\
    \vdots \\
 \tilde{R}(\bar{x})
    \end{bmatrix} > (<) \mathbf{0}_{J}.\\
\end{align*}

\noindent
This is particularly useful for applied work that utilizes regression kink designs, as the "plateau"-schedules that often form the basis for kink designs tend to generate empirical CEFs that plausibly satisfy shape constraints. Note that this applies directly only to the sharp setting, as the properties of $\theta_{\tau}(c)$ depend on both $\mu_{Y}$ and $\mu_{T}$ and even $c$. In the fuzzy setting one therefore needs to be more careful when thinking about imposing shape constraints in this fashion.

\newpage

\subsection{Runtime Estimates}
\label{appendix:G}

Table A.1 shows average runtimes of the procedure proposed in this paper for 100 simulated runs for each of the sample sizes reported in the first column. The data was drawn from a continuous uniform distribution on $[-1,1]$ with $L=2$ and $\sigma^2=0.1^2$. The computations were conducted on a standard desktop computer using \texttt{R} 4.1.0. and the implementation provided in the replication files.

\begin{table*}[h!]\centering
  \begin{threeparttable}
  
\caption*{Table G.1. Runtime Estimates}
\ra{1.2}
\begin{tabular*}{0.6\textwidth}{ccc}\toprule
Sample Size & UMSE-optimal & HL-optimal \\ \toprule
500 & 0.115 & 2.738 \\
1000 & 0.109 & 2.813 \\
1500 & 0.108 & 2.849 \\
2000 & 0.116 & 3.076 \\
2500 & 0.120 & 3.030 \\
3000 & 0.117 & 3.141 \\
3500 & 0.120 & 3.226 \\
4000 & 0.129 & 3.286 \\
4500 & 0.129 & 3.115 \\
5000 & 0.127 & 3.304 \\
5500 & 0.130 & 3.568  \\
6000 & 0.129 & 3.590 \\
6500 & 0.133 & 3.419 \\
7000 & 0.132 & 3.458 \\
7500 & 0.136 & 3.365 \\
8000 & 0.132 & 3.320 \\
8500 & 0.138 & 3.419 \\
9000 & 0.141 & 3.641 \\
9500 & 0.145 & 3.806 \\
10000 &  0.144 & 3.899 \\
\bottomrule           
\end{tabular*}

\begin{tablenotes}
\small
   \item \textit{Note:} Runtime estimates in seconds for UMSE and length optimized linear confidence intervals for different sample sizes.
\end{tablenotes}
\end{threeparttable}
\end{table*}

\subsection{Additional Simulation Results}
\label{appendix:I}

\noindent
Tables \ref{Tab:one2} and \ref{Tab:two2} contain the results of a more extensive Monte Carlo study based on 20.000 Monte Carlo runs. The setup is the same as in Section 4.2 and the results remain qualitatively the same.

\newpage

% TBL App. 1

\newgeometry{top=0.75cm,bottom=1.25cm,footskip=0cm}

\begin{table}[h!] %\centering

\hspace*{-1.50cm}  \resizebox{1.20\textwidth}{!}{  \begin{threeparttable}
  
\caption{Monte Carlo Results: Coverage and Relative Length I}
\ra{1.1}
\begin{tabular}{@{}lllcclcccl@{}}\toprule
$\mu_{1}(x)$ & & & \multicolumn{3}{c}{$L = 2$} & \phantom{abc}& \multicolumn{3}{c}{$L = 6$} \\
\cmidrule{4-6} \cmidrule{8-10}  
Method & Tuning & \phantom{abc} & Cov. & RL & $h / \kappa$ &&  Cov. & RL & $h / \kappa$ \\ \midrule
\multicolumn{2}{c}{\textit{Cont. Design}}\\
Conv. & $h_{PMSE}$ 				&&  43.4  &  0.567 &  0.225			&& 	18.4  &  0.398 &  0.183 \\
US  & $h_{US}$ 					&&  87.1  &  1.000 &  0.154 		&& 	64.0  &  0.704 &  0.125 \\
RBC & $b_{PMSE},h_{PMSE}$ 		&& 54.3   & 0.850 &  0.225 (0.462)		&& 	38.0  &  0.583 &  0.183 (0.388)\\
RBC & $h=b=h_{PMSE}$ 			&& 95.2  &  2.150 &  0.225			&& 	95.0  &  1.510 &  0.183 \\
RBC & $b_{CE},h_{CE}$ 			&& 84.6  &  1.160 &  0.154 (0.462)			&&	69.6  &  0.808 &  0.125 (0.388)\\
RBC & $h=b=h_{UMSE,L=\hat{L}}$	&& 95.3  &  4.590 &  0.166			&& 	94.7  &  4.030 &  0.104\\
LL-FL & $h_{UMSE,L=\hat{L}}$ 	&& 95.1  &  1.770 &  0.166			&& 	94.9  &  1.560 &  0.104\\
LL-FL & $h_{HL,L=\hat{L}}$ 		&&  96.6 &   1.610 &  0.219			&& 	93.1  &  1.470 &  0.138\\
Opt & $\kappa_{UMSE,L=\hat{L}}$ 	&& 95.0  &  1.750 &  1.000		&& 	95.1  &  1.550 &  1.000\\
Opt & $\kappa_{HL,L=\hat{L}}$		&& 96.6  &  1.650 &  0.246		&& 	93.1  &  1.460 &  0.246\\
\midrule
LL-FL & $h_{UMSE,L=2}$ 		&& 95.1  &  1.060 &  0.187				&& 	21.1  &  0.550 &  0.187\\
LL-FL & $h_{HL,L=2}$ 		&& 96.7  &  1.000 &  0.247				&& 	 0.38  &  0.517 &  0.247\\
Opt & $\kappa_{UMSE,L=2}$ 	&& 95.0  &  1.060 &  1.000				&& 	21.0  &  0.549 &  1.000\\
Opt & $\kappa_{HL,L=2}$ 	&& 96.7  &  1.000 &  0.246				&&  0.34   & 0.517  & 0.246\\
\midrule
LL-FL & $h_{UMSE,L=6}$ 		&& 99.2  &  2.060  & 0.121				&& 	94.9  &  1.060 &  0.121\\
LL-FL & $h_{HL,L=6}$ 		&& 100.0 &   1.940 &  0.160				&& 	94.8  &  1.000 &  0.160\\
Opt & $\kappa_{UMSE,L=6}$ 	&& 99.3  &  2.050  & 1.000				&& 	95.0  &  1.060 &  1.000\\
Opt & $\kappa_{HL,L=6}$ 	&& 100.0 &   1.940 &  0.244				&& 	94.8  &  1.000 &  0.244\\
\midrule
\multicolumn{2}{c}{\textit{Disc. Design }}\\
Conv. & $h_{PMSE}$ 					&& 45.2  &  0.579 &  0.224		&& 	18.0  &  0.404 &  0.183 \\
US  & $h_{US}$ 						&& 87.6  &  1.040 &  0.153 		&& 	65.9  &  0.725 &  0.126 \\
RBC & $b_{PMSE},h_{PMSE}$ 			&& 54.7  &  0.870 &  0.224 (0.438)		&& 	37.5  &  0.594 &  0.183 (0.389) \\
RBC & $h=b=h_{PMSE}$ 				&& 95.4  &  2.320 &  0.224		&& 	95.2  &  1.670 &  0.183 \\
RBC & $b_{CE},h_{CE}$ 				&& 85.1  &  1.210 &  0.153 (0.438)		&&	70.4  &  0.834 &  0.126 (0.389)\\
RBC & $h=b=h_{UMSE,L=\hat{L}}$		&& 95.1  &  4.600 &  0.169		&& 	94.9  &  4.420 &  0.108\\
LL-FL & $h_{UMSE,L=\hat{L}}$ 		&& 95.4  &  1.890 &  0.169		&& 	95.4  &  1.750 &  0.108\\
LL-FL & $h_{HL,L=\hat{L}}$ 			&& 97.2  &  1.790 &  0.223		&& 	93.5  &  1.660 &  0.141\\
Opt & $\kappa_{UMSE,L=\hat{L}}$ 	&& 95.4  &  1.850 &  1.000 		&& 	95.3  &  1.710 &  1.000\\
Opt & $\kappa_{HL,L=\hat{L}}$		&& 97.1  &  1.750 &  0.245		&& 	93.5  &  1.620 &  0.243\\
\midrule
LL-FL & $h_{UMSE,L=2}$ 				&& 95.0  &  1.070 &  0.189		&& 	19.8  &  0.548  & 0.189\\
LL-FL & $h_{HL,L=2}$ 				&& 96.8 &   1.000 &  0.251		&& 	0.28   & 0.515  & 0.251\\
Opt & $\kappa_{UMSE,L=2}$ 			&& 95.1  &  1.060 &  1.000		&& 	20.9  &  0.546 &  1.000\\
Opt & $\kappa_{HL,L=2}$ 			&& 96.9  &  1.000 &  0.242		&& 	0.32   & 0.513 &  0.242\\
\midrule
LL-FL & $h_{UMSE,L=6}$ 		&& 99.4  &  2.100  & 0.125				&& 	94.9  &  1.080 &  0.125\\
LL-FL & $h_{HL,L=6}$ 		&& 100.0  & 1.960  & 0.163				&& 	94.9  &  1.010 &  0.163\\
Opt & $\kappa_{UMSE,L=6}$ 	&& 99.3   &  2.080 &  1.000			&& 	95.0  &  1.070 &  1.000\\
Opt & $\kappa_{HL,L=6}$ 	&& 100.0  &  1.950  & 0.237				&& 	94.8  &  1.000 &  0.237\\
\bottomrule

\end{tabular}
\label{Tab:one2}

\begin{tablenotes}
\small
   \item \textit{Note:} Empirical coverage rate (Cov.), average length relative to optimized interval with true smoothness bound (RL) and tuning parameter choice ($h / \kappa$) of conventional (Conv.), undersmoothed (US), robust bias-corrected (RBC), fixed length (LL-FL) and optimized linear (Opt) 95\% confidence intervals over 20.000 Monte Carlo draws. The pilot RBC bandwidh is reported in parentheses if it is selected separately from $h$
\end{tablenotes}
\end{threeparttable}}
\end{table}

\restoregeometry

%%%% TBL App 2

\newpage

\newgeometry{top=0.75cm,bottom=1.25cm,footskip=0cm}

\begin{table*}[h!]\centering
\hspace*{-1.50cm}  \resizebox{1.20\textwidth}{!}{  \begin{threeparttable}
  
\caption{Monte Carlo Results: Coverage and Relative Length II}
\ra{1.1}
\begin{tabular}{@{}lllcclcccl@{}}\toprule
$\mu_{2}(x)$ & & & \multicolumn{3}{c}{$L = 2$} & \phantom{abc}& \multicolumn{3}{c}{$L = 6$} \\
\cmidrule{4-6} \cmidrule{8-10}  
Method & Tuning & \phantom{abc} & Cov. & RL & $h / \kappa$ &&  Cov. & RL & $h / \kappa$ \\ \midrule
\multicolumn{2}{c}{\textit{Cont. Design}}\\
Conv. & $h_{PMSE}$ 					&&  43.1  &  0.600 &  0.221 		&& 	56.0  &  0.595 & 0.143 \\
US  & $h_{US}$ 						&&  84.6  &  1.060 &  0.151			&& 	80.2  &  1.050 & 0.098 \\
RBC & $b_{PMSE},h_{PMSE}$ 			&& 62.5  &  0.911 &  0.221 (0.439)			&& 	83.2  &  0.781 & 0.143 (0.336)\\
RBC & $h=b=h_{PMSE}$  				&& 95.4  &  2.270 &  0.221			&& 	94.0  &  2.260 & 0.143\\
RBC & $b_{CE},h_{CE}$ 				&& 86.8  &  1.240 &  0.151 (0.439)			&&	86.6  &  1.150 & 0.098 (0.336)\\
RBC & $h=b=h_{UMSE,L=\hat{L}}$		&& 95.1  &  5.710 &  0.134			&& 	94.4  &  4.970 & 0.088\\
LL-FL & $h_{UMSE,L=\hat{L}}$ 		&& 98.3  &  2.200 &  0.134			&& 	98.7  &  1.920 & 0.088\\
LL-FL & $h_{HL,L=\hat{L}}$ 			&& 99.3  &  2.070 &  0.177			&& 	99.6  &  1.180 & 0.115\\
Opt & $\kappa_{UMSE,L=\hat{L}}$ 	&& 98.3  &  2.190 &  1.000			&& 	98.7  &  1.910 & 1.000\\
Opt & $\kappa_{HL,L=\hat{L}}$		&& 99.3  &  2.060 &  0.246			&& 	99.5  &  1.800 & 0.246\\
\midrule
LL-FL & $h_{UMSE,L=2}$ 		&& 95.0  &  1.060 &  0.187				&& 	20.8  &  0.550 & 0.187\\
LL-FL & $h_{HL,L=2}$ 		&& 95.1  &  1.000 &  0.247				&& 	 0.04  &  0.517 & 0.247\\
Opt & $\kappa_{UMSE,L=2}$ 	&& 95.0  &  1.060 &  1.000				&& 	20.7 &   0.549 & 1.000\\
Opt & $\kappa_{HL,L=2}$ 	&& 95.0  &  1.000 &  0.246				&& 	0.05  &  0.517  & 0.246\\
\midrule
LL-FL & $h_{UMSE,L=6}$ 		&& 99.2  &  2.060 &  0.121				&& 	94.9 &   1.060 & 0.121\\
LL-FL & $h_{HL,L=6}$ 		&& 100.0 &   1.940 &  0.160				&& 	94.8 &   1.000 & 0.160\\
Opt & $\kappa_{UMSE,L=6}$ 	&& 99.3  &  2.050 &  1.000				&& 	95.1 &   1.060 & 1.000\\
Opt & $\kappa_{HL,L=6}$ 	&& 100.0 &   1.940 &  0.244				&& 	94.9 &   1.000 & 0.244\\
\midrule
\multicolumn{2}{c}{\textit{Disc. Design }}\\
Conv. & $h_{PMSE}$ 					&& 44.8  &  0.622 &  0.218			&& 	54.1 &   0.608 & 0.143 \\
US  & $h_{US}$ 						&& 84.7  &  1.120 &  0.149 			&& 	81.1 &   1.000 & 0.106\\
RBC & $b_{PMSE},h_{PMSE}$ 			&& 62.3  &  0.941 &  0.218 (0.436)			&& 	83.4  &  0.803 & 0.143 (0.334) \\
RBC & $h=b=h_{PMSE}$ 				&& 95.5  &  2.530 &  0.218 			&& 	94.1 &   2.690 & 0.143 \\
RBC & $b_{CE},h_{CE}$ 				&& 86.0  &  1.300 &  0.149 (0.436)			&&	87.8 &   1.120 & 0.106 (0.334)\\
RBC & $h=b=h_{UMSE,L=\hat{L}}$		&& 95.2  &  6.010 &  0.137			&& 	94.9 &   5.740 & 0.092\\
LL-FL & $h_{UMSE,L=\hat{L}}$ 		&& 98.4  &  2.430 &  0.137			&& 	99.0 &   2.060 & 0.092\\
LL-FL & $h_{HL,L=\hat{L}}$ 			&& 99.4  &  2.310 &  0.180			&& 	99.8 &   1.940 & 0.121\\
Opt & $\kappa_{UMSE,L=\hat{L}}$ 	&& 98.4  &  2.380 &  1.000			&& 	98.9 &   2.010 & 1.000\\
Opt & $\kappa_{HL,L=\hat{L}}$		&& 99.4  &  2.250 &  0.244			&& 	99.8 &   1.890 & 0.237 \\
\midrule
LL-FL & $h_{UMSE,L=2}$ 				&& 94.9  &  1.070 &  0.189			&& 	19.0 &   0.548 & 0.189\\
LL-FL & $h_{HL,L=2}$ 				&& 95.2  &  1.000 &  0.251			&& 	0.03  &  0.515 & 0.251\\
Opt & $\kappa_{UMSE,L=2}$ 			&& 95.0  &  1.060 &  1.000			&& 	20.4 &   0.546 & 1.000\\
Opt & $\kappa_{HL,L=2}$ 			&& 95.2  &  1.000 &  0.242		&& 	0.04 &   0.513 & 0.242\\
\midrule
LL-FL & $h_{UMSE,L=6}$ 		&& 99.4  &  2.100  & 0.125				&& 	94.9  &  1.080 &  0.125\\
LL-FL & $h_{HL,L=6}$ 		&& 100.0 &   1.960 &  0.163				&& 	94.9  &  1.010 & 0.163\\
Opt & $\kappa_{UMSE,L=6}$ 	&& 99.3  &  2.080  & 1.000			&& 	95.0 &   1.070  & 1.000\\
Opt & $\kappa_{HL,L=6}$ 	&& 100.0 &   1.950 &  0.237			&& 	94.9 &   1.000  & 0.237\\
\bottomrule

\end{tabular}
\label{Tab:two2}

\begin{tablenotes}
\small
   \item \textit{Note:} Empirical coverage rate (Cov.), average length relative to optimized interval with true smoothness bound (RL) and tuning parameter choice ($h / \kappa$) of conventional (Conv.), undersmoothed (US), robust bias-corrected (RBC), fixed length (LL-FL) and optimized linear (Opt) 95\% confidence intervals over 20.000 Monte Carlo draws. The pilot RBC bandwidh is reported in parentheses if it is selected separately from $h$
\end{tablenotes}
\end{threeparttable}}
\end{table*}

\restoregeometry

\clearpage
\newpage

\printbibliography

\end{document}